\documentclass[%
 reprint,
superscriptaddress,
 amsmath,amssymb,
 aps,showpacs
]{revtex4-1}
\usepackage{amssymb}
\usepackage{lineno}
\usepackage{hyperref}
\usepackage{siunitx}
\usepackage{multirow}
\usepackage{wasysym}
\usepackage{gensymb}
\usepackage{tabularx}
\usepackage{braket}
\usepackage{url}
\usepackage{todonotes}
\usepackage{comment}
\usepackage{xcolor}
\usepackage{booktabs,multirow}%

\usepackage{graphicx}% Include figure files
\usepackage{dcolumn}% Align table columns on decimal point
\usepackage{bm}% bold math
\begin{document}
%\linenumbers

\preprint{APS/123-QED}

\title{Teleportation Systems Towards a Quantum Internet}

%\thanks{DO NOT DISTRIBUTE/CONFIDENTIAL/DRAFT }%
%%% primaries non alphabetic -- 
\author{Raju Valivarthi}
\author{Samantha Davis}
\affiliation{Division of Physics, Mathematics and Astronomy, California Institute of Technology, Pasadena, CA 91125, USA}
\affiliation{Alliance for Quantum Technologies (AQT), California Institute of Technology, Pasadena, CA 91125, USA}
\author{Cristi\'{a}n Pe\~{n}a}
\affiliation{Division of Physics, Mathematics and Astronomy, California Institute of Technology, Pasadena, CA 91125, USA}
\affiliation{Alliance for Quantum Technologies (AQT), California Institute of Technology, Pasadena, CA 91125, USA}
\affiliation{Fermi National Accelerator Laboratory, Batavia, IL 60510, USA}
\author{Si Xie}
\affiliation{Division of Physics, Mathematics and Astronomy, California Institute of Technology, Pasadena, CA 91125, USA}
\affiliation{Alliance for Quantum Technologies (AQT), California Institute of Technology, Pasadena, CA 91125, USA}
\author{Nikolai Lauk}
\affiliation{Division of Physics, Mathematics and Astronomy, California Institute of Technology, Pasadena, CA 91125, USA}
\affiliation{Alliance for Quantum Technologies (AQT), California Institute of Technology, Pasadena, CA 91125, USA}

\author{Lautaro Narv\'{a}ez}
\affiliation{Division of Physics, Mathematics and Astronomy, California Institute of Technology, Pasadena, CA 91125, USA}
\affiliation{Alliance for Quantum Technologies (AQT), California Institute of Technology, Pasadena, CA 91125, USA}
%%% secondaries, alphabetic
\author{Jason P. Allmaras}
\author{Andrew D. Beyer}
\affiliation{Jet Propulsion Laboratory, California Institute of Technology, Pasadena, CA 91109, USA}
\author{Yewon Gim}
\affiliation{Alliance for Quantum Technologies (AQT), California Institute of Technology, Pasadena, CA 91125, USA}
\affiliation{AT\&T Foundry, Palo Alto, CA 94301, USA}
\author{Meraj Hussein}
\affiliation{Alliance for Quantum Technologies (AQT), California Institute of Technology, Pasadena, CA 91125, USA}
\author{George Iskander}
\affiliation{Division of Physics, Mathematics and Astronomy, California Institute of Technology, Pasadena, CA 91125, USA}
\author{Hyunseong Linus Kim}
\affiliation{Division of Physics, Mathematics and Astronomy, California Institute of Technology, Pasadena, CA 91125, USA}
\affiliation{Alliance for Quantum Technologies (AQT), California Institute of Technology, Pasadena, CA 91125, USA}
\author{Boris Korzh}
\affiliation{Jet Propulsion Laboratory, California Institute of Technology, Pasadena, CA 91109, USA}
\author{Andrew Mueller}
\affiliation{Division of Physics, Mathematics and Astronomy, California Institute of Technology, Pasadena, CA 91125, USA}
\author{Mandy Rominsky}
\affiliation{Fermi National Accelerator Laboratory, Batavia, IL 60510, USA}
\author{Matthew Shaw}
\affiliation{Jet Propulsion Laboratory, California Institute of Technology, Pasadena, CA 91109, USA}
\author{Dawn Tang}
\affiliation{Division of Physics, Mathematics and Astronomy, California Institute of Technology, Pasadena, CA 91125, USA}
\affiliation{Alliance for Quantum Technologies (AQT), California Institute of Technology, Pasadena, CA 91125, USA}
\author{Emma E. Wollman}
\affiliation{Jet Propulsion Laboratory, California Institute of Technology, Pasadena, CA 91109, USA}
%%% tertiary oversight
\author{Christoph Simon}
\affiliation{Institute for Quantum Science and Technology, and Department of Physics \& Astronomy, University of Calgary, Calgary, AB T2N 1N4, Canada}
\author{Panagiotis Spentzouris}
\affiliation{Fermi National Accelerator Laboratory, Batavia, IL 60510, USA}
\author{Neil Sinclair}
\affiliation{Division of Physics, Mathematics and Astronomy, California Institute of Technology, Pasadena, CA 91125, USA}
\affiliation{Alliance for Quantum Technologies (AQT), California Institute of Technology, Pasadena, CA 91125, USA}
\affiliation{John A. Paulson School of Engineering and Applied Sciences, Harvard University, Cambridge, MA 02138, USA}
\author{Daniel Oblak}
\affiliation{Institute for Quantum Science and Technology, and Department of Physics \& Astronomy, University of Calgary, Calgary, AB T2N 1N4, Canada}
%%% last PI's oversight
\author{Maria Spiropulu}
\affiliation{Division of Physics, Mathematics and Astronomy, California Institute of Technology, Pasadena, CA 91125, USA}
\affiliation{Alliance for Quantum Technologies (AQT), California Institute of Technology, Pasadena, CA 91125, USA}
%updated affiliations, formatting was wrong

%Raju Valivarthi$^{1,2}$, Samantha Davis$^{1,2}$, Cristi\'{a}n Pe\~{n}a$^{1,2,3}$, Si Xie$^{1,2}$, Nikolai Lauk$^{1,2}$, Neil Sinclair$^{1,2,4}$, Lautaro Narváez$^{1,2}$,
%%% secondaries, alphabetic
%Jason P. Allmaras$^{5}$, Andrew D. Beyer$^{5}$, Yewon Gim$^{2,6}$, Meraj Hussein$^{2}$, George Iskander$^{1}$,  Hyunseong Linus Kim$^{1,2}$, Boris Korzh$^{5}$, Andrew Mueller$^{1}$, Daniel Oblak$^{7}$, Mandy Rominsky$^{3}$, Matthew Shaw$^{5}$, Christoph Simon$^{7}$, Panagiotis Spentzouris$^{3}$, Dawn Tang$^{1,2}$, Emma E. Wollman$^{5}$,  
%%% tertiary PI's oversight
%and Maria Spiropulu$^{1,2}$
%\affiliation{1. Division of Physics, Mathematics and Astronomy, California Institute of Technology, Pasadena, CA 91125, USA.\\
%2. Alliance for Quantum Technologies (AQT), California Institute of Technology, Pasadena, CA 91125, USA. \\
%3. Fermi National Accelerator Laboratory.\\
%4. John A. Paulson School of Engineering and Applied Sciences, Harvard University, Cambridge, MA 02138, USA.\\
%5. Jet Propulsion Laboratory, California Institute of Technology, Pasadena, CA 91109, USA.\\
%6. AT\&T Palo Alto Foundry, Palo Alto, CA 94301, USA.\\
%7. Institute for Quantum Science and Technology, and Department of Physics \& Astronomy, University of Calgary, Calgary, AB T2N 1N4, Canada.\\
%}

\date{\today}% It is always \today, today,

\begin{abstract}
Quantum teleportation is essential for many quantum information technologies including long-distance quantum networks.  
Using fiber-coupled devices, including  state-of-the-art low-noise superconducting nanowire single photon detectors and off-the-shelf optics, we achieve quantum teleportation of time-bin qubits at the telecommunication wavelength of 1536.5 nm. 
We measure teleportation fidelities of $\geq90$\% that are consistent with an analytical model of our system, which includes realistic imperfections.
To demonstrate the compatibility of our setup with deployed quantum networks, we teleport qubits over 22 km of single-mode fiber while transmitting qubits over an additional 22 km of fiber.
Our systems, which are compatible with emerging solid-state quantum devices, provide a realistic foundation for a high-fidelity quantum internet with practical devices. 
\end{abstract}

\maketitle

\section{Introduction}\label{sec:intro}

Quantum teleportation~\cite{bennett1993teleporting}, one of the most captivating predictions of quantum theory, has been widely investigated since its seminal demonstrations over 20 years ago~\cite{bouwmeester1997experimental,boschi1998experimental,furusawa1998unconditional}.
This is due to its connections to fundamental physics \cite{hensen2015loophole,Shalm2015,Giustina2015,Rosenfeld2017,gao2017traversable,yoshida2019disentangling,landsman2019verified,lloyd2011closed, aspelmeyer2014cavity,hou2016quantum},
%\cite{hensen2015loophole,gao2017traversable,hayden2007black,Kitaev,shenker2014black,maldacena2016bound,yoshida2017efficient,yoshida2019disentangling,landsman2019verified}
and its central role in the realization of quantum information technology such as quantum computers and networks \cite{nielsen2001quantum,ladd2010quantum,gisin2007quantum,pirandola2015advances,briegel1998quantum}. 
The goal of a quantum network is to distribute qubits between different locations, a key task for quantum cryptography, distributed quantum computing and sensing. 
A quantum network is expected to form part of a future quantum internet ~\cite{kimble2008quantum,simon2017towards,wehner2018quantum}: a globally distributed set of quantum processors, sensors, or users there-of that are mutually connected over a network capable of allocating quantum resources (e.g. qubits and entangled states) between locations.
Many architectures for quantum networks require quantum teleportation, such as star-type networks that distribute entanglement from a central location or quantum repeaters that overcome the rate-loss trade-off of direct transmission of qubits~\cite{briegel1998quantum,sangouard2011quantum,dias2017quantum,lucamarini2018overcoming,bhaskar2020experimental}.

Quantum teleportation of a qubit can be achieved by performing a Bell-state measurement (BSM) between the qubit and another that forms one member of an entangled Bell state \cite{bennett1993teleporting,pirandola2015advances,xia2017long}.
The quality of the teleportation is often characterized by the fidelity $F=\bra{\psi} \rho \ket{\psi}$ of the teleported state $\rho$ with respect to the state $\ket{\psi}$ accomplished by ideal generation and teleportation \cite{nielsen2001quantum}.
This metric is becoming increasingly important as quantum networks move beyond specific applications, such as quantum key distribution, and towards the quantum internet.

Qubits encoded by the time-of-arrival of individual photons, i.e. time-bin qubits~\cite{brendel1999pulsed}, are useful for networks due to their simplicity of generation, interfacing with quantum devices, as well as independence of dynamic transformations of real-world fibers. 
Individual telecom-band photons (around 1.5 $\mu$m wavelength) are ideal carriers of qubits in networks due to their ability to rapidly travel over long distances in deployed optical fibers \cite{gisin2007quantum,Valivarthi2016,Sun2016,Takesue2009} or atmospheric channels~\cite{liao2017long}, among other properties.
Moreover, the improvement and growing availability of sources and detectors of individual telecom-band photons has accelerated progress towards workable quantum networks and associated technologies, such as quantum memories~\cite{lvovsky2009optical}, transducers~\cite{lauk2020perspectives,lambert2020coherent}, or quantum non-destructive measurement devices~\cite{braginsky1996quantum}.

Teleportation of telecom-band photonic time-bin qubits has been performed inside and outside the laboratory with impressive results~\cite{marcikic2003long,de2004long, takesue2015quantum,landry2007quantum,halder2007entangling,Sun2016,Valivarthi2016,bussieres2014quantum,Takesue2009}.
Despite this, there has been little work to increase $F$ beyond $\sim90$\% for these qubits, in particular using practical devices that allow straightforward replication and deployment of quantum networks (e.g. using fiber-coupled and commercially available devices).
Moreover, it is desirable to develop  teleportation systems that are forward-compatible with emerging quantum devices for the quantum internet.

In the context of Caltech's  multi-disciplinary multi-institutional collaborative public-private  research program on Intelligent Quantum Networks and Technologies (IN-Q-NET) founded with AT\&T as well as Fermi National Accelerator Laboratory and Jet Propulsion  Laboratory in 2017, we designed, built, commissioned and deployed two quantum teleportation systems: one at Fermilab, the Fermilab Quantum Network (FQNET),  and one at  Caltech's Lauritsen Laboratory for High Energy Physics, the Caltech Quantum Network (CQNET). The CQNET system serves as an R\&D, prototyping, and commissioning system, while FQNET serves as an expandable system, for scaling up to long distances and is used in multiple projects funded currently by DOE's Office of High Energy Physics (HEP) and Advanced Scientific Research Computing (ASCR). Material and devices level R\&D  in both systems is facilitated and funded by the Office of Basic Energy Sciences (BES). Both systems are accessible to quantum researchers for R\&D purposes as well as testing and integration of various novel devices, such as for example on-chip integrated nanophotonic devices and quantum memories,  needed to upgrade such systems towards a realistic quantum internet.Importantly both systems are also used  for  improvements of the entanglement quality and distribution with emphasis on implementation of protocols with complex entangled states towards  advanced and complex quantum communications channels. These will assist in studies of systems that implement new teleportation protocols whose gravitational duals correspond to wormholes \cite{gao2019traversable},  error correlation properties of wormhole teleportation, on-chip codes as well as possible implementation of protocols on quantum optics communication platforms.   Hence the systems serve both fundamental quantum information science as well as quantum  technologies. 

Here we perform quantum teleportation of time-bin qubits at a wavelength of 1536.5 nm with an average $F\geq90$\%. 
This is accomplished using a compact setup of fiber-coupled devices, including low-dark-count single photon detectors and  off-the-shelf optics,  allowing straight-forward reproduction for multi-node networks.
To illustrate network compatibility, teleportation is performed with up to 44 km of single-mode fiber between the qubit generation and the measurement of the teleported qubit, and is facilitated using semi-autonomous control, monitoring, and synchronization systems, with results collected using scalable acquisition hardware.
Our systems, which operates at a clock rate of 90 MHz, can be run remotely for several days without interruption and yield teleportation rates of a few Hz using the full length of fiber.
Our qubits are also compatible with erbium-doped crystals, e.g. Er:Y$_2$SiO$_5$, that are used to develop quantum network devices like memories and transducers~\cite{miyazono2016coupling,lauritzen2010, welinski2019}. 
Finally, we develop an analytical model of our system, which includes experimental imperfections, predicting that the fidelity can be improved further towards unity by well-understood methods (such as improvement in photon indistinguishability).
Our demonstrations provide a step towards a workable quantum network with practical and replicable nodes, such as the ambitious U.S. Department of Energy quantum research network envisioned to link the U.S. National Laboratories.

In the following we describe the components of our systems as well as characterization measurements that support our teleportation results, including the fidelity of our entangled Bell state and Hong-Ou-Mandel (HOM) interference \cite{hong1987oumandel} that underpins the success of the BSM.
We then present our teleportation results using both quantum state tomography (QST) \cite{altepeter2005photonic} and projection measurements based on a decoy state method \cite{ma2005practical}, followed by a discussion of our model.
We conclude  by considering improvements towards near-unit fidelity and GHz level teleportation rates.

\section{Setup}\label{sec:setup}

\begin{figure*}[htbp!]
  \centering
   \includegraphics[width=0.9\linewidth]{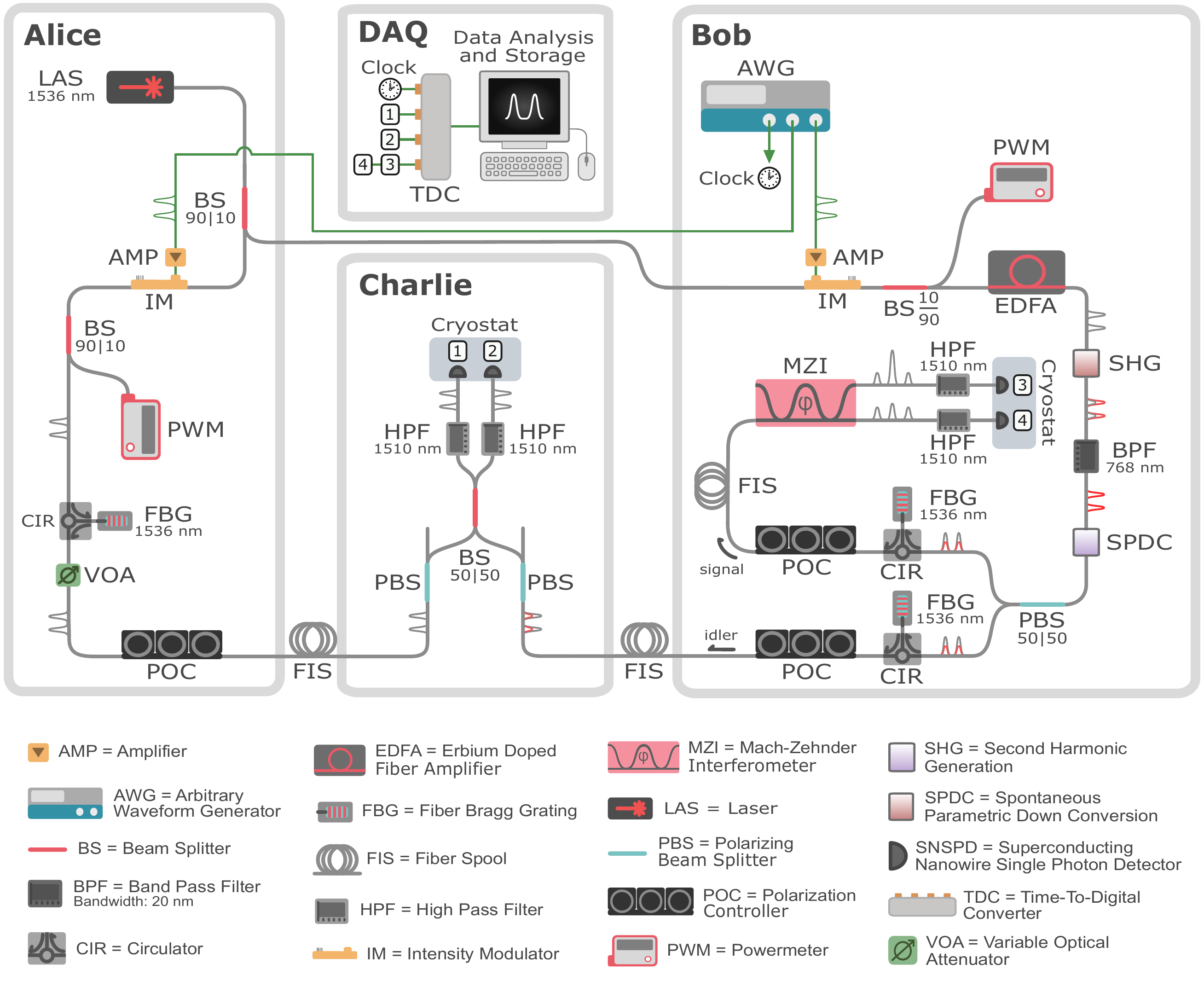}
   \caption{Schematic diagram of the quantum teleportation system consisting of Alice, Bob, Charlie, and the data acquisition (DAQ) subsystems.
   See the main text for descriptions of each subsystem.
   One cryostat is used to house all SNSPDs, it is drawn as two for ease of explanation.
   Detection signals generated by each of the SNSPDs are labelled 1-4 and collected at the TDC, with 3 and 4 being time-multiplexed.
   All individual components are labeled in the legend, with single-mode optical fibers (electronic cables) in grey (green), and with uni- and bi-chromatic (i.e. unfiltered) optical pulses indicated.
   }
   \label{fig:setup}
 \end{figure*}

Our fiber-based experimental system is summarized in the diagram of Fig.~\ref{fig:setup}.
It allow us to demonstrate a quantum teleportation protocol in which a photonic qubit (provided by Alice) is interfered with one member of an entangled photon-pair (from Bob) and projected (by Charlie) onto a Bell-state whereby the state of Alice's qubit can be transferred to the remaining member of Bob's entangled photon pair.
Up to 22 (11) km of single mode fiber is introduced between Alice and Charlie (Bob and Charlie), as well as up to another 11 km at Bob, depending on the experiment (see Sec. \ref{sec:results}).
All qubits are generated at the clock rate, with all of their measurements collected using a data acquisition (DAQ) system.
Each of the Alice, Bob, Charlie subsystems are further detailed in the following subsections, with the DAQ subsystem described in Appendix \ref{subsec:daq_and_sync}.
%fiber lengths AC=22 km, BC=11 km, signal=11 km.

\subsection{Alice: single-qubit generation}\label{subsec:alice}

To generate the time-bin qubit that Alice will teleport to Bob, light from a fiber-coupled 1536.5~nm continuous wave (CW) laser is input into a lithium niobate intensity modulator (IM). 
We drive the IM with one pulse, or two pulses separated by 2~ns.
Each pulse is of $\sim$65~ps full width at half maximum (FWHM) duration. 
The pulses are produced by an arbitrary waveform generator (AWG) and amplified by a 27~dB-gain high-bandwidth amplifier to generate optical pulses that have an extinction ratio of up to 22~dB.
We note that this method of creating time-bin qubits offers us flexibility not only in terms of choosing a suitable time-bin separation, but also for synchronizing qubits originating from different nodes in a network. 
A 90/10 polarization-maintaining fiber beam splitter combined with a power monitor (PWM) is used to apply feedback to the DC-bias port of the IM so as to maintain a constant 22~dB extinction ratio \cite{iskander2019}. 
In order to successfully execute the quantum teleportation protocol, photons from Alice and Bob must be indistinguishable in all degrees of freedom (see Sec. \ref{subsec:HOM}). 
Hence, the optical pulses at the output of the IM are band-pass filtered using a 2~GHz-bandwidth (FWHM) fiber Bragg grating (FBG) centered at 1536.5~nm to match the spectrum of the photons from the entangled pair-source (described in Sec.~\ref{subsec:bob}).
Furthermore, the polarization of Alice's photons is determined by a manual polarization controller (POC) in conjunction with a polarizing beam splitter (PBS) at Charlie.
Finally, the optical pulses from Alice are attenuated to the single photon level by a variable optical attenuator (VOA), to approximate photonic time-bin qubits of the form $\ket{A}=\gamma\ket{e}_{A}+\sqrt{1-\gamma^2}\ket{l}_{A}$, where the late state $\ket{l}_{A}$ arrives 2 ns after the early state $\ket{e}_{A}$, $\gamma$ is real and set to be either 1, 0, or $1/\sqrt{2}$ to generate $\ket{e}_{A}$, $\ket{l}_{A}$, or $\ket{+}_{A}=(\ket{e}_{A}+\ket{l}_{A})/\sqrt{2}$, respectively, depending on the experiment.
The complex relative phase is absorbed into the definition of $\ket{l}_{A}$.
The duration of each time bin is 800~ps.

\subsection{Bob: entangled qubit generation and teleported-qubit measurement}\label{subsec:bob}

Similar to Alice, one (two) optical pulse(s) with a FWHM of $\sim$ 65~ps is (and separated by 2~ns are) created using a 1536.5~nm CW laser in conjunction with a lithium niobate IM driven by an AWG, while the 90/10 beam splitter and PWM are used to maintain an extinction ratio of at least 20~dB. 
An Erbium-Doped Fiber Amplifier (EDFA) is used after the IM to boost the pulse power and thus maintain a high output rate of photon pairs.

The output of the EDFA is sent to a Type-0 periodically poled lithium niobate (PPLN) waveguide for second harmonic generation (SHG), upconverting the pulses to 768.25~nm.
The residual light at 1536.5~nm is removed by a 768~nm band-pass filter with an extinction ratio $\geq$ 80 dB.
These pulses undergo spontaneous parametric down-conversion (SPDC) using a Type-II PPLN waveguide coupled to a polarization-maintaining fiber (PMF), approximately producing either a photon pair $\ket{pair}_B=\ket{ee}_B$, or the time-bin entangled state $\ket{\phi^+}_B=(\ket{ee}_B+\ket{ll}_B)/\sqrt{2}$, if one or two pulses, respectively, are used to drive the IM.

The ordering of the states refers to so-called signal and idler modes of the pair of which the former has parallel, and the latter orthogonal, polarization with respect to the axis of the PMF.
As before, the relative phase is absorbed into the definition of $\ket{ll}_B$. 
Each photon is separated into different fibers using a PBS and spectrally filtered with FBGs akin to that at Alice. Note the bandwidth of the FBG is chosen as a trade-off between spectral purity and generation rate of Bob's photons \cite{rarity1995interference}. 

The photon in the idler mode is sent to Charlie for teleportation or HOM measurements (see Sec. \ref{subsec:HOM}), or to the MZI (see below) for characterizations of the entangled state (see Sec. \ref{subsec:entanglement}), with its polarization determined using a POC.The photon in the signal mode is sent to a Mach Zehnder interferometer (MZI) by way of a POC (and an additional 11 km of single-mode fiber for some measurements), and is detected by superconducting nanowire single photon detectors (SNSPDs) \cite{marsili2013detecting} after high-pass filtering (HPF) to reject any remaining 768.25~nm light.
The MZI and detectors are used for projection measurements of the teleported state, characterization of the time-bin entangled state, or measuring HOM interference at Charlie.
The time-of-arrival of the photons is recorded by the DAQ subsystem using a time-to-digital converter (TDC) referenced to the clock signal from the AWG.

All SNSPDs are installed in a compact sorption fridge cryostat  \cite{photonspot}, which operates at a temperature of 0.8~K for typically 24~h before a required 2~h downtime.
Our SNSPDs are developed at the Jet Propulsion Laboratory and have detection efficiencies between 76 and 85$\%$, with low dark count rates of 2-3 Hz. 
The FWHM temporal resolution of all detectors is between 60 and 90~ps while their recovery time is $\sim$50~ns.
A detailed description of the SNSPDs and associated setup is provided in Appendix ~\ref{subsec:det}.

The MZI has a path length difference of 2~ns and is used to perform projection measurements of $\ket{e}_B$, $\ket{l}_B$, and $(\ket{e}_B+e^{i\varphi} \ket{l}_B)/\sqrt{2}$, by detecting photons at three distinct arrival times in one of the outputs, and varying the relative phase $\varphi$ \cite{brendel1999pulsed}.
Detection at the other output yields the same measurements except with a relative phase of $\varphi+\pi$.
Using a custom temperature-feedback system, we slowly vary $\varphi$ for up to 15 hour time intervals to collect all measurements, which is within the cryostat hold time.
Further details of the MZI setup is described in Appendix \ref{subsec:MZItempstab}.

\subsection{Charlie: Bell-state measurement}\label{subsec:charlie}

Charlie consists of a 50/50 polarization-maintaining fiber beam splitter (BS), with relevant photons from the Alice and Bob subsystems directed to each of its inputs via a PBSs and optical fiber.
The photons are detected at each output with an SNSPD after HPFs, with their arrival times recorded using the DAQ as was done at Bob.
Teleportation is facilitated by measurement of the $\ket{\Psi^-}_{AB}=(\ket{el}_{AB}-\ket{le}_{AB})/\sqrt{2}$ Bell state, which corresponds to the detection of a photon in  $\ket{e}$ at one detector followed by the detection of a photon in $\ket{l}$ at the other detector after Alice and Bob's (indistinguishable) qubits arrive at the BS \cite{lutkenhaus1999bell}.
Projection on the $\ket{\Psi^-}_{AB}$ state corresponds to teleportation of $\ket{A}$ up to a known local unitary transformation, i.e. our system produces $-i \sigma_y \ket{A}$, with $\sigma_y$ being the Pauli $y$-matrix.

\section{Experimental Results} \label{sec:results}

Prior to performing quantum teleportation, we measure some key parameters of our system that underpin the teleportation fidelity.
Specifically, we determine the fidelity of the entangled state produced by Bob by measuring the entanglement visibility $V_{ent}$ \cite{marcikic2002femtosecond}, and also determine to what extent Alice and Bob's photons are indistinguishable at Charlie's BS using the HOM effect \cite{hong1987oumandel}.

\subsection{Entanglement visibility}\label{subsec:entanglement}

The state $\ket{pair}_B$ (and hence the entangled state $\ket{\phi^+}_B$) described in Sec. \ref{subsec:bob} is idealized. 
In reality, the state produced by Bob is better approximated by a two-mode squeezed vacuum state $\ket{\textrm{TMSV}}_B = \sqrt{1-p}\sum_{n = 0}^{\infty} \sqrt{p}^n \ket{nn}_B$ after the FBG filter and neglecting loss \cite{mandel1995optical}. 
Here, $n$ is the number of photons per temporal mode (or qubit), $p$ is the emission probability of a single pair per mode (or qubit), with state
ordering referring to signal and idler modes. 
However, $\ket{\textrm{TMSV}}_B$ approximates a photon pair for $p<<1$, with $p\approx\mu_B$ mean number of pairs per mode (or qubit), conditioned on measurement of a pair such that the $n=0$ term is eliminated.
As a compromise between the pair-creation rate $\propto p$ and the quality of entanglement, here and henceforth we set the mean photon number of our pair source to be $\mu_B = (8.0 \pm 0.4)\times10^{-3}$ per time bin, which is feasible because of the exceptionally low dark counts of our SNSPDs.
Measurement of $\mu_B$ is outlined in Appendix \ref{sec:pairsource}.

We generate $\ket{\phi^+}_B$ and measure $V_{ent}$ by directing the idler photon to the second input port of the MZI, slightly modifying the setup of Fig. \ref{fig:setup}.
The idler photon is delayed compared to the signal, allowing unambiguous measurement of each qubit.
We vary $\varphi$ and project each qubit of the entangled state onto phase-varied superpositions of $\ket{e}$ and $\ket{l}$ by accumulating coincidence events of photons at both the outputs of the interferometer \cite{marcikic2002femtosecond}. 

\begin{figure}[htb]
   \centering
   \includegraphics[width=0.95\columnwidth]{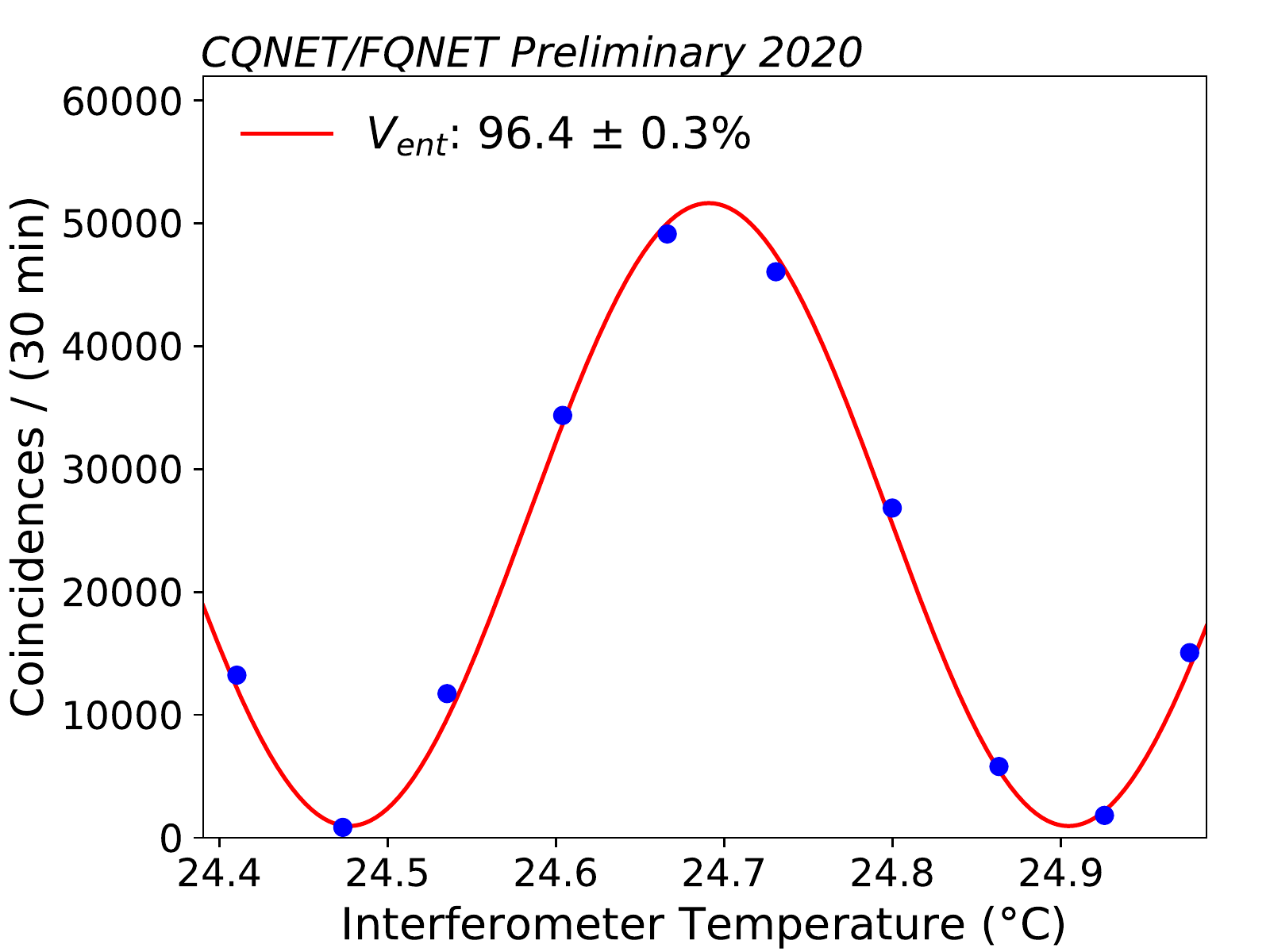}
   \caption{Entanglement visibility. The temperature of the interferometer is varied to reveal the expected sinusoidal variations in the rate of coincidence events. 
   A fit reveals the entanglement visibility $V_{ent}=96.4 \pm 0.3 \%$, see main text for details. Uncertainties here and in all measurements are calculated assuming Poisson statistics.
   }
\label{fig:entanglement_visibility}
\end{figure}

The results shown in Fig.~\ref{fig:entanglement_visibility} are fit proportional to $1 + V_{ent}\sin{(\omega T + \Phi)}$, where $V_{ent}=(R_{x}-R_{n})/(R_{x}+R_{n})$, with $R_{x(n)}$ denoting the maximum (minimum) rate of coincidence events \cite{marcikic2002femtosecond}, $\omega$ and $\Phi$ are unconstrained constants, and $T$ is the temperature of the MZI, finding $V_{ent}=96.4 \pm 0.3 \%$.

The deviation from unit visibility is mainly due to non-zero multi photon emissions~\cite{zhong2013nonlocal}, which is supported by an analytical model that includes experimental imperfections \cite{theory_nikolai}.
Nonetheless, this visibility is far beyond the $1/3$ required for non-separability of a Werner state \cite{werner1989quantum} and the locality bound of $1/\sqrt{2}$ \cite{clauser1969proposed,marcikic2002femtosecond}. Furthermore, it predicts a fidelity $F_{ent}=(3 V_{ent}+1)/4= 97.3 \pm .2$\% with respect to $\ket{\phi^+}$ \cite{werner1989quantum}, and hence is sufficient for quantum teleportation.%\textcolor{red}{SD: calculate with errors carried}.

\subsection{HOM interference visibility}\label{subsec:HOM}

The BSM relies on quantum interference of photons from Alice and Bob.
This is ensured by the BS at Charlie, precise control of the arrival time of photons with IMs, identical FBG filters, and POCs (with PBSs) to provide the required indistinguishabiliy.
The degree of interference is quantified by way of the HOM interference visibility $V_{HOM}=(R_{d}-R_{i})/R_{d}$, with $R_{d(i)}$ denoting the rate of coincident detections of photons after the BS when the photons are rendered as distinguishable (indistinguishable) as possible \cite{hong1987oumandel}.
Completely indistinguishable single photons from Alice and Bob may yield $V_{HOM}=1$. 
However in our system, Alice's qubit is approximated from a coherent state $\ket{\alpha}_A=\mathrm{e}^{-|\alpha| ^2/2}\sum_{n=0}^{\infty}\frac{\alpha^n}{\sqrt{n!}}\ket{n}_A$ with $\alpha<<1$, akin to how Bob's pair is approximated from $\ket{\textrm{TMSV}}_B$ (see Sec. \ref{subsec:entanglement}), with $\mu_A=|\alpha|^2$ being Alice's mean photon number per mode (or qubit) \cite{mandel1995optical}.
Therefore, the contribution of undesired photons from Alice and Bob lowers the maximum achievable $V_{HOM}$ below unity, with a further reduction if the interfering photons are not completely indistinguishable.
The dependence of $V_{HOM}$ with varied $\mu_A$ and $\mu_B$, including effects of losses or distinguishable photons in our system is analytically modelled in Ref. \cite{theory_nikolai}, and briefly discussed in Sec. ~\ref{sec:modeling}.

\begin{figure}[htbp!]
  \centering
    \includegraphics[width=0.48\textwidth]{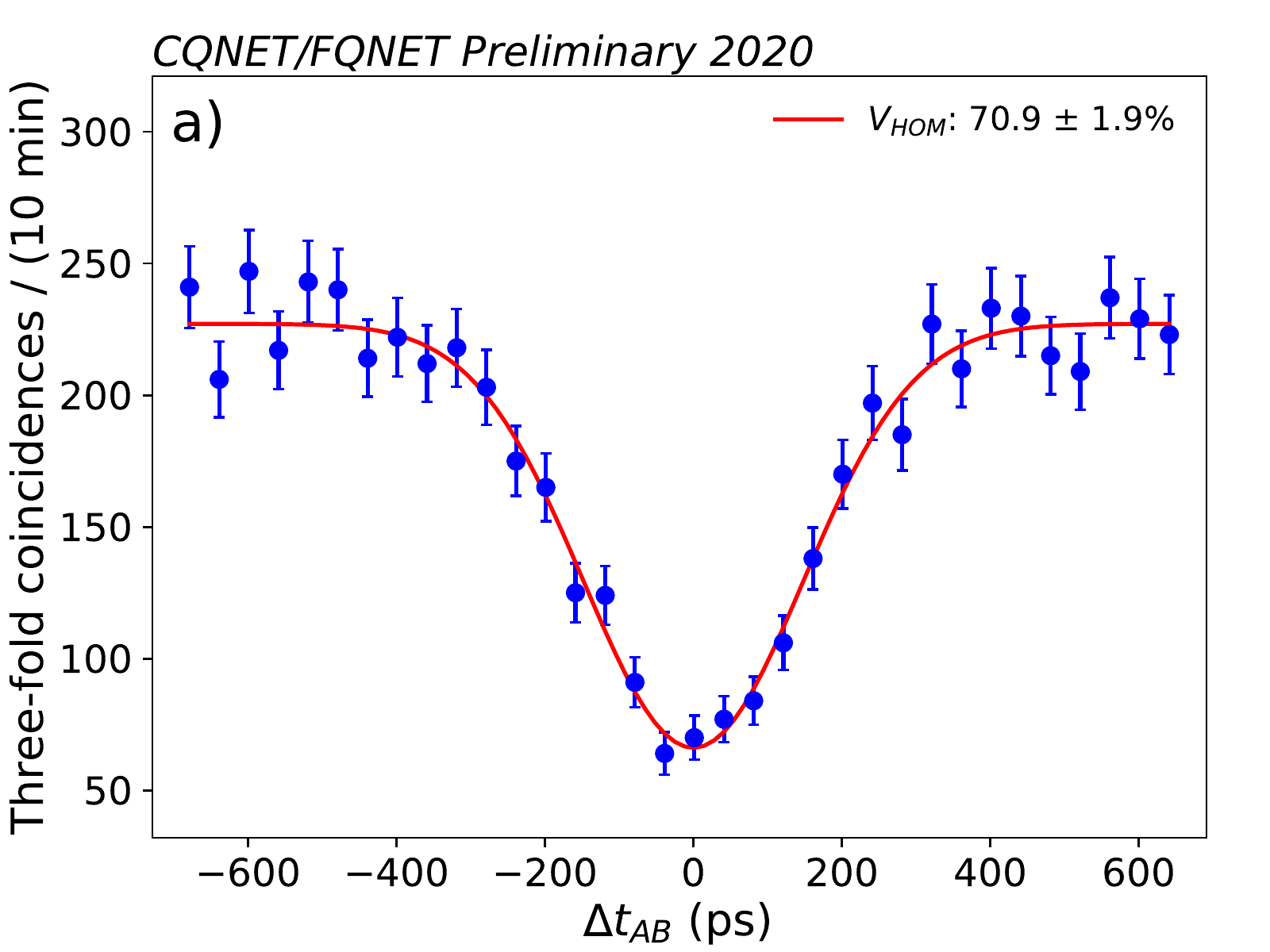}
  \includegraphics[width=0.48\textwidth]{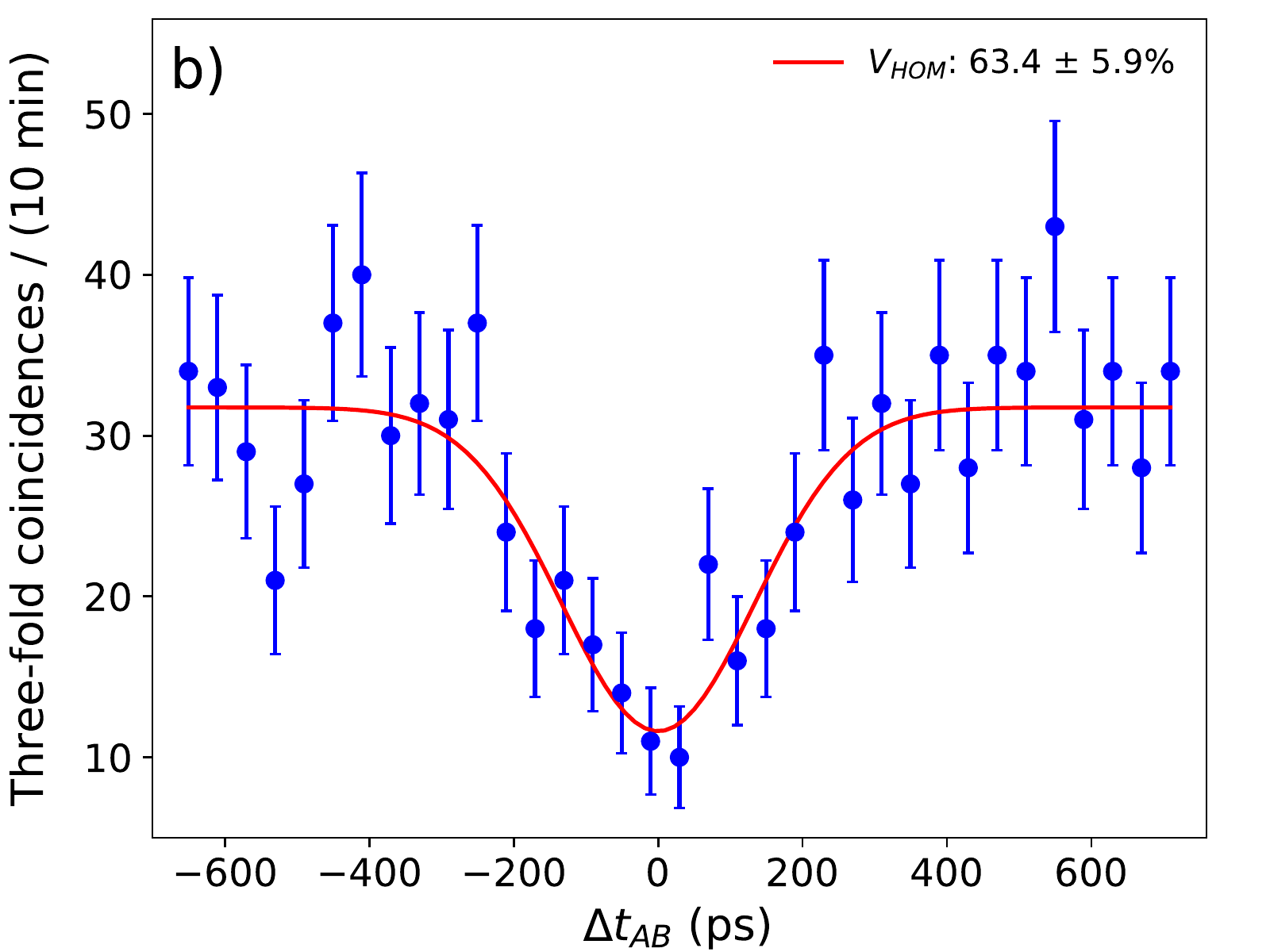}
  \caption{Hong-Ou-Mandel (HOM) interference. 
  A relative difference in arrival time is introduced between photons from Alice and Bob at Charlie's BS.
  HOM interference produces a reduction of the three-fold coincidence detection rate of photons as measured with SNSPDs after Charlie's BS and at Bob.
  A fit reveals a) $V_{HOM}=70.9\pm1.9 \%$ and b) $V_{HOM}=63.4\pm5.9 \%$ when lengths of fiber are added, see main text for details. 
 }
  \label{fig:hom}
\end{figure}

We measure $V_{HOM}$ by slightly modifying the setup of Fig. \ref{fig:setup}:
We prepare  $\ket{A}=\ket{e}_A$ with $\mu_A=2.6 \times 10^{-3}$ and Bob as $\ket{pair}_B$ and direct Alice's photon and Bob's idler to Charlie, with Bob's signal bypassing the MZI to be directly measured by an SNSPD.
Alice's IM is used to introduce distinguishability by way of a relative difference in arrival time $\Delta t_{AB}$ of Alice and Bob's photons at Charlie's BS.
Using Charlie's SNSPDs and the third detector at Bob, a three-fold coincidence detection rate is measured for varying $\Delta t_{AB}$, with results shown in Fig. \ref{fig:hom}a.
Since the temporal profiles of our photons are approximately Gaussian, we fit our results to $A[1-V_{HOM}\exp(-\frac{\Delta t_{AB}^2}{2\sigma^2})]$, 
%https://www.osapublishing.org/oe/abstract.cfm?uri=oe-15-12-7591
%https://doi.org/10.1063/1.2739077
where A is the maximum coincidence rate when the photons are completely distinguishable and $\sigma=300$~ps is the $1/e$ temporal duration of the optical pulses \cite{hong1987oumandel,takesue2007hom}, finding $V_{HOM}=70.9 \pm 1.9\%$.
The maximum $V_{HOM}$ for this experiment is $83.5\%$ if the photons were completely indistinguishable \cite{theory_nikolai}, with the difference ascribed to slight distinguishability between our photons as supported by the further measurements and analytical modelling in Sec. \ref{sec:modeling}.
Improvements to our system to remove this distinguishability is discussed in Sec. \ref{sec:discussion}.

To test our system for quantum teleportation over long distances, we introduce the aforementioned 22, 11, and 11~km lengths of single-mode fiber between Alice and Charlie, Bob and Charlie, and in the path of Bob's signal photon, respectively, repeat our measurement of $V_{HOM}$ and fit the results as before (see Fig. \ref{fig:hom}b).
We find $V_{HOM}=63.4\pm5.9 \%$, which is consistent with the maximum $V_{HOM}$ we expect when including the impact of the additional 5.92 (2.56) dB loss between Charlie and Alice (Bob) as well as the effect of photon distinguishability (analyzed in Sec. \ref{sec:modeling}).
This suggests that the additional fiber importantly does not introduce any further distinguishability (that we cannot account for), thereby supporting our system's use in quantum networking.
Overall, the presence of clear HOM interference suggests our system (with or without the additional fiber) introduces relatively little imperfections that can negatively impact the BSM and hence the fidelity of quantum teleportation.

\subsection{Quantum teleportation}
\label{subsec:teleportation}

We now perform quantum teleportation of the time-bin qubit basis states $\ket{e}$, $\ket{l}$ and $\ket{+}$, so as to measure the teleportation fidelities, $F_e$, $F_l$, an $F_+$, respectively, of the teleported states with respect to their ideal counterparts, up to the local unitary introduced by the BSM (see Sec. \ref{subsec:charlie}).
Since measurement of $\ket{+}$ in our setup by symmetry is equivalent to any state of the form $(\ket{e}+e^{i\varphi}\ket{l})/\sqrt{2}$ (and in particular the remaining three basis states $(\ket{e}-\ket{l})/\sqrt{2}$ and $(\ket{e}\pm i\ket{l})/\sqrt{2}$), we may determine the average teleportation fidelity $F_{avg}=(F_e + F_l + 4F_+)/6$ of any time-bin qubit.

First, we prepare $\ket{e}_A$ and $\ket{l}_A$ with $\mu_A=3.53 \times 10^{-2}$, with Bob's idler bypassing the MZI to be detected by a single SNSPD. 
We measure  $F_e= 95 \pm 1\%$ and $F_l= 96 \pm 1\%$, conditioned on a successful measurement of $|\Psi^-\rangle_{AB}$ at Charlie, with fidelity limited by multiphoton events in Alice and Bob's qubits and dark counts of the SNSPDs~\cite{theory_nikolai}.
We then repeat the measurement with $\mu_A=9.5 \times 10^{-3}$ after inserting the aforementioned 44 km length of fiber as before to emulate Alice, Charlie and parts of Bob being separated by long distances.
This gives $F_e= 98 \pm 1\%$ and $F_l= 98 \pm 2\%$, with no reduction from the additional fiber loss owing to our low noise SNSPDs.

\begin{figure}[htbp!]
  \centering
  %\begin{minipage}[b]{.5\textwidth}
  \includegraphics[width=0.9\columnwidth]{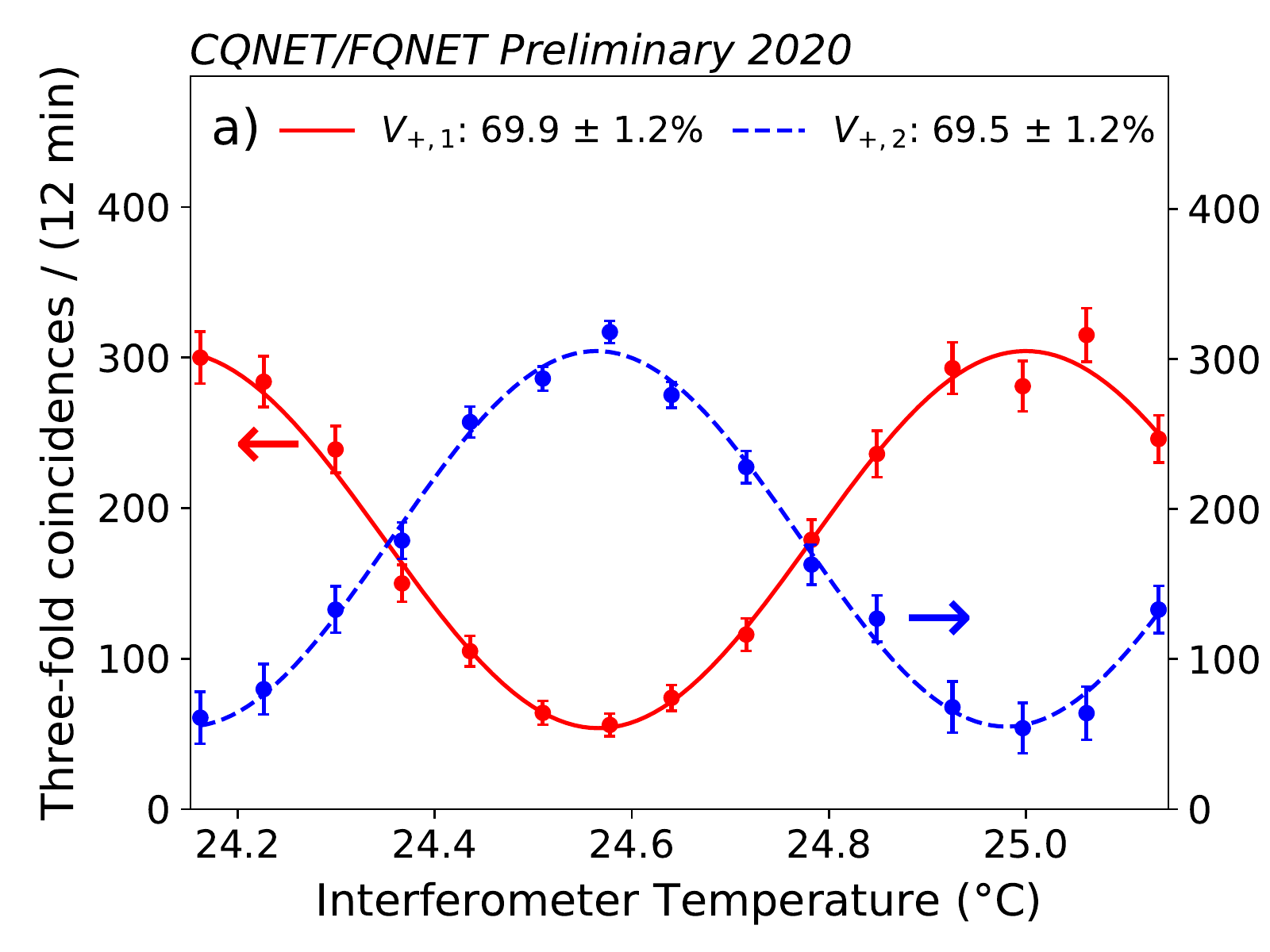}
  %\end{minipage}
  %\begin{minipage}[b]{.5\textwidth}
  \includegraphics[width=0.9\columnwidth]{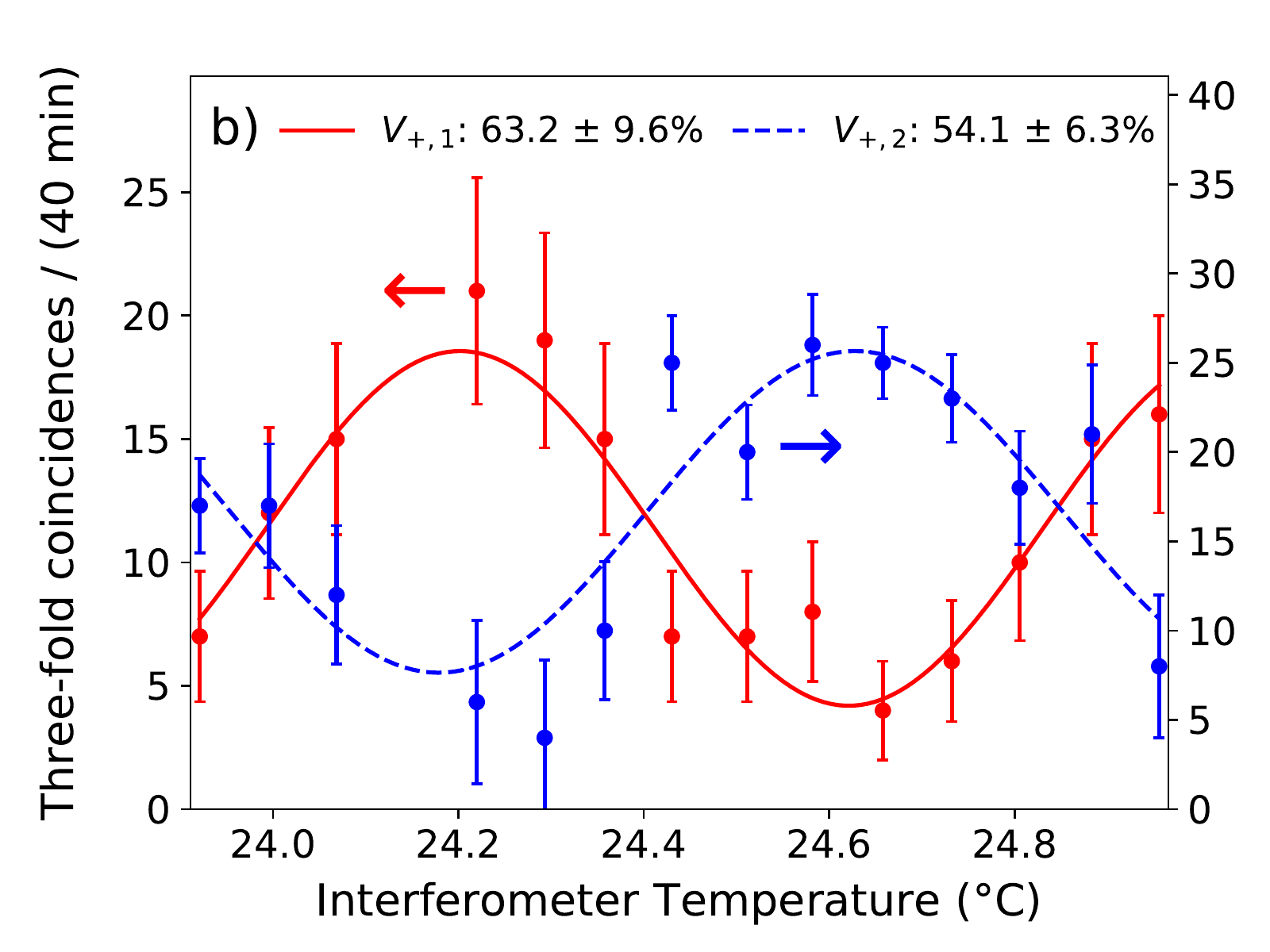}
  %\end{minipage}
  \caption{Quantum teleportation of $\ket{+}$.
  Teleportation is performed b) with and a) without an additional 44 km of single-mode fiber inserted into the system.
  The temperature of the inteferometer is varied to yield a sinusoidal variation of the three-fold coincidence rate at each output of the MZI (blue and red points).
  A fit of the visibilities (see Sec. \ref{subsec:entanglement}) measured at each output ($V_{+,1}$, $V_{+,2})$ of the MZI gives an average visibility $V_+ =(V_{+,1} +V_{+,2})/2$ of a) $69.7 \pm 0.91\%$ without the additional fiber and b) $58.6 \pm 5.7\%$ with the additional fiber.
  }
  \label{fig:x_teleportation_visibility}
\end{figure}

Next, we prepare $\ket{+}_A$ with $\mu_A=9.38 \times 10^{-3}$, insert the MZI and, conditioned on the BSM, we measure $F_+=(1+V_+)/2= 84.9 \pm 0.5\%$ by varying $\varphi$.
Here, $V_+= 69.7 \pm 0.9\%$ is the average visibility obtained by fits to the resultant interference measured at each output of the MZI, as shown in Fig. \ref{fig:x_teleportation_visibility}a.
The reduction in fidelity from unity is due to multiphoton events and distinguishability, consistent with that inferred from HOM interference, as supported by further measurements and analytical modelling in Sec. \ref{sec:modeling}.

The measurement is repeated with the additional long fiber, giving $V_+= 58.6 \pm 5.7\%$ and $F_+=79.3 \pm 2.9\%$ with results and corresponding fit shown in Fig. \ref{fig:x_teleportation_visibility}b.
The reduced fidelity is likely due to aforementioned polarization variations over the long fibers, consistent with the reduction in HOM interference visibility, and exacerbated here owing to the less than ideal visibility of the MZI over long measurement times (see Sec.~\ref{subsec:MZItempstab}).

The results yield $F_{avg}= 89 \pm 1 \%$ $(86 \pm 3\%)$ without (with) the additional fiber, which is significantly above the classical bound of $2/3$, implying strong evidence of quantum teleportation \cite{massar2005optimal}, and limited from unity by multiphotons events, distinguishability, and polarization variations, as mentioned \cite{theory_nikolai}.

\begin{figure}[htbp]
  \centering
    \includegraphics[width=0.95\columnwidth]{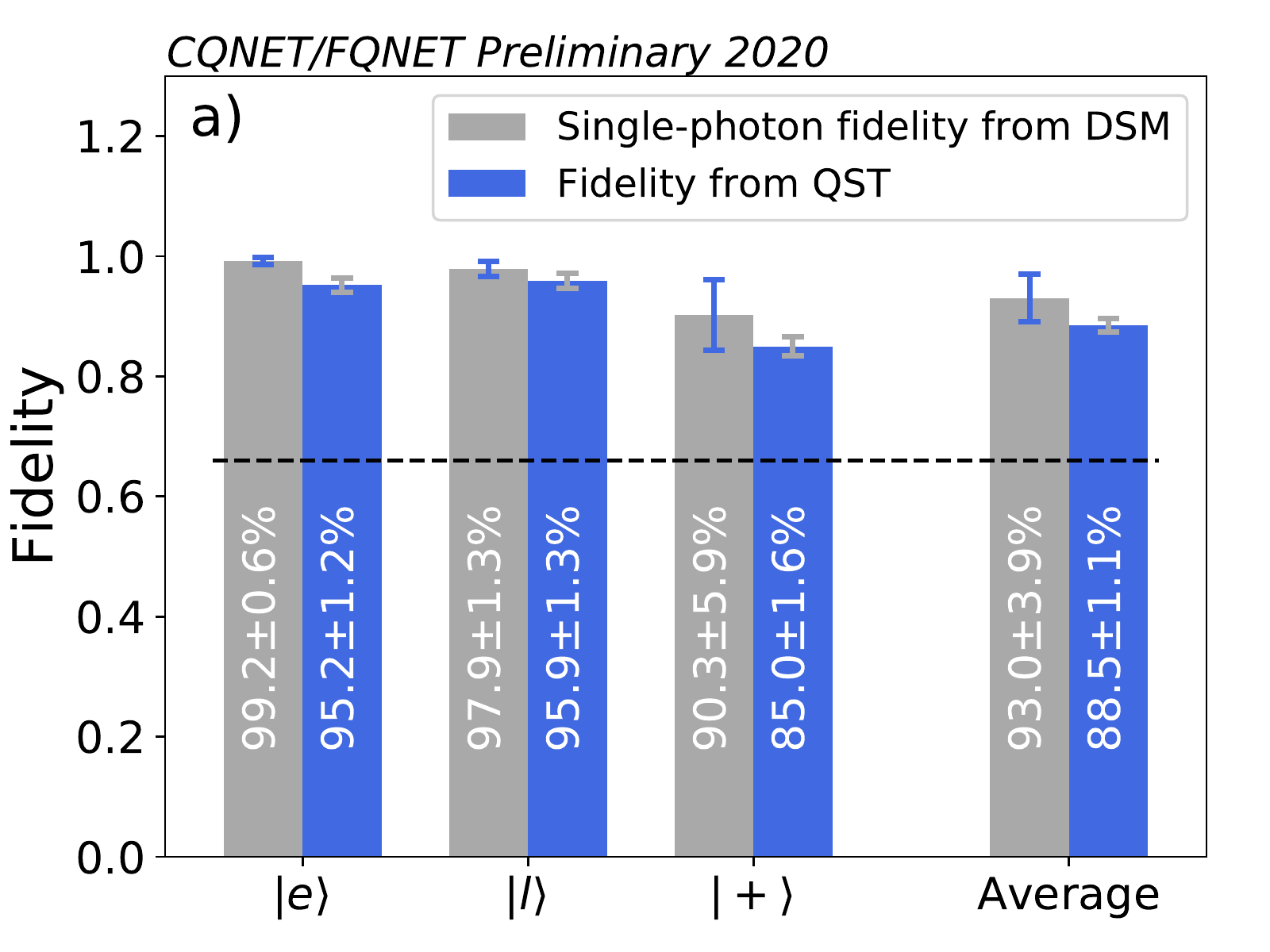}
  \includegraphics[width=0.95\columnwidth]{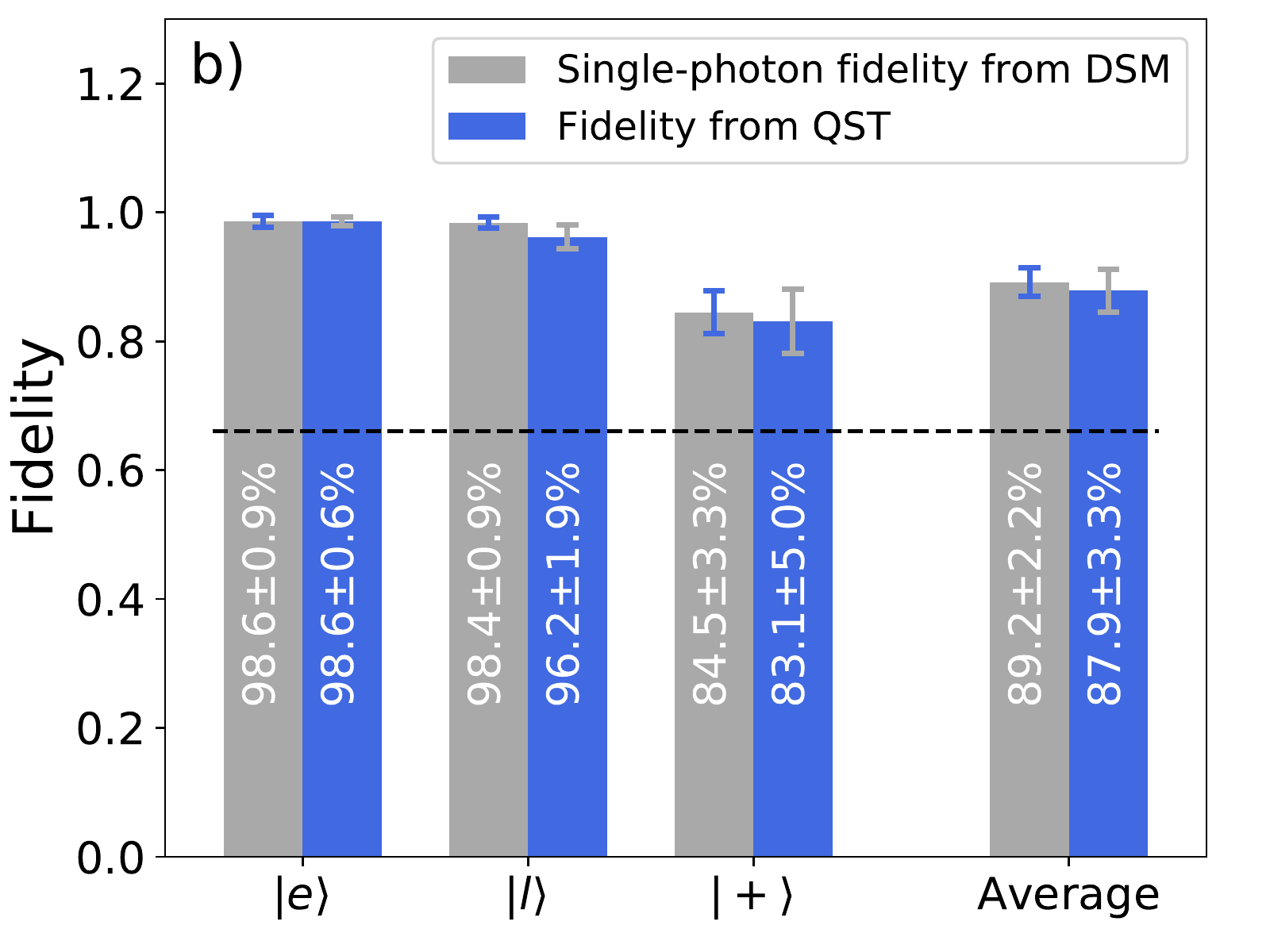}
  \caption{Quantum teleportation fidelities for $\ket{e}_A$, $\ket{l}_A$, and $\ket{+}_A$, including the average fidelity.
  The dashed line represents the classical bound. 
  Fidelities using quantum state tomography (QST) are shown using blue bars while the minimum fidelities for qubits prepared using $\ket{n=1}$, $F^d_e$, $F^d_l$, and $F^d_+$, including the associated average fidelity $F^d_{avg}$, respectively, using a decoy state method (DSM) is shown in grey. 
  Panels a) and b) depicts the results without and with additional fiber, respectively. 
  Uncertainties are calculated using Monte-Carlo simulations with Poissonian statistics.
  }
  \label{fig:summary_fidelity}
\end{figure}

To glean more information about our teleportation system beyond the fidelity, we reconstruct the density matrices of the teleported states using a maximum-likelihood QST~\cite{altepeter2005photonic} described in Appendix \ref{sec:qst}. 
The results of the QST with and without the additional fiber lengths are summarized in Figs. \ref{fig:tomography_withspools} and \ref{fig:tomography_withoutspools}, respectively. 
As can be seen, the diagonal elements for $\ket{+}$ are very close to the expected value indicating the preservation of probabilities for the basis states of $\ket{e}$ and $\ket{l}$ after teleportation, while the deviation of the off-diagonal elements indicate the deterioration of coherence between the basis states. 
The decoherence is attributed to multiphoton emissions from our entangled pair source and distinguishability, consistent with the aforementioned teleportation fidelities of $\ket{+}_A$, and further discussed in Sec. \ref{sec:modeling}.
Finally, we do also extract the teleportation fidelity from these density matrices, finding the results shown in Fig. \ref{fig:summary_fidelity}, and $F_{avg}= 89 \pm 1\%$ $(88 \pm 3\%)$ without (with) the fiber spools, which are consistent with previous measurements given the similar $\mu_A$ used for QST.

We point out that the $2/3$ classical bound may only be applied if Alice prepares her qubits using genuine single photons, i.e. $\ket{n=1}$, rather than using $\ket{\alpha<<1}$ as we do in this work \cite{specht2011single}.
As a way to account for the photon statistics of Alice's qubits we turn to an analysis using decoy states.

\subsubsection{Teleportation fidelity using decoy states}\label{subsubsec:decoy}

To determine the minimum teleportation fidelity of qubits prepared using single photons, we use a decoy state method \cite{ma2005practical} and follow the approach of Refs.  \cite{sinclair2014spectral,Valivarthi2016}.
Decoy states, which are traditionally used in quantum key distribution to defend against photon-number splitting attacks, are qubits encoded into coherent states $\ket{\alpha}$ with varying mean photon number $\mu_A=|\alpha|^2$. 
Measuring fidelities of the teleported qubits for different $\mu_A$, the decoy-state method allows us to calculate a lower bound $F^d_A$ on the teleportation fidelity if Alice had encoded her qubits using $\ket{n=1}$. 

We prepare decoy states $\ket{e}_A$, $\ket{l}_A$, and $\ket{+}_A$ with varying $\mu_A$, as listed in Table~\ref{tab:fid_summary}, and perform quantum teleportation both with and without the added fiber, with teleportation fidelities shown in Table~\ref{tab:fid_summary}.
From these results we calculate $F^d_A$ as shown in Fig.~\ref{fig:summary_fidelity}, with $F^d_{avg}\geq93 \pm4\%$ ($F^d_{avg}\geq 89\pm2\%$) without (with) the added fiber, which significantly violate the classical bound and the bound of $5/6$ given by an optimal symmetric universal cloner \cite{bruss1998optimal,buvzek1996quantum}, clearly demonstrating the capability of our system for high-fidelity teleportation.
As depicted in Fig. \ref{fig:summary_fidelity} these fidelities nearly match the results we obtained without decoy states within statistical uncertainty.
This is due to the suitable $\mu_A$, as well as low $\mu_B$ and SNSPD dark counts in our previous measurements \cite{theory_nikolai}.

\begin{table}[ht]
    %\centering
    \begin{tabular}{|c|cc|cc|}
   % \hline
    \hline
    qubit & \multicolumn{2}{c|}{without long fiber} & \multicolumn{2}{c|}{with long fiber} \\
         & $\mu_A$ $(\times10^{-3})$ & $F^d_A$ (\%) & $\mu_A$ $(\times10^{-3})$ & $F^d_A$ (\%)\\
    \hline
       $\ket{e}_A$ & 3.53 & 95.2 $\pm$ 1  & 26.6 & 95.7 $\pm$ 1.5 \\
                 & 1.24  & 86.7 $\pm$ 2  & 9.01 & 98.4 $\pm$ 1.1 \\
                 & 0       & 52.8 $\pm$ 3.4  & - & - \\
    \hline
           $\ket{l}_A$ & 3.53 & 95.9 $\pm$ 1  & 32.9 & 98.6 $\pm$ 0.7 \\
                 & 1.24  & 90.5 $\pm$ 2  &  9.49 & 98.4 $\pm$ 1.6\\
                 & 0       & 52.8 $\pm$ 3.4  & - & -\\
    \hline
           $\ket{+}_A$ & 9.38  & 84.7 $\pm$ 1.1  & 29.7 & 73.6 $\pm$ 3.0  \\
                 & 2.01 & 83.2 $\pm$ 3.6  & 10.6 & 82.21 $\pm$ 3.9 \\
                 & 0       & 52.8 $\pm$ 3.4  & - & - \\
    \hline
 %   \hline
    \end{tabular}
    \caption{Teleportation fidelities with (right column) and without (center column) the 44 km-length of fiber for Alice's qubit states prepared with varying $\mu_A$.
    Mean photon numbers and fidelities for vacuum states with fiber are assumed to be zero and 50\%, respectively. }
    \label{tab:fid_summary}
\end{table}

\section{Analytical model and simulation}\label{sec:modeling}

As our measurements have suggested, multi-photon components in, and distinguishability between, Alice and Bob's qubits reduce the values of key metrics including HOM interference visibility and, consequently, quantum teleportation fidelity.
To capture these effects in our model, we employ a Gaussian-state characteristic-function method developed in Ref.~\cite{theory_nikolai}, which was enabled by work in Ref.~\cite{Takeoka2015}.
This approach is well-suited to analyze our system because the quantum states, operations, and imperfections (including losses, dark counts, etc.) of the experiment can be fully described using Gaussian operators, see e.g. Ref. ~\cite{Weedbrook2012}. 
We now briefly outline the model of Ref. \cite{theory_nikolai}, and employ it to estimate the amount of indistinguishability $\zeta$ between Alice and Bob's qubits in our measurements of HOM interference and quantum teleportation. 

\begin{figure}[b]
    \centering
    \includegraphics[width=0.48\textwidth]{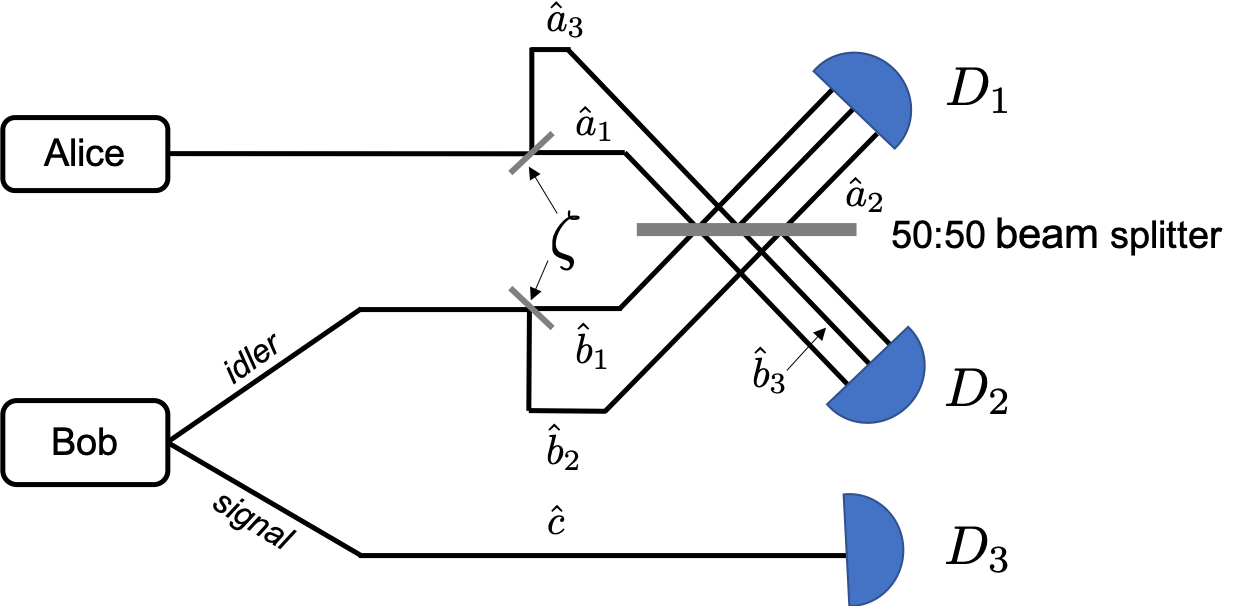}
    \caption{Schematic depiction of distingushability between Alice and Bob's photons at Charlie's BS. 
    Distinguishability is modeled by means of a virtual beam splitter with a transmittance $\zeta$.
    Indistinguishable photons contribute to interference at the Charlie's BS while distinguishable photons are mixed with vacuum, leading to a reduction of HOM visibility and teleportation fidelity.
    See main text for further details.} \label{fig:HOMmodel}
\end{figure}

Distinguishability in any degree-of-freedom may be modelled by introducing a virtual beam splitter of transmittance $\zeta$ into the paths of Alice and Bob's relevant photons.
As shown in Fig. ~\ref{fig:HOMmodel}, indistinguishable components of incoming photon modes are directed towards Charlie's BS where they interfere, whereas distinguishable components are mixed with vacuum at the BS and do not contribute to interference.
Here $\zeta=1$ ($\zeta=0$) corresponds to the case when both incoming photons are perfectly indistinguishable (distinguishable). 
Now we may calculate the probability of a three-fold coincidence detection event $P_{3f}$ between $D_1$, $D_2$ (Charlies' detectors), and $D_3$ (detects Bob's signal photon) for a given qubit state $\rho_{AB}$ from Alice and Bob:
\begin{align}
P_{3f} =&
\ \textrm{Tr} \{\rho_{AB}(\mathbb{I}-(\ket{0}\bra{0})^{\otimes^3}_{\hat{a}_1,\hat{a}_2,\hat{a}_3}) \nonumber\\ &\otimes(\mathbb{I}-(\ket{0}\bra{0})^{\otimes^3}_{\hat{b}_1,\hat{b}_2,\hat{b}_3})\otimes(\mathbb{I}-(\ket{0}\bra{0})_{\hat{c}})\}, 
\end{align}
where the $\hat{a}$ and $\hat{b}$ operators refer to modes, which originate from Alice and Bob's virtual beam splitters and are directed to $D_1$ and $D_2$, respectively, and $\hat{c}$ corresponds to Bob's idler mode, which is directed to $D_3$, see Fig. \ref{fig:HOMmodel}. 
This allows the derivation of an expression for the HOM interference visibility 
\begin{align}
V_{HOM}(\zeta)= [P_{3f}(0)-P_{3f}(\zeta)]/P_{3f}(0),
\end{align}
consistent with that introduced in Sec. \ref{subsec:HOM}.
Since Alice and Bob ideally produce $\rho_{AB} = (\ket{\alpha}\bra{\alpha})\otimes(\ket{\textrm{TMSV}}\bra{\textrm{TMSV}})$, and recognizing that all operators in $P_{3f}$ are Gaussian, we analytically derive 
% \begin{align}
% P_{3f}(\zeta) &= 1-2\frac{\textrm{e}^{-\frac{\mu_A/2[1+(1-\zeta^2)\eta_i\mu_B/2]}{1+\eta_i\mu_B/2}}}{1+\eta_i\mu_B/2}  - \frac{1}{1+\eta_s\mu_B}\nonumber\\
% & + \frac{\textrm{e}^{-\mu_A}}{1+\eta_i\mu_B}- \frac{\textrm{e}^{-\mu_A}}{1+(1-\eta_s)\eta_i\mu_B + \eta_s\mu_B}\nonumber\\
% & +2\frac{\textrm{e}^{-\frac{\mu_A/2[1+ (1-\zeta^2)(1-\eta_s)\eta_i\mu_B/2+\eta_s\mu_B]}{1+(1-\eta_s)\eta_i\mu_B/2+\eta_s\mu_B}}}{1+(1-\eta_s)\eta_i\mu_B/2+\eta_s\mu_B},
% \end{align}
\begin{widetext}
\begin{align}
P_{3f}(\zeta) = 1-2\frac{\exp({-\frac{\mu_A/2[1+(1-\zeta^2)\eta_i\mu_B/2]}{1+\eta_i\mu_B/2}})}{1+\eta_i\mu_B/2} - \frac{1}{1+\eta_s\mu_B}%\nonumber\\
&+ \frac{\exp({-\mu_A})}{1+\eta_i\mu_B}\nonumber\\ -\frac{\exp({-\mu_A})}{1+(1-\eta_s)\eta_i\mu_B + \eta_s\mu_B}
&+2\frac{\exp({-\frac{\mu_A/2[1+ (1-\zeta^2)(1-\eta_s)\eta_i\mu_B/2+\eta_s\mu_B]}{1+(1-\eta_s)\eta_i\mu_B/2+\eta_s\mu_B}})}{1+(1-\eta_s)\eta_i\mu_B/2+\eta_s\mu_B},
\end{align}
\end{widetext}
for varied $\zeta$, where $\eta_i$ and $\eta_s$ are the transmission efficiencies of the signal and idler photons, including detector efficiencies.
We similarly calculate the impact of distinguishability on the teleportation fidelity of $\ket{+}$: 
\begin{align}
F(\zeta)= P_{3f}(\zeta,\varphi_{max})/[P_{3f}(\zeta,\varphi_{max})+P_{3f}(\zeta, \varphi_{min})], 
\end{align}
where $\varphi_{max}$ ($\varphi_{min}$) is the phase of the MZI added into the path of the signal photon, corresponding to maximum (minimum) three-fold detection rates.

\begin{figure}[t]
    \centering
    \includegraphics[width=\columnwidth]{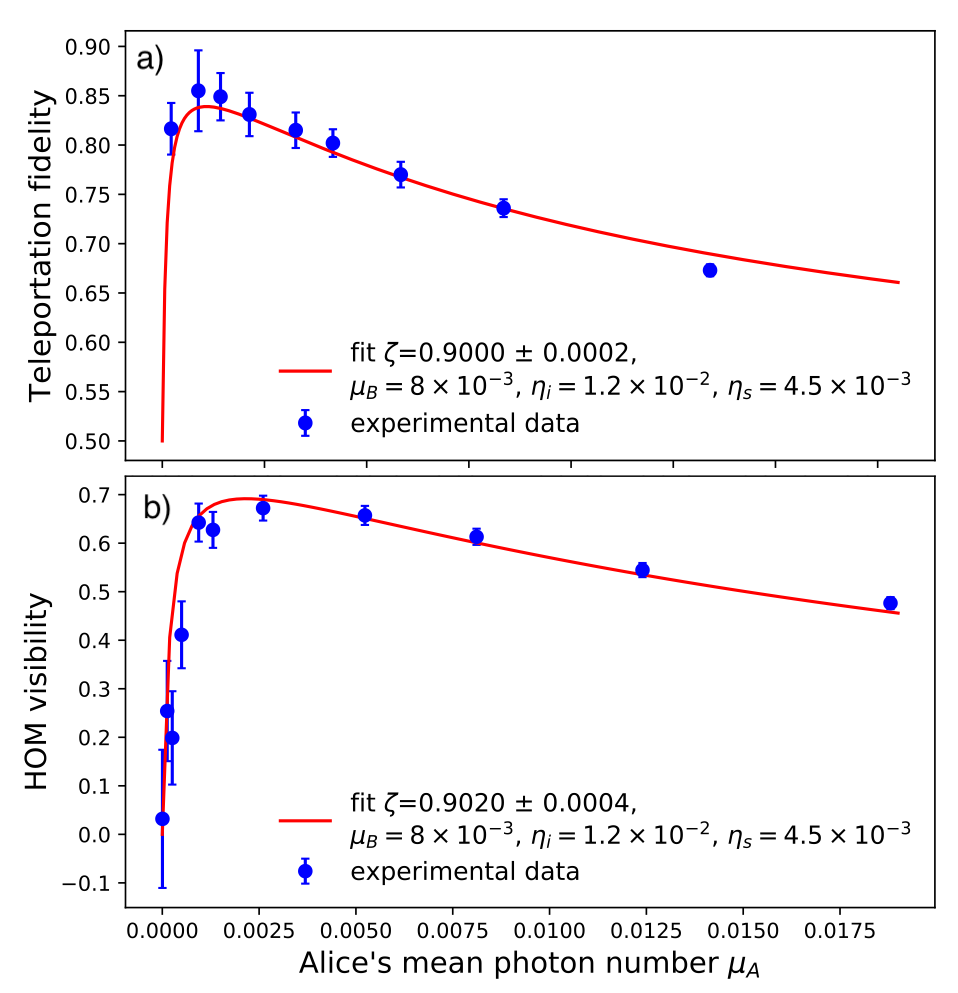}
    \caption{Evaluation of photon indistinguishability using an analytical model. Panel a) depicts the quantum teleportation fidelity of $\ket{+}$ while panel b) shows the HOM interference visibility, each with varied mean photon number $\mu_A$ of Alice's qubits.
    Fits of analytical models the data reveal $\zeta=90\%$ indistinguishability between Alice and Bob's photons at Charlie's BS.
    Bob produces $\mu_B$ photon pairs on average, $\eta_i$ and $\eta_s$ are the probabilities for an individual idler (signal) photon to arrive at Charlie's BS and be detected at Bob's detector, respectively.}
    \label{fig:HOMandFidelity}
\end{figure}

To compare the model to our measurements, we use the experimental mean photon numbers for the photon-pair source $\eta_i=1.2\times10^{-2}$ and $\eta_s=4.5\times10^{-3}$ as determined by the method described in Appendix \ref{sec:pairsource}. 
We then measure the teleportation fidelity of $\ket{+}$ and HOM interference visibility (keeping the MZI in the system to ensure $\eta_s$ remains unchanged) for different values $\mu_A$. 
The results are plotted in Fig.~\ref{fig:HOMandFidelity}. 
The data is then fitted to the expressions $V_{HOM}(\zeta)$ and $F(\zeta)$ derived in our model and graphed in Fig.~\ref{fig:HOMandFidelity}. 
The fitted curves are in very good agreement with our experimental values and consistently yield a value of $\zeta=90\%$ for both measurements types.
This implies that we have only a small amount of residual distinguishability between Alice and Bob's photons.
Potential effects leading to this distinguishability are discussed in Sec. \ref{sec:discussion}.

Overall, our analytic model is consistent with our experimental data \cite{theory_nikolai} in the regime of $\mu_A<<1$, which is the parameter space most often used in quantum networking protocols (e.g. key distribution). 
Our model, thus, offers a practical way to determine any underlying distinguishability in a deployed network where a full characterization of the properties of Alice and Bob's photons may not be possible.

\section{Discussion and Outlook}\label{sec:discussion}
%NS: writing this section without many abbreviations for casual reader

We have demonstrated  quantum teleportation systems for photonic time-bin qubits at a quantum channel- and device-compatible wavelength of 1536.5~nm using a fiber-based setup comprising state-of-the-art SNSPDs and off-the-shelf components. We measure an average fidelity above 90\% using QST and a decoy state analysis with up to 44~km of single mode fiber in our setup.
Our results are further supported by an analytical model in conjunction with measurements of the entanglement and HOM interference visibilities.

The decoy state analysis indicates that the maximum teleportation fidelity is currently restricted by that of the teleported qubits prepared in superposition states, for which a $10\%$ distinguishability between the qubits undergoing BSM and the contribution of multiple photon pairs play the largest role.
Our model predicts that the average fidelity will increase to 95\% with completely indistinguishable photons, while fidelities close to unity can be achieved with lowered mean number of photon pairs.
Alternatively, we may replace our SNSPDs with efficient photon-number resolving (PNR) SNSPDs~\cite{PMID:32271591} to allow postselection of multiphoton events at the MZI or BSM \cite{krovi2016practical}.
Both approaches must be accompanied by increased coupling efficiencies of the pair source beyond the current $\sim1\%$ either to maintain realistic teleportation rates (above the system noise and current rate of phase drift of the MZI), or to derive any advantage from PNR capability.
%Ways to improve the indistinguishability and the coupling efficiencies are briefly discussed below.

As suggested by the width of our HOM interference fringe -- which predicts an average photon bandwidth of $0.44/\sigma\sim$1.5 GHz (see Sec. \ref{subsec:HOM}), i.e. less than the 2~GHz bandwidth of our FBGs -- the indistinguishability in our system could be limited by the large difference in the bandwidth between the photons originating from the SPDC ($>$100 GHz) and those generated at Alice by the IM (15 GHz), leading to nonidentical filtering by the FBG.
This can be improved by narrower FBGs or by using a more broadband pump at Alice (e.g. using a mode locked laser or a higher bandwidth IM, e.g $>$ 50 GHz, which is commercially available).
Alternatively, pure photon pairs may be generated by engineered phase matching, see e.g. Ref. \cite{mosley2008heralded}. 
Distinguishability owing to nonlinear modulation during the SHG process could also play a role~\cite{rassoul1997second}. 
The origin of distinguishability in our system, whether due to imperfect filtering or other device imperfections (e.g. PBS or BS) will be studied in future work.
Coupling loss can be minimized to less than a few dB overall by improved fiber-to-chip coupling, lower-loss components of the FBGs (e.g. the required isolator), spliced fiber connections, and reduced losses within our MZI.
Note that our current coupling efficiency is equivalent to $\sim$50~km of single mode fiber, suggesting that our system is well-suited for quantum networks provided loss is reduced.

While the fidelities we demonstrate are sufficient for several applications, the current $\sim$Hz teleportation rates with the 44 km length of fiber are still low.
Higher repetition rates (e.g. using high-bandwidth modulators with wide-band wavelength division multiplexed filters and low-jitter SNSPDs~\cite{korzh2020demonstration}), improvements to coupling and detector efficiencies, enhanced BSM efficiency with fast-recovery SNSPDs \cite{valivarthi2014efficient}, or multiplexing in frequency~\cite{sinclair2014spectral} will all yield substantial increases in teleportation rate.
Note that increased repetition rates permits a reduction in time bin separation which will allow constructing the MZI on chip, providing exceptional phase stability and hence, achievable fidelity.
Importantly, the aforementioned increases in repetition rate and efficiency are afforded by improvements in SNSPD technology that are currently being pursued with our JPL, NIST and other academic partners.

Upcoming system-level improvements we plan to investigate and implement include further automation by the implementation of free-running temporal and polarization feedback schemes to render the photons indistinguishable at the BSM~\cite{Valivarthi2016,Sun2016}. Furthermore, several electrical components can be miniaturized, scaled, and made more cost effective (e.g. field-programmable gate arrays can replace the AWG).We note that our setup prototype  will be easily extended to independent lasers at different locations, also with appropriate feedback mechanisms for spectral overlap \cite{valivarthi2017cost,sun2017entanglement}.  These planned improvements are compatible with the data acquisition and control systems that were built for the systems and experiments at FQNET and CQNET presented in this work.

Overall, our high-fidelity teleportation systems — achieving state-of-the-art teleporation fidelities of time-bin qubits— serve as a blueprint for the construction of quantum network test-beds and eventually global quantum networks towards the quantum internet. In this work, we present a complete analytical model of the teleporation system that includes imperfections, and compare it with our measurements. Our implementation, using approaches from High Energy Physics experimental systems and real-world quantum networking, features near fully-automated data acquisition, monitoring, and real-time data analysis. In this regard our Fermilab and Caltech Quantum Networks  serve as R\& D laboratories and prototypes towards real-world quantum networks.
%Overall, our high-fidelity teleportation systems -- achieving state-of-the-art teleporation fidelities in time-bin qubits and record uptime operations, serve as a blueprint for the construction of quantum network test-beds and eventually global quantum networks towards the quantum internet. In this work, we present for the first time a complete analytical model with imperfections  of the full teleporation system and comparisons with the data. Moreover, our implementation, inspired by High Energy Physics experimental systems, features near fully-automated data acquisition, monitoring, and real-time data analysis. In this regard our Fermilab and Caltech Quantum Networks  serve as R\& D laboratories and prototypes towards real-world quantum networks. %s including the Illinois-Express Quantum Network~\cite{IEQNET}.
The high fidelities achieved in our experiments using practical and replicable devices are essential when expanding a quantum network to many nodes, and  enable the realization of  more advanced protocols, e.g.~\cite{pirandola2015advances,braunstein2012side,gottesman1999demonstrating}.

\section*{Acknowledgements}\label{sec:ack}
%\todo[inline]{write this section in as compact way as possible, make sure names are ordered according to author list}
R.V., N.L., L.N., C.P., N.S., M.S. and S.X. acknowledge partial and S.D. full support from the Alliance for Quantum Technologies’ (AQT) Intelligent Quantum Networks and Technologies  (IN-Q-NET) research program.
R.V., N.L., L.N., C.P., N.S., M.S.  S.X. and A.M.  acknowledge partial support from the U.S. Department of Energy, Office of Science, High Energy Physics, QuantISED program grant, under award number DE-SC0019219. A.M. is supported  in part by the JPL President and Director’s Research and Development Fund (PDRDF). C.P. further acknowledges partial support from the Fermilab's Lederman Fellowship and LDRD. D.O. and N.S. acknowledge partial support from the Natural Sciences and Research Council of Canada (NSERC). D.O. further acknowledges the Canadian Foundation for Innovation, Alberta Innovates, and Alberta Economic Development, Trade and Tourism’s Major Innovation Fund. J.A. acknowledges support by a NASA Space Technology Research Fellowship.   Part of the research was carried out at the Jet Propulsion Laboratory, California Institute of Technology, under a contract with the National Aeronautics and Space Administration (80NM0018D0004).
We thank   Jason Trevor (Caltech Lauritsen Laboratory for High Energy Physics),  Nigel Lockyer and Joseph Lykken (Fermilab), Vikas Anant (PhotonSpot), Aaron Miller (Quantum Opus), Inder Monga and his ESNET group at LBNL, the groups of Wolfgang Tittel and Christoph Simon at the University of Calgary, the groups of Nick Hutzler, Oskar Painter, Andrei Faraon, Manuel Enders and Alireza Marandi at Caltech, Marko Loncar's group at Harvard, Artur Apresyan and the HL-LHC USCMS-MTD Fermilab group; Marco Colangelo (MIT); Tian Zhong (Chicago); AT\&T's Soren Telfer, Rishi Pravahan, Igal Elbaz, Andre Feutch and John Donovan.  We acknowledge the enthusiastic support of the Kavli Foundation on funding QIS\&T workshops and events and the Brinson Foundation support especially for students working at FQNET and CQNET. 
M.S. is especially grateful to  Norm Augustine (Lockheed Martin), Carl Williams (NIST) and Joe Broz (SRI, QED-C); Hartmut Neven (Google Venice); Amir Yacoby and Misha Lukin (Harvard);  Ned Allen (Lockheed Martin);  Larry James and  Ed Chow (JPL);  the QCCFP wormhole teleportation team especially Daniel Jafferis (Harvard)  and Alex Zlokapa (Caltech),  Mark Kasevich (Stanford), Ronald Walsworth (Maryland), Jun Yeh and Sae Woo Nam (NIST);  Irfan Siddiqi (Berkeley); Prem Kumar (Northwestern), Saikat Guha (Arizona), Paul Kwiat (UIUC), Mark Saffman (Wisconcin), Jelena Vuckovic (Stanford) Jack Hidary (X),  and the quantum networking teams at ORNL, ANL, and BNL, for productive  discussions and interactions on quantum networks and communications. 

\appendix

%\todo[inline]{check consistency of acronym use and formatting of references}

\section{Detailed description of experimental components}\label{sec:experimentaldetails}

\subsection{Control systems and data acquisition}\label{subsec:daq_and_sync}
%NS modified to be consistent with main text

Our system is built with a vision towards future replicability, with particular emphasis on systems integration. %'scalability' is often reserved only for on-chip applications
Each of the Alice, Bob and Charlie subsystems is equipped with monitoring and active feedback stabilization systems (e.g. for IM extinction ratio), or has capability for remote control of critical network parameters (e.g. varying the qubit generation time).
Each subsystem has a central classical processing unit with the following functions: oversight of automated functions and workflows within the subsystem, data acquisition and management, and handling of input and output synchronization streams. As the quantum information is encoded in the time domain  the correct operation of the classical processing unit depends critically on the recorded time-of-arrival of the photons at the SNSPDs.
Thus significant effort was dedicated to build a robust DAQ subsystem capable of recording and processing large volumes of time-tagged signals from the SNSPDs and recorded by our TDCs at a high rate.
The DAQ is designed to enable both real-time data analysis for prompt data quality monitoring as well as post-processing data analysis that allows to achieve the best understanding of the data. 

The DAQ system is built on top of the standalone Linux library of our commercial  TDC. 
It records time tags whenever a signal is detected in any channel in coincidence with the reference 90 MHz clock. 
Time tags are streamed to a PC where they are processed in real-time and stored to disk for future analysis. 
A graphical user interface has been developed, capable of real-time visualization and monitoring of photons detected while executing teleportation. 
It also allows for easy control of the time-intervals used for each channel and to configure relevant coincidences between different photon detection events across all TDC channels.
%An automated control system monitors the operation of the DAQ subsystem and performs all relevant measurements including teleportation and HOM measurements. %this sentence has already been stated in main text and earlier in this section.
We expect our DAQ subsystem to serve as the foundation for future real-world time-bin quantum networking experiments (see Sec. \ref{sec:discussion}).

\subsection{Superconducting nanowire single photon detectors}\label{subsec:det}

We employ amorphous tungsten silicide SNSPDs manufactured in the JPL Microdevices Laboratory for all measurements at the single photon level (see Sec. \ref{subsec:bob}) ~\cite{marsili2013detecting}. The entire detection system is customized for optimum autonomous operation in a quantum network. The SNSPDs are operated at 0.8~K in a closed-cycle sorption fridge \cite{photonspot}.
The detectors have nanowire widths between 140 to 160~nm and are biased at a current current of 8 to 9~$\mu$A. The full-width at half maximum (FWHM) timing jitter (i.e. temporal resolution) for all detectors is between 60 and 90~ps (measured using a Becker \& Hickl SPC-150NXX time-tagging module). The system detection efficiencies (as measured from the fiber bulkhead of the cryostat) are between 76 and 85~$\%$.
The SNSPDs feature low dark count rates between 2 and 3 Hz, achieved by short-pass filtering of background black-body radiation through coiling of optical fiber to a 3~cm diameter within the 40~K cryogenic environment, and an additional band-pass filter coating deposited on the detector fiber pigtails (by Andover Corporation).
Biasing of the SNSPDs is facilitated by cryogenic bias-Ts with inductive shunts to prevent latching, thus enabling uninterrupted operation. 
The detection signals are amplified using Mini-Circuits ZX60-P103LN+ and ZFL-1000LN+ amplifiers at room temperature, achieving a total noise figure of 0.61~dB and gain of 39~dB at 1~GHz, which enables the low system jitter. Note that FWHM jitter as low as 45~ps is achievable with the system, by biasing the detectors at approximately 10~$\mu$A, at the cost of an elevated DCR on the order of 30~cps. Using commercially available components, the system is readily scalable to as many as 64 channels per cryostat, ideal for star-type quantum networks, with uninterrupted 24/7 operation. The bulkiest component of the current system is an external helium compressor, however, compact rack-mountable versions are readily available \cite{photonspot}.

\subsection{Interferometer and phase stabilization}\label{subsec:MZItempstab}

We use a commercial Kylia 04906-MINT MZI, which is constructed of free-space devices (e.g mirrors, beam spliters) with small form-factor that fits into a hand-held box.
Light is coupled into and out of the MZI using polarization maintaining fiber with loss of $\sim$2.5 dB. 
The interferometer features an average visibility of 98.5\% that was determined by directing $\ket{+}$ with $\mu_A=0.07$ into one of the input ports, measuring the fringe visibility on each of the outputs using an SNSPD. 
The relative phase $\varphi$ is controlled by a voltage-driven heater that introduces a small change in refractive index in one arm of the MZI.
However, this built-in heater did not permit phase stability sufficient to measure high-fidelity teleportation, with the relative phase following the slowly-varying ambient temperature of the room.
To mitigate this instability, we built another casing, thermally isolating the MZI enclosure from the laboratory environment and controlled the temperature via a closed-loop feedback control system based on a commercial thermoelectric cooler and a LTC1923 PID-controller.
The temperature feedback is provided by a $10$~k$\Omega$ NTC thermistor while the set-point is applied with a programmable power supply. 
This control system permits us to measure visbilities by slowly varying $\varphi$ over up to 15 hour timescales.
We remark that no additional methods of phase control were used beyond that of temperature.
% [sixie: this needs to be measured at the end, when we are done with teleporation measurements]

\section{Estimation of mean number of photon pairs and transmission efficiencies of signal and idler photons}\label{sec:pairsource}

Using a method described in Ref. \cite{marcikic2002femtosecond}, we measure the mean number of photon pairs produced by Bob $\mu_B$ as a function of laser excitation power before the PPLN waveguide used for SHG.
To this end, we modify the setup of Fig. \ref{fig:setup} and direct each of Bob's signal and idler photons to a SNSPD.
We then measure detection events while varying the amplification of our EDFA by way of an applied current.
We extract events when photon pairs which originated from the same clock cycle are measured in coincidence, and when one photon originating from a cycle is measured in coincidence with a photons originated from a preceding or following clock cycle, in other words we measure the so-called coincidence and accidental rates.
The ratio of accidentals to coincidences approximates $\mu_B<<1$, with all results shown in Table \ref{tab:car}.
For all measurements we use $\mu_B = (8.0 \pm 0.4)\times10^{-3}$ per time bin, which corresponds to an EDFA current of 600 mA.

The transmission efficiencies of the signal and idler photons, $\eta_s$ and $\eta_i$, respectively, mentioned in Sec. \ref{sec:modeling} are determined by calculating the ratio of the independent rates of detection of the idler and signal photons, respectively, with the coincidence rate of the photons pairs (in the same clock cycle) \cite{marcikic2002femtosecond}.
We repeat the measurements using the setup shown in Fig. \ref{fig:setup}, which is that used to generate the results of Fig. \ref{fig:HOMandFidelity} (i.e. we direct the signal and idler photons through the setup as if we are to perform teleportation).
We find $\eta_s=4.5\times10^{-3}$ and $\eta_i=1.2\times10^{-2}$, which take into account losses between when the photons are produced to when they are detected by their respective SNSPDs.

\begin{table}[ht]
\centering
\begin{tabular}[t]{ccccccccc}
\hline
 EDFA current &{}& Coincidences &{}& Accidentals &{}& $\mu_B$  \\
 (mA) &{}& (per 10 s)  &{}& (per 10 s) &{}& ($\times 10^{-3}$) \\
 \hline
 400 &{}& 469.2 $\pm$ 3.6 &{}& 1.8 $\pm$ 0.3 &{}& 3.9 $\pm$ 0.7\\
 450 &{}& 1156.3 $\pm$ 4.9 &{}& 6.1 $\pm$ 0.5 &{}& 5.3 $\pm$ 0.4\\
 500 &{}& 1653.9 $\pm$ 5.9 &{}& 9.5 $\pm$ 0.6 &{}& 5.8 $\pm$ 0.4\\
 550 &{}& 2095.8 $\pm$ 6.6 &{}& 13.7 $\pm$ 0.8 &{}& 6.5 $\pm$ 0.4\\
 575 &{}& 2343.2 $\pm$ 7.0 &{}& 17.7 $\pm$ 0.9 &{}& 7.5 $\pm$ 0.4\\
 600 &{}& 2548.7 $\pm$ 7.3 &{}& 18.5 $\pm$ 0.9 &{}& 8.0 $\pm$ 0.4\\
\hline
\end{tabular}
\caption{Bob's photon pair source is characterized by the measured mean photon number per time bin $\mu_B$, and the rate of accidental and true coincidence detections with varied EDFA current.}
\label{tab:car}
\end{table}

%%%%%%%%%%%%%%commented%%%%%%%%%%
\begin{comment}
\begin{table}[ht]
\centering
\caption{entangled source  characterization. Repetition rate is 90 MHz}
\begin{tabular}[t]{cccccc}
\hline
\hline
 EDFA current [mA] & $\ket{e}$ counts & $\ket{\ell}$ counts & noise counts & qubit $\mu$ & extinction ratio (db) \\ 
 \multirow{2}{*}{597 } &18300 & 17750 & 75 & \multirow{2}{*}{0.0009} & \multirow{2}{*}{23} \\ 
 & 22900 & 22400 & 120 &  & \\
 \hline
 \multirow{2}{*}{497} & 13173 & 12493 & 65 & \multirow{2}{*}{0.00064} & \multirow{2}{*}{23.04} \\ 
 & 16419 & 15749 & 82 &  &  \\
 \hline
\end{tabular}
\label{tab:bob_extintion_ratio}
\end{table}
\begin{table*}[ht]
\centering
\caption{The entanglement source is characterized by the measured mean photon number, the coincidence counts, and the accidental counts as a function of the EDFA current per each time-bin. The counts are normalized to an integration window of 10~s, with a repetition rate of 90 MHz.}
\begin{tabular}[t]{cccc}
\hline
 EDFA current [mA] & Coincidence counts &  Accidentals &  Mean photon $\mu$  \\ 
 400 & 469.2 $\pm$ 3.6 & 1.8 $\pm$ 0.3 & 0.0039 $\pm$ 0.0007\\
 450 & 1156.3 $\pm$ 4.9 & 6.1 $\pm$ 0.5 & 0.0053 $\pm$ 0.0004\\
 500 & 1653.9 $\pm$ 5.9 & 9.5 $\pm$ 0.6 & 0.0058 $\pm$ 0.0004\\
 550 & 2095.8 $\pm$ 6.6 & 13.7 $\pm$ 0.8 & 0.0065 $\pm$ 0.0004\\
 575 & 2343.2 $\pm$ 7.0 & 17.7 $\pm$ 0.9 & 0.0075 $\pm$ 0.0004\\
 600 & 2548.7 $\pm$ 7.3 & 18.5 $\pm$ 0.9 & 0.0073 $\pm$ 0.0004\\
 \hline
\end{tabular}
\label{tab:car}
\end{table*}
\end{comment}
%%%%%%%%%%%%%%commented%%%%%%%%%%
\section{Quantum State tomography}\label{sec:qst}
We perform projection measurements on the teleported states $\ket{e}_A$, $\ket{l}_B$, and $\ket{+}_A$, in all three of the qubit bases formed by the Pauli matrices $\sigma_x, \sigma_y, \textrm{and } \sigma_z$, i.e. measuring photons at each of the arrival times after the MZI and varying $\varphi$.
These results allow reconstructing the density matrix of each teleported state, both with and without the additional 44 km fiber, using maximum likelihood estimation~\cite{altepeter2005photonic}.
Our resultant matrices clearly match the expected teleported state, with the calculated high teleportation fidelities in Sec. \ref{subsec:teleportation}, up to the aforementioned effects due to multiple photons and distinguishability.
\begin{figure}[htbp!]
  \includegraphics[width=0.45\textwidth]{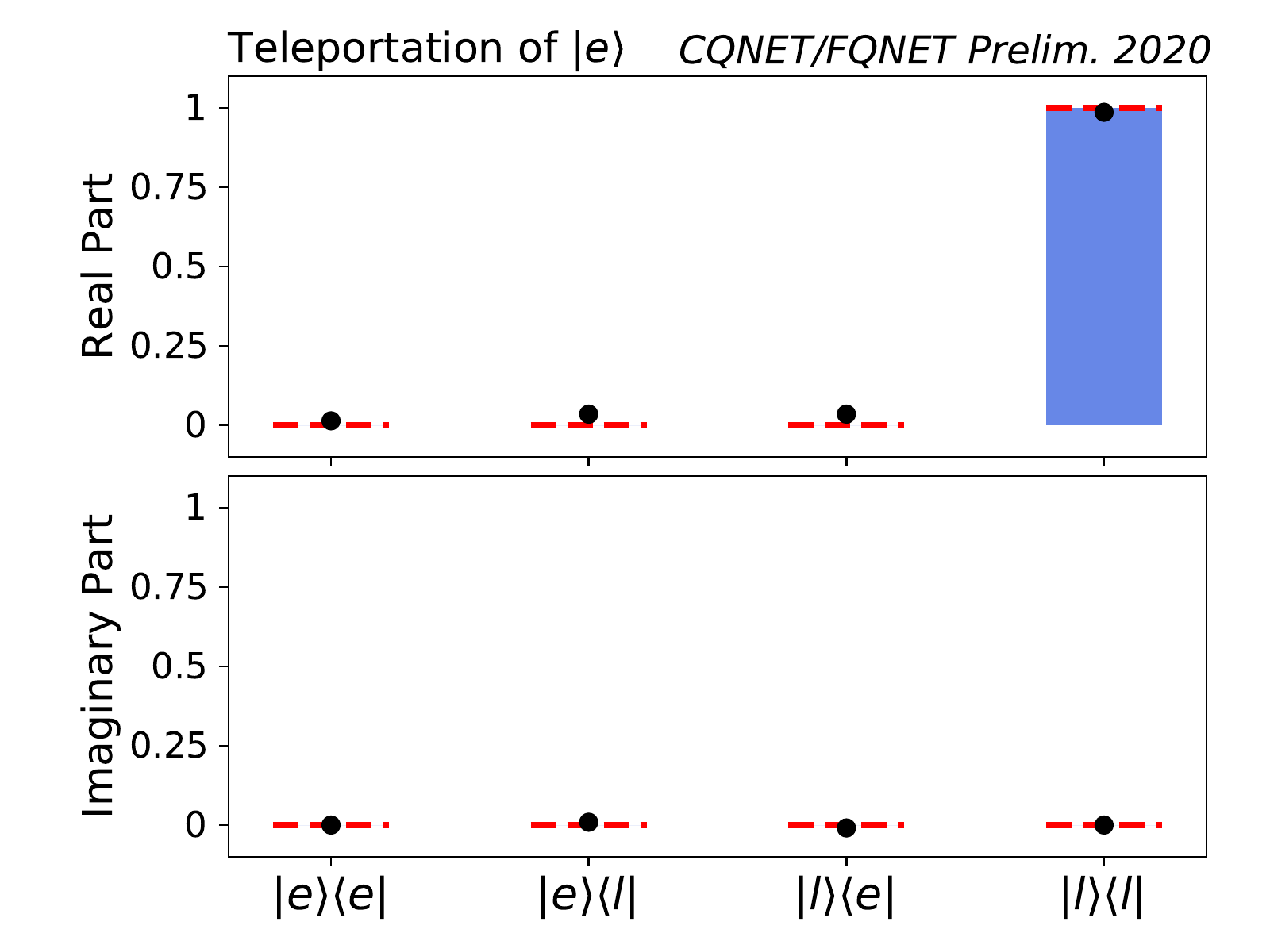}
  \includegraphics[width=0.45\textwidth]{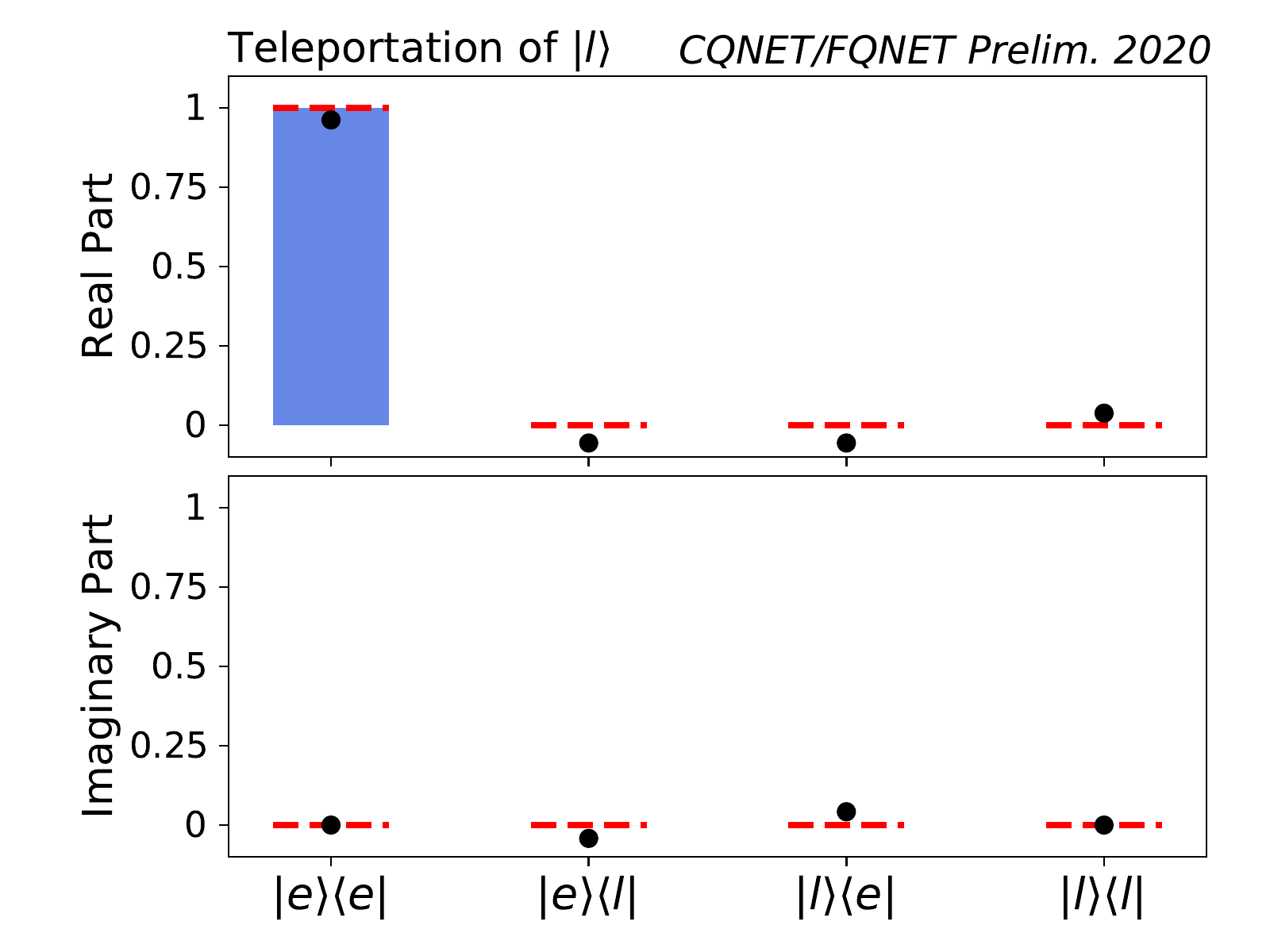}
%  \newline
%  \centering
  \includegraphics[width=0.45\textwidth]{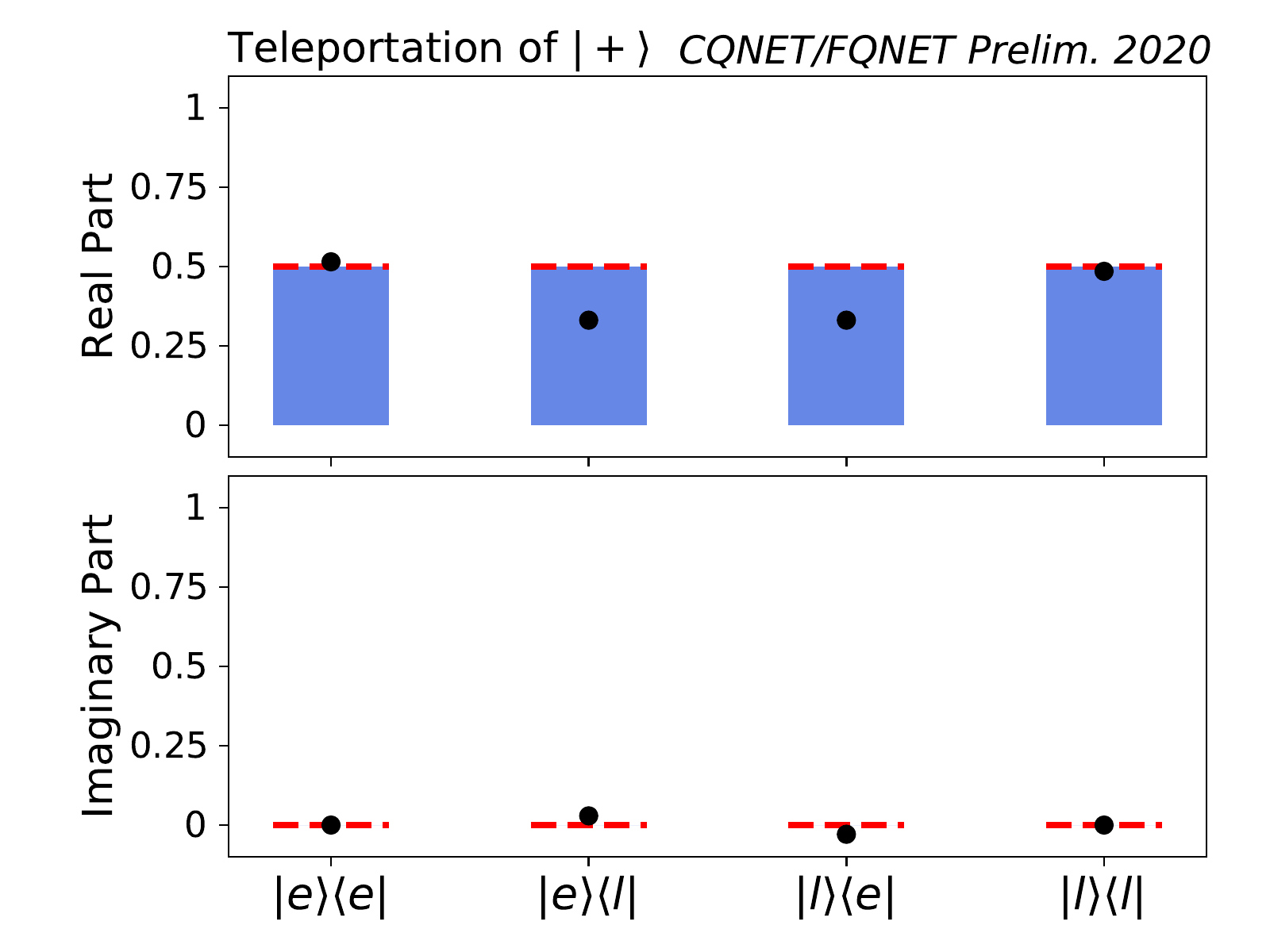}
  \caption{Elements of the density matrices of teleported $\ket{e}$, $\ket{\l}$, and $\ket{+}$ states with the additional 44 km of fiber in the system.The black points are generated by our teleportation system and the blue bars with red dashed lines are the values assuming ideal teleportation.}
  \label{fig:tomography_withspools}
\end{figure}

\begin{figure}[htbp!]
  \includegraphics[width=0.45\textwidth]{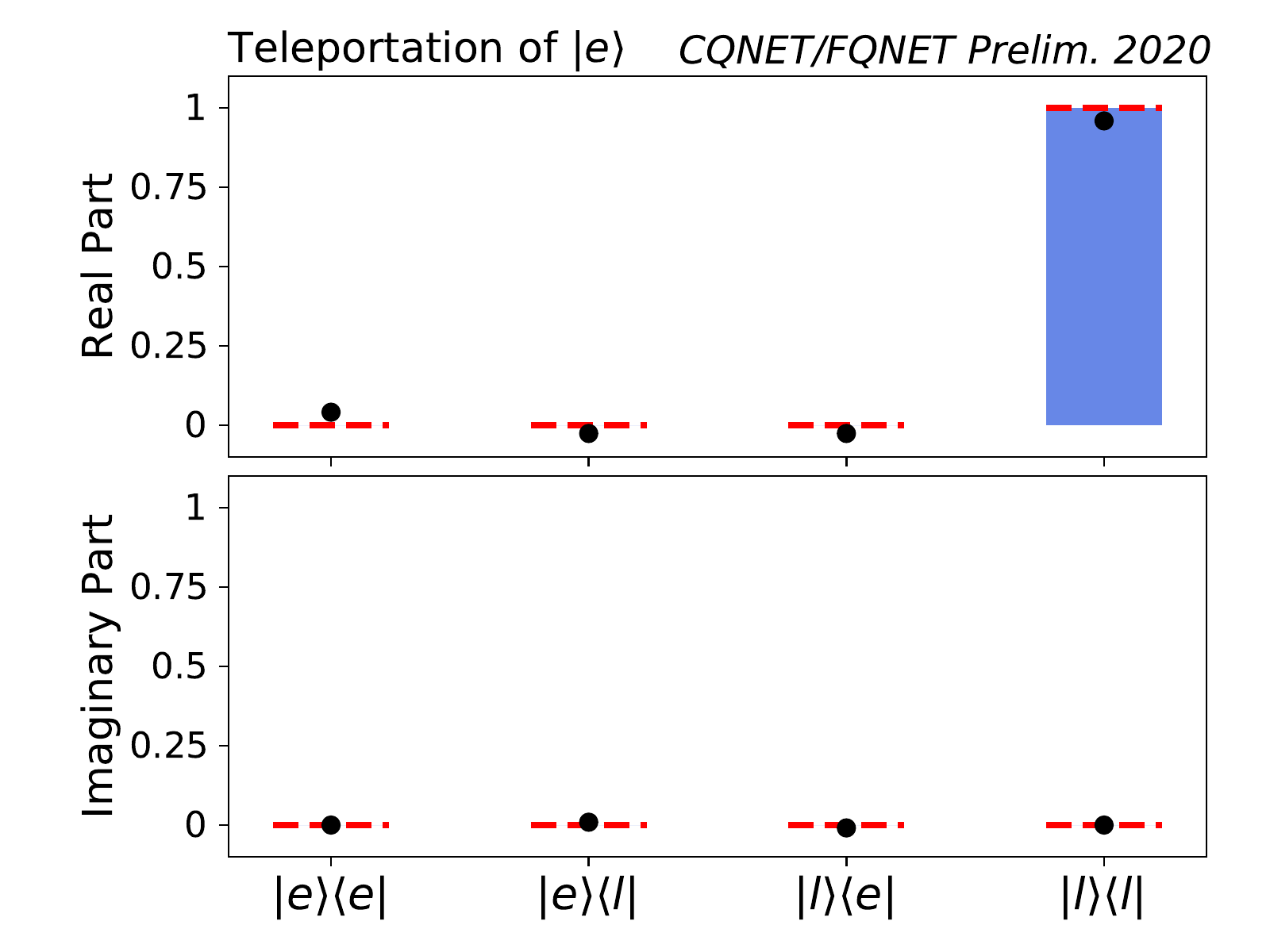}
  \includegraphics[width=0.45\textwidth]{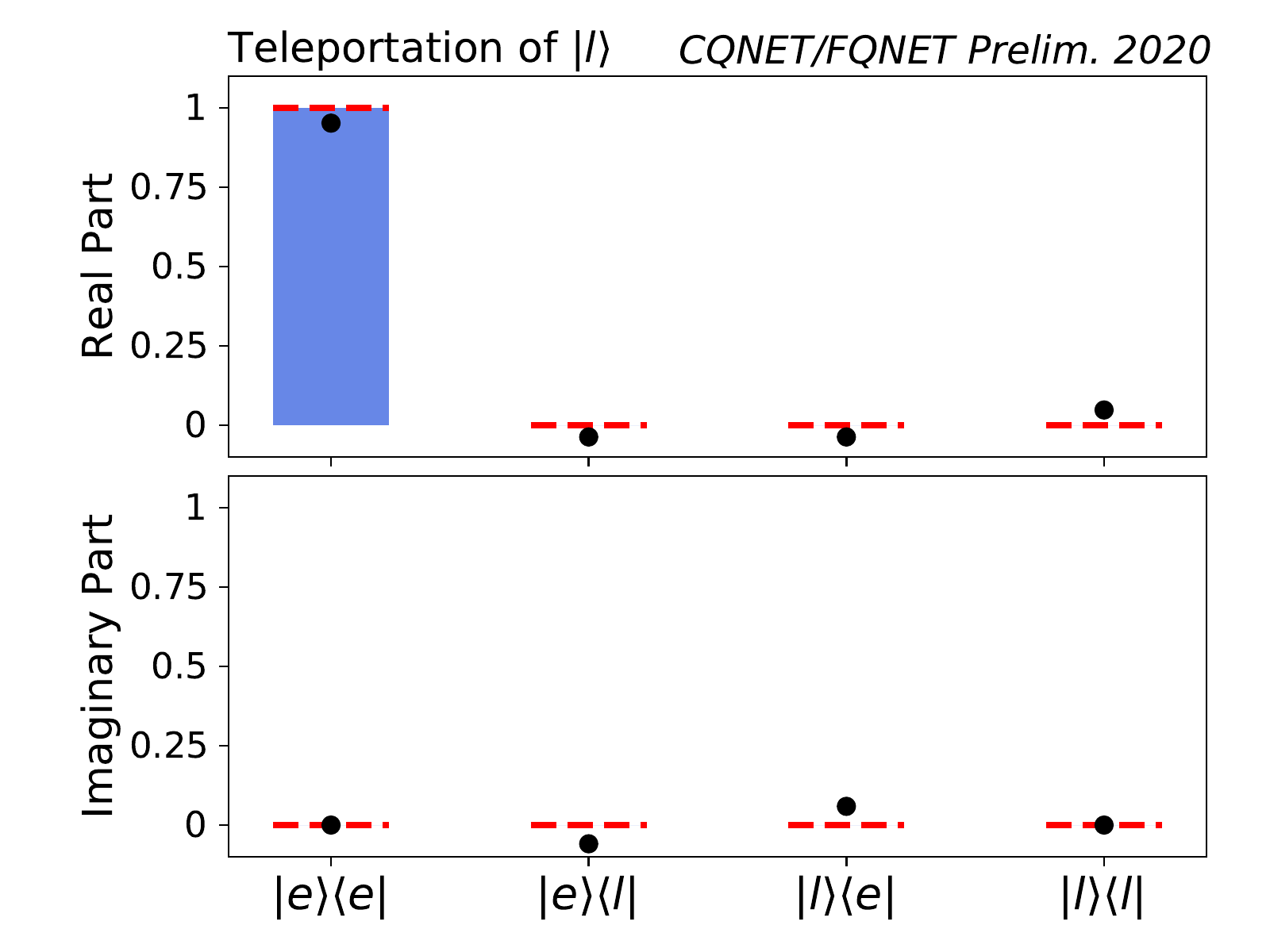}
%  \newline
%  \centering
  \includegraphics[width=0.45\textwidth]{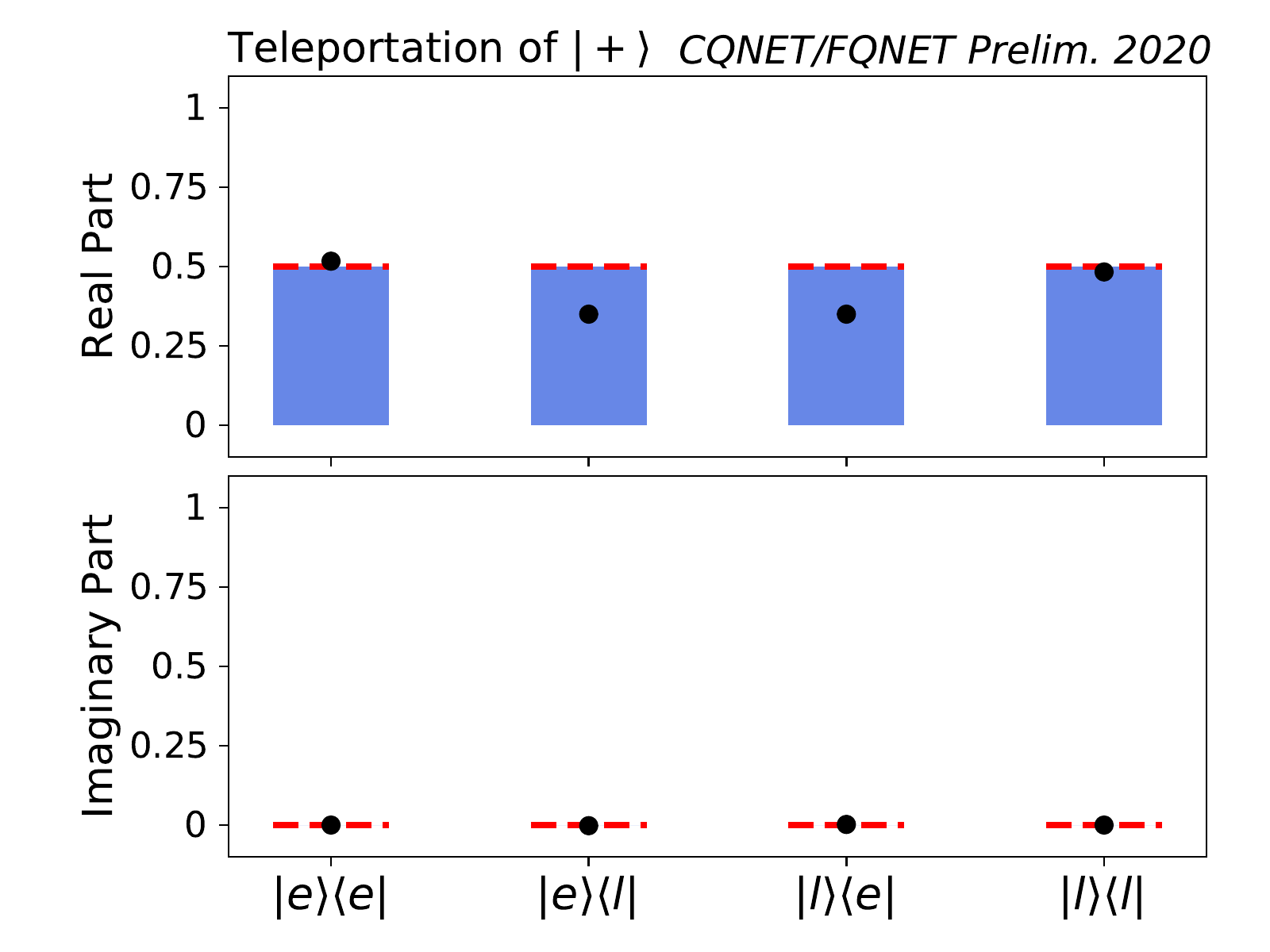}
  \caption{Elements of the density matrices of teleported $\ket{e}$, $\ket{\l}$, and $\ket{+}$ states. The black points are generated by our teleportation system and the blue bars with red dashed lines are the values assuming ideal teleportation.}
  \label{fig:tomography_withoutspools}
\end{figure}

%%%%%%%%%%%%%%%%%%%%%%% References %%%%%%%%%%%%%%%%%%%%%%%%%
%%%%%%%%%% If using BibTeX:
\bibliography{Referencesteleport}

\begin{thebibliography}{78}%
\makeatletter
\providecommand \@ifxundefined [1]{%
 \@ifx{#1\undefined}
}%
\providecommand \@ifnum [1]{%
 \ifnum #1\expandafter \@firstoftwo
 \else \expandafter \@secondoftwo
 \fi
}%
\providecommand \@ifx [1]{%
 \ifx #1\expandafter \@firstoftwo
 \else \expandafter \@secondoftwo
 \fi
}%
\providecommand \natexlab [1]{#1}%
\providecommand \enquote  [1]{``#1''}%
\providecommand \bibnamefont  [1]{#1}%
\providecommand \bibfnamefont [1]{#1}%
\providecommand \citenamefont [1]{#1}%
\providecommand \href@noop [0]{\@secondoftwo}%
\providecommand \href [0]{\begingroup \@sanitize@url \@href}%
\providecommand \@href[1]{\@@startlink{#1}\@@href}%
\providecommand \@@href[1]{\endgroup#1\@@endlink}%
\providecommand \@sanitize@url [0]{\catcode `\\12\catcode `\$12\catcode
  `\&12\catcode `\#12\catcode `\^12\catcode `\_12\catcode `\%12\relax}%
\providecommand \@@startlink[1]{}%
\providecommand \@@endlink[0]{}%
\providecommand \url  [0]{\begingroup\@sanitize@url \@url }%
\providecommand \@url [1]{\endgroup\@href {#1}{\urlprefix }}%
\providecommand \urlprefix  [0]{URL }%
\providecommand \Eprint [0]{\href }%
\providecommand \doibase [0]{https://doi.org/}%
\providecommand \selectlanguage [0]{\@gobble}%
\providecommand \bibinfo  [0]{\@secondoftwo}%
\providecommand \bibfield  [0]{\@secondoftwo}%
\providecommand \translation [1]{[#1]}%
\providecommand \BibitemOpen [0]{}%
\providecommand \bibitemStop [0]{}%
\providecommand \bibitemNoStop [0]{.\EOS\space}%
\providecommand \EOS [0]{\spacefactor3000\relax}%
\providecommand \BibitemShut  [1]{\csname bibitem#1\endcsname}%
\let\auto@bib@innerbib\@empty
%</preamble>
\bibitem [{\citenamefont {Bennett}\ \emph {et~al.}(1993)\citenamefont
  {Bennett}, \citenamefont {Brassard}, \citenamefont {Cr{\'e}peau},
  \citenamefont {Jozsa}, \citenamefont {Peres},\ and\ \citenamefont
  {Wootters}}]{bennett1993teleporting}%
  \BibitemOpen
  \bibfield  {author} {\bibinfo {author} {\bibfnamefont {C.~H.}\ \bibnamefont
  {Bennett}}, \bibinfo {author} {\bibfnamefont {G.}~\bibnamefont {Brassard}},
  \bibinfo {author} {\bibfnamefont {C.}~\bibnamefont {Cr{\'e}peau}}, \bibinfo
  {author} {\bibfnamefont {R.}~\bibnamefont {Jozsa}}, \bibinfo {author}
  {\bibfnamefont {A.}~\bibnamefont {Peres}},\ and\ \bibinfo {author}
  {\bibfnamefont {W.~K.}\ \bibnamefont {Wootters}},\ }\bibfield  {title}
  {\bibinfo {title} {Teleporting an unknown quantum state via dual classical
  and einstein-podolsky-rosen channels},\ }\href@noop {} {\bibfield  {journal}
  {\bibinfo  {journal} {Physical review letters}\ }\textbf {\bibinfo {volume}
  {70}},\ \bibinfo {pages} {1895} (\bibinfo {year} {1993})}\BibitemShut
  {NoStop}%
\bibitem [{\citenamefont {Bouwmeester}\ \emph {et~al.}(1997)\citenamefont
  {Bouwmeester}, \citenamefont {Pan}, \citenamefont {Mattle}, \citenamefont
  {Eibl}, \citenamefont {Weinfurter},\ and\ \citenamefont
  {Zeilinger}}]{bouwmeester1997experimental}%
  \BibitemOpen
  \bibfield  {author} {\bibinfo {author} {\bibfnamefont {D.}~\bibnamefont
  {Bouwmeester}}, \bibinfo {author} {\bibfnamefont {J.-W.}\ \bibnamefont
  {Pan}}, \bibinfo {author} {\bibfnamefont {K.}~\bibnamefont {Mattle}},
  \bibinfo {author} {\bibfnamefont {M.}~\bibnamefont {Eibl}}, \bibinfo {author}
  {\bibfnamefont {H.}~\bibnamefont {Weinfurter}},\ and\ \bibinfo {author}
  {\bibfnamefont {A.}~\bibnamefont {Zeilinger}},\ }\bibfield  {title} {\bibinfo
  {title} {Experimental quantum teleportation},\ }\href@noop {} {\bibfield
  {journal} {\bibinfo  {journal} {Nature}\ }\textbf {\bibinfo {volume} {390}},\
  \bibinfo {pages} {575} (\bibinfo {year} {1997})}\BibitemShut {NoStop}%
\bibitem [{\citenamefont {Boschi}\ \emph {et~al.}(1998)\citenamefont {Boschi},
  \citenamefont {Branca}, \citenamefont {De~Martini}, \citenamefont {Hardy},\
  and\ \citenamefont {Popescu}}]{boschi1998experimental}%
  \BibitemOpen
  \bibfield  {author} {\bibinfo {author} {\bibfnamefont {D.}~\bibnamefont
  {Boschi}}, \bibinfo {author} {\bibfnamefont {S.}~\bibnamefont {Branca}},
  \bibinfo {author} {\bibfnamefont {F.}~\bibnamefont {De~Martini}}, \bibinfo
  {author} {\bibfnamefont {L.}~\bibnamefont {Hardy}},\ and\ \bibinfo {author}
  {\bibfnamefont {S.}~\bibnamefont {Popescu}},\ }\bibfield  {title} {\bibinfo
  {title} {Experimental realization of teleporting an unknown pure quantum
  state via dual classical and einstein-podolsky-rosen channels},\ }\href@noop
  {} {\bibfield  {journal} {\bibinfo  {journal} {Physical Review Letters}\
  }\textbf {\bibinfo {volume} {80}},\ \bibinfo {pages} {1121} (\bibinfo {year}
  {1998})}\BibitemShut {NoStop}%
\bibitem [{\citenamefont {Furusawa}\ \emph {et~al.}(1998)\citenamefont
  {Furusawa}, \citenamefont {S{\o}rensen}, \citenamefont {Braunstein},
  \citenamefont {Fuchs}, \citenamefont {Kimble},\ and\ \citenamefont
  {Polzik}}]{furusawa1998unconditional}%
  \BibitemOpen
  \bibfield  {author} {\bibinfo {author} {\bibfnamefont {A.}~\bibnamefont
  {Furusawa}}, \bibinfo {author} {\bibfnamefont {J.~L.}\ \bibnamefont
  {S{\o}rensen}}, \bibinfo {author} {\bibfnamefont {S.~L.}\ \bibnamefont
  {Braunstein}}, \bibinfo {author} {\bibfnamefont {C.~A.}\ \bibnamefont
  {Fuchs}}, \bibinfo {author} {\bibfnamefont {H.~J.}\ \bibnamefont {Kimble}},\
  and\ \bibinfo {author} {\bibfnamefont {E.~S.}\ \bibnamefont {Polzik}},\
  }\bibfield  {title} {\bibinfo {title} {Unconditional quantum teleportation},\
  }\href@noop {} {\bibfield  {journal} {\bibinfo  {journal} {Science}\ }\textbf
  {\bibinfo {volume} {282}},\ \bibinfo {pages} {706} (\bibinfo {year}
  {1998})}\BibitemShut {NoStop}%
\bibitem [{\citenamefont {Hensen}\ \emph {et~al.}(2015)\citenamefont {Hensen},
  \citenamefont {Bernien}, \citenamefont {Dr{\'e}au}, \citenamefont {Reiserer},
  \citenamefont {Kalb}, \citenamefont {Blok}, \citenamefont {Ruitenberg},
  \citenamefont {Vermeulen}, \citenamefont {Schouten}, \citenamefont
  {Abell{\'a}n}, \citenamefont {Amaya}, \citenamefont {Pruneri}, \citenamefont
  {Mitchell}, \citenamefont {Markham}, \citenamefont {Twitchen}, \citenamefont
  {Elkouss}, \citenamefont {Wehner}, \citenamefont {Taminiau},\ and\
  \citenamefont {Hanson}}]{hensen2015loophole}%
  \BibitemOpen
  \bibfield  {author} {\bibinfo {author} {\bibfnamefont {B.}~\bibnamefont
  {Hensen}}, \bibinfo {author} {\bibfnamefont {H.}~\bibnamefont {Bernien}},
  \bibinfo {author} {\bibfnamefont {A.~E.}\ \bibnamefont {Dr{\'e}au}}, \bibinfo
  {author} {\bibfnamefont {A.}~\bibnamefont {Reiserer}}, \bibinfo {author}
  {\bibfnamefont {N.}~\bibnamefont {Kalb}}, \bibinfo {author} {\bibfnamefont
  {M.~S.}\ \bibnamefont {Blok}}, \bibinfo {author} {\bibfnamefont
  {J.}~\bibnamefont {Ruitenberg}}, \bibinfo {author} {\bibfnamefont {R.~F.~L.}\
  \bibnamefont {Vermeulen}}, \bibinfo {author} {\bibfnamefont {R.~N.}\
  \bibnamefont {Schouten}}, \bibinfo {author} {\bibfnamefont {C.}~\bibnamefont
  {Abell{\'a}n}}, \bibinfo {author} {\bibfnamefont {W.}~\bibnamefont {Amaya}},
  \bibinfo {author} {\bibfnamefont {V.}~\bibnamefont {Pruneri}}, \bibinfo
  {author} {\bibfnamefont {M.~W.}\ \bibnamefont {Mitchell}}, \bibinfo {author}
  {\bibfnamefont {M.}~\bibnamefont {Markham}}, \bibinfo {author} {\bibfnamefont
  {D.~J.}\ \bibnamefont {Twitchen}}, \bibinfo {author} {\bibfnamefont
  {D.}~\bibnamefont {Elkouss}}, \bibinfo {author} {\bibfnamefont
  {S.}~\bibnamefont {Wehner}}, \bibinfo {author} {\bibfnamefont {T.~H.}\
  \bibnamefont {Taminiau}},\ and\ \bibinfo {author} {\bibfnamefont
  {R.}~\bibnamefont {Hanson}},\ }\bibfield  {title} {\bibinfo {title}
  {Loophole-free bell inequality violation using electron spins separated by
  1.3 kilometres},\ }\href {https://doi.org/10.1038/nature15759} {\bibfield
  {journal} {\bibinfo  {journal} {Nature}\ }\textbf {\bibinfo {volume} {526}},\
  \bibinfo {pages} {682} (\bibinfo {year} {2015})}\BibitemShut {NoStop}%
\bibitem [{\citenamefont {Shalm}\ \emph {et~al.}(2015)\citenamefont {Shalm},
  \citenamefont {Meyer-Scott}, \citenamefont {Christensen}, \citenamefont
  {Bierhorst}, \citenamefont {Wayne}, \citenamefont {Stevens}, \citenamefont
  {Gerrits}, \citenamefont {Glancy}, \citenamefont {Hamel}, \citenamefont
  {Allman}, \citenamefont {Coakley}, \citenamefont {Dyer}, \citenamefont
  {Hodge}, \citenamefont {Lita}, \citenamefont {Verma}, \citenamefont
  {Lambrocco}, \citenamefont {Tortorici}, \citenamefont {Migdall},
  \citenamefont {Zhang}, \citenamefont {Kumor}, \citenamefont {Farr},
  \citenamefont {Marsili}, \citenamefont {Shaw}, \citenamefont {Stern},
  \citenamefont {Abell\'an}, \citenamefont {Amaya}, \citenamefont {Pruneri},
  \citenamefont {Jennewein}, \citenamefont {Mitchell}, \citenamefont {Kwiat},
  \citenamefont {Bienfang}, \citenamefont {Mirin}, \citenamefont {Knill},\ and\
  \citenamefont {Nam}}]{Shalm2015}%
  \BibitemOpen
  \bibfield  {author} {\bibinfo {author} {\bibfnamefont {L.~K.}\ \bibnamefont
  {Shalm}}, \bibinfo {author} {\bibfnamefont {E.}~\bibnamefont {Meyer-Scott}},
  \bibinfo {author} {\bibfnamefont {B.~G.}\ \bibnamefont {Christensen}},
  \bibinfo {author} {\bibfnamefont {P.}~\bibnamefont {Bierhorst}}, \bibinfo
  {author} {\bibfnamefont {M.~A.}\ \bibnamefont {Wayne}}, \bibinfo {author}
  {\bibfnamefont {M.~J.}\ \bibnamefont {Stevens}}, \bibinfo {author}
  {\bibfnamefont {T.}~\bibnamefont {Gerrits}}, \bibinfo {author} {\bibfnamefont
  {S.}~\bibnamefont {Glancy}}, \bibinfo {author} {\bibfnamefont {D.~R.}\
  \bibnamefont {Hamel}}, \bibinfo {author} {\bibfnamefont {M.~S.}\ \bibnamefont
  {Allman}}, \bibinfo {author} {\bibfnamefont {K.~J.}\ \bibnamefont {Coakley}},
  \bibinfo {author} {\bibfnamefont {S.~D.}\ \bibnamefont {Dyer}}, \bibinfo
  {author} {\bibfnamefont {C.}~\bibnamefont {Hodge}}, \bibinfo {author}
  {\bibfnamefont {A.~E.}\ \bibnamefont {Lita}}, \bibinfo {author}
  {\bibfnamefont {V.~B.}\ \bibnamefont {Verma}}, \bibinfo {author}
  {\bibfnamefont {C.}~\bibnamefont {Lambrocco}}, \bibinfo {author}
  {\bibfnamefont {E.}~\bibnamefont {Tortorici}}, \bibinfo {author}
  {\bibfnamefont {A.~L.}\ \bibnamefont {Migdall}}, \bibinfo {author}
  {\bibfnamefont {Y.}~\bibnamefont {Zhang}}, \bibinfo {author} {\bibfnamefont
  {D.~R.}\ \bibnamefont {Kumor}}, \bibinfo {author} {\bibfnamefont {W.~H.}\
  \bibnamefont {Farr}}, \bibinfo {author} {\bibfnamefont {F.}~\bibnamefont
  {Marsili}}, \bibinfo {author} {\bibfnamefont {M.~D.}\ \bibnamefont {Shaw}},
  \bibinfo {author} {\bibfnamefont {J.~A.}\ \bibnamefont {Stern}}, \bibinfo
  {author} {\bibfnamefont {C.}~\bibnamefont {Abell\'an}}, \bibinfo {author}
  {\bibfnamefont {W.}~\bibnamefont {Amaya}}, \bibinfo {author} {\bibfnamefont
  {V.}~\bibnamefont {Pruneri}}, \bibinfo {author} {\bibfnamefont
  {T.}~\bibnamefont {Jennewein}}, \bibinfo {author} {\bibfnamefont {M.~W.}\
  \bibnamefont {Mitchell}}, \bibinfo {author} {\bibfnamefont {P.~G.}\
  \bibnamefont {Kwiat}}, \bibinfo {author} {\bibfnamefont {J.~C.}\ \bibnamefont
  {Bienfang}}, \bibinfo {author} {\bibfnamefont {R.~P.}\ \bibnamefont {Mirin}},
  \bibinfo {author} {\bibfnamefont {E.}~\bibnamefont {Knill}},\ and\ \bibinfo
  {author} {\bibfnamefont {S.~W.}\ \bibnamefont {Nam}},\ }\bibfield  {title}
  {\bibinfo {title} {Strong loophole-free test of local realism},\ }\href
  {https://doi.org/10.1103/PhysRevLett.115.250402} {\bibfield  {journal}
  {\bibinfo  {journal} {Phys. Rev. Lett.}\ }\textbf {\bibinfo {volume} {115}},\
  \bibinfo {pages} {250402} (\bibinfo {year} {2015})}\BibitemShut {NoStop}%
\bibitem [{\citenamefont {Giustina}\ \emph {et~al.}(2015)\citenamefont
  {Giustina}, \citenamefont {Versteegh}, \citenamefont {Wengerowsky},
  \citenamefont {Handsteiner}, \citenamefont {Hochrainer}, \citenamefont
  {Phelan}, \citenamefont {Steinlechner}, \citenamefont {Kofler}, \citenamefont
  {Larsson}, \citenamefont {Abell\'an}, \citenamefont {Amaya}, \citenamefont
  {Pruneri}, \citenamefont {Mitchell}, \citenamefont {Beyer}, \citenamefont
  {Gerrits}, \citenamefont {Lita}, \citenamefont {Shalm}, \citenamefont {Nam},
  \citenamefont {Scheidl}, \citenamefont {Ursin}, \citenamefont {Wittmann},\
  and\ \citenamefont {Zeilinger}}]{Giustina2015}%
  \BibitemOpen
  \bibfield  {author} {\bibinfo {author} {\bibfnamefont {M.}~\bibnamefont
  {Giustina}}, \bibinfo {author} {\bibfnamefont {M.~A.~M.}\ \bibnamefont
  {Versteegh}}, \bibinfo {author} {\bibfnamefont {S.}~\bibnamefont
  {Wengerowsky}}, \bibinfo {author} {\bibfnamefont {J.}~\bibnamefont
  {Handsteiner}}, \bibinfo {author} {\bibfnamefont {A.}~\bibnamefont
  {Hochrainer}}, \bibinfo {author} {\bibfnamefont {K.}~\bibnamefont {Phelan}},
  \bibinfo {author} {\bibfnamefont {F.}~\bibnamefont {Steinlechner}}, \bibinfo
  {author} {\bibfnamefont {J.}~\bibnamefont {Kofler}}, \bibinfo {author}
  {\bibfnamefont {J.-A.}\ \bibnamefont {Larsson}}, \bibinfo {author}
  {\bibfnamefont {C.}~\bibnamefont {Abell\'an}}, \bibinfo {author}
  {\bibfnamefont {W.}~\bibnamefont {Amaya}}, \bibinfo {author} {\bibfnamefont
  {V.}~\bibnamefont {Pruneri}}, \bibinfo {author} {\bibfnamefont {M.~W.}\
  \bibnamefont {Mitchell}}, \bibinfo {author} {\bibfnamefont {J.}~\bibnamefont
  {Beyer}}, \bibinfo {author} {\bibfnamefont {T.}~\bibnamefont {Gerrits}},
  \bibinfo {author} {\bibfnamefont {A.~E.}\ \bibnamefont {Lita}}, \bibinfo
  {author} {\bibfnamefont {L.~K.}\ \bibnamefont {Shalm}}, \bibinfo {author}
  {\bibfnamefont {S.~W.}\ \bibnamefont {Nam}}, \bibinfo {author} {\bibfnamefont
  {T.}~\bibnamefont {Scheidl}}, \bibinfo {author} {\bibfnamefont
  {R.}~\bibnamefont {Ursin}}, \bibinfo {author} {\bibfnamefont
  {B.}~\bibnamefont {Wittmann}},\ and\ \bibinfo {author} {\bibfnamefont
  {A.}~\bibnamefont {Zeilinger}},\ }\bibfield  {title} {\bibinfo {title}
  {Significant-loophole-free test of bell's theorem with entangled photons},\
  }\href {https://doi.org/10.1103/PhysRevLett.115.250401} {\bibfield  {journal}
  {\bibinfo  {journal} {Phys. Rev. Lett.}\ }\textbf {\bibinfo {volume} {115}},\
  \bibinfo {pages} {250401} (\bibinfo {year} {2015})}\BibitemShut {NoStop}%
\bibitem [{\citenamefont {Rosenfeld}\ \emph {et~al.}(2017)\citenamefont
  {Rosenfeld}, \citenamefont {Burchardt}, \citenamefont {Garthoff},
  \citenamefont {Redeker}, \citenamefont {Ortegel}, \citenamefont {Rau},\ and\
  \citenamefont {Weinfurter}}]{Rosenfeld2017}%
  \BibitemOpen
  \bibfield  {author} {\bibinfo {author} {\bibfnamefont {W.}~\bibnamefont
  {Rosenfeld}}, \bibinfo {author} {\bibfnamefont {D.}~\bibnamefont
  {Burchardt}}, \bibinfo {author} {\bibfnamefont {R.}~\bibnamefont {Garthoff}},
  \bibinfo {author} {\bibfnamefont {K.}~\bibnamefont {Redeker}}, \bibinfo
  {author} {\bibfnamefont {N.}~\bibnamefont {Ortegel}}, \bibinfo {author}
  {\bibfnamefont {M.}~\bibnamefont {Rau}},\ and\ \bibinfo {author}
  {\bibfnamefont {H.}~\bibnamefont {Weinfurter}},\ }\bibfield  {title}
  {\bibinfo {title} {Event-ready bell test using entangled atoms simultaneously
  closing detection and locality loopholes},\ }\href
  {https://doi.org/10.1103/PhysRevLett.119.010402} {\bibfield  {journal}
  {\bibinfo  {journal} {Phys. Rev. Lett.}\ }\textbf {\bibinfo {volume} {119}},\
  \bibinfo {pages} {010402} (\bibinfo {year} {2017})}\BibitemShut {NoStop}%
\bibitem [{\citenamefont {Gao}\ \emph {et~al.}(2017)\citenamefont {Gao},
  \citenamefont {Jafferis},\ and\ \citenamefont {Wall}}]{gao2017traversable}%
  \BibitemOpen
  \bibfield  {author} {\bibinfo {author} {\bibfnamefont {P.}~\bibnamefont
  {Gao}}, \bibinfo {author} {\bibfnamefont {D.~L.}\ \bibnamefont {Jafferis}},\
  and\ \bibinfo {author} {\bibfnamefont {A.~C.}\ \bibnamefont {Wall}},\
  }\bibfield  {title} {\bibinfo {title} {Traversable wormholes via a double
  trace deformation},\ }\href@noop {} {\bibfield  {journal} {\bibinfo
  {journal} {Journal of High Energy Physics}\ }\textbf {\bibinfo {volume}
  {2017}},\ \bibinfo {pages} {151} (\bibinfo {year} {2017})}\BibitemShut
  {NoStop}%
\bibitem [{\citenamefont {Yoshida}\ and\ \citenamefont
  {Yao}(2019)}]{yoshida2019disentangling}%
  \BibitemOpen
  \bibfield  {author} {\bibinfo {author} {\bibfnamefont {B.}~\bibnamefont
  {Yoshida}}\ and\ \bibinfo {author} {\bibfnamefont {N.~Y.}\ \bibnamefont
  {Yao}},\ }\bibfield  {title} {\bibinfo {title} {Disentangling scrambling and
  decoherence via quantum teleportation},\ }\href@noop {} {\bibfield  {journal}
  {\bibinfo  {journal} {Physical Review X}\ }\textbf {\bibinfo {volume} {9}},\
  \bibinfo {pages} {011006} (\bibinfo {year} {2019})}\BibitemShut {NoStop}%
\bibitem [{\citenamefont {Landsman}\ \emph {et~al.}(2019)\citenamefont
  {Landsman}, \citenamefont {Figgatt}, \citenamefont {Schuster}, \citenamefont
  {Linke}, \citenamefont {Yoshida}, \citenamefont {Yao},\ and\ \citenamefont
  {Monroe}}]{landsman2019verified}%
  \BibitemOpen
  \bibfield  {author} {\bibinfo {author} {\bibfnamefont {K.~A.}\ \bibnamefont
  {Landsman}}, \bibinfo {author} {\bibfnamefont {C.}~\bibnamefont {Figgatt}},
  \bibinfo {author} {\bibfnamefont {T.}~\bibnamefont {Schuster}}, \bibinfo
  {author} {\bibfnamefont {N.~M.}\ \bibnamefont {Linke}}, \bibinfo {author}
  {\bibfnamefont {B.}~\bibnamefont {Yoshida}}, \bibinfo {author} {\bibfnamefont
  {N.~Y.}\ \bibnamefont {Yao}},\ and\ \bibinfo {author} {\bibfnamefont
  {C.}~\bibnamefont {Monroe}},\ }\bibfield  {title} {\bibinfo {title} {Verified
  quantum information scrambling},\ }\href@noop {} {\bibfield  {journal}
  {\bibinfo  {journal} {Nature}\ }\textbf {\bibinfo {volume} {567}},\ \bibinfo
  {pages} {61} (\bibinfo {year} {2019})}\BibitemShut {NoStop}%
\bibitem [{\citenamefont {Lloyd}\ \emph {et~al.}(2011)\citenamefont {Lloyd},
  \citenamefont {Maccone}, \citenamefont {Garcia-Patron}, \citenamefont
  {Giovannetti}, \citenamefont {Shikano}, \citenamefont {Pirandola},
  \citenamefont {Rozema}, \citenamefont {Darabi}, \citenamefont {Soudagar},
  \citenamefont {Shalm},\ and\ \citenamefont {Steinberg}}]{lloyd2011closed}%
  \BibitemOpen
  \bibfield  {author} {\bibinfo {author} {\bibfnamefont {S.}~\bibnamefont
  {Lloyd}}, \bibinfo {author} {\bibfnamefont {L.}~\bibnamefont {Maccone}},
  \bibinfo {author} {\bibfnamefont {R.}~\bibnamefont {Garcia-Patron}}, \bibinfo
  {author} {\bibfnamefont {V.}~\bibnamefont {Giovannetti}}, \bibinfo {author}
  {\bibfnamefont {Y.}~\bibnamefont {Shikano}}, \bibinfo {author} {\bibfnamefont
  {S.}~\bibnamefont {Pirandola}}, \bibinfo {author} {\bibfnamefont {L.~A.}\
  \bibnamefont {Rozema}}, \bibinfo {author} {\bibfnamefont {A.}~\bibnamefont
  {Darabi}}, \bibinfo {author} {\bibfnamefont {Y.}~\bibnamefont {Soudagar}},
  \bibinfo {author} {\bibfnamefont {L.~K.}\ \bibnamefont {Shalm}},\ and\
  \bibinfo {author} {\bibfnamefont {A.~M.}\ \bibnamefont {Steinberg}},\
  }\bibfield  {title} {\bibinfo {title} {Closed timelike curves via
  postselection: Theory and experimental test of consistency},\ }\href
  {https://doi.org/10.1103/PhysRevLett.106.040403} {\bibfield  {journal}
  {\bibinfo  {journal} {Phys. Rev. Lett.}\ }\textbf {\bibinfo {volume} {106}},\
  \bibinfo {pages} {040403} (\bibinfo {year} {2011})}\BibitemShut {NoStop}%
\bibitem [{\citenamefont {Aspelmeyer}\ \emph {et~al.}(2014)\citenamefont
  {Aspelmeyer}, \citenamefont {Kippenberg},\ and\ \citenamefont
  {Marquardt}}]{aspelmeyer2014cavity}%
  \BibitemOpen
  \bibfield  {author} {\bibinfo {author} {\bibfnamefont {M.}~\bibnamefont
  {Aspelmeyer}}, \bibinfo {author} {\bibfnamefont {T.~J.}\ \bibnamefont
  {Kippenberg}},\ and\ \bibinfo {author} {\bibfnamefont {F.}~\bibnamefont
  {Marquardt}},\ }\bibfield  {title} {\bibinfo {title} {Cavity optomechanics},\
  }\href@noop {} {\bibfield  {journal} {\bibinfo  {journal} {Reviews of Modern
  Physics}\ }\textbf {\bibinfo {volume} {86}},\ \bibinfo {pages} {1391}
  (\bibinfo {year} {2014})}\BibitemShut {NoStop}%
\bibitem [{\citenamefont {Hou}\ \emph {et~al.}(2016)\citenamefont {Hou},
  \citenamefont {Huang}, \citenamefont {Yuan}, \citenamefont {Chang},
  \citenamefont {Zu}, \citenamefont {He},\ and\ \citenamefont
  {Duan}}]{hou2016quantum}%
  \BibitemOpen
  \bibfield  {author} {\bibinfo {author} {\bibfnamefont {P.-Y.}\ \bibnamefont
  {Hou}}, \bibinfo {author} {\bibfnamefont {Y.-Y.}\ \bibnamefont {Huang}},
  \bibinfo {author} {\bibfnamefont {X.-X.}\ \bibnamefont {Yuan}}, \bibinfo
  {author} {\bibfnamefont {X.-Y.}\ \bibnamefont {Chang}}, \bibinfo {author}
  {\bibfnamefont {C.}~\bibnamefont {Zu}}, \bibinfo {author} {\bibfnamefont
  {L.}~\bibnamefont {He}},\ and\ \bibinfo {author} {\bibfnamefont {L.-M.}\
  \bibnamefont {Duan}},\ }\bibfield  {title} {\bibinfo {title} {Quantum
  teleportation from light beams to vibrational states of a macroscopic
  diamond},\ }\href@noop {} {\bibfield  {journal} {\bibinfo  {journal} {Nature
  communications}\ }\textbf {\bibinfo {volume} {7}},\ \bibinfo {pages} {11736}
  (\bibinfo {year} {2016})}\BibitemShut {NoStop}%
\bibitem [{\citenamefont {Nielsen}\ and\ \citenamefont
  {Chuang}(2011)}]{nielsen2001quantum}%
  \BibitemOpen
  \bibfield  {author} {\bibinfo {author} {\bibfnamefont {M.~A.}\ \bibnamefont
  {Nielsen}}\ and\ \bibinfo {author} {\bibfnamefont {I.~L.}\ \bibnamefont
  {Chuang}},\ }\href@noop {} {\emph {\bibinfo {title} {Quantum Computation and
  Quantum Information}}},\ \bibinfo {edition} {10th}\ ed.\ (\bibinfo
  {publisher} {Cambridge University Press},\ \bibinfo {address} {USA},\
  \bibinfo {year} {2011})\BibitemShut {NoStop}%
\bibitem [{\citenamefont {Ladd}\ \emph {et~al.}(2010)\citenamefont {Ladd},
  \citenamefont {Jelezko}, \citenamefont {Laflamme}, \citenamefont {Nakamura},
  \citenamefont {Monroe},\ and\ \citenamefont {O’Brien}}]{ladd2010quantum}%
  \BibitemOpen
  \bibfield  {author} {\bibinfo {author} {\bibfnamefont {T.~D.}\ \bibnamefont
  {Ladd}}, \bibinfo {author} {\bibfnamefont {F.}~\bibnamefont {Jelezko}},
  \bibinfo {author} {\bibfnamefont {R.}~\bibnamefont {Laflamme}}, \bibinfo
  {author} {\bibfnamefont {Y.}~\bibnamefont {Nakamura}}, \bibinfo {author}
  {\bibfnamefont {C.}~\bibnamefont {Monroe}},\ and\ \bibinfo {author}
  {\bibfnamefont {J.~L.}\ \bibnamefont {O’Brien}},\ }\bibfield  {title}
  {\bibinfo {title} {Quantum computers},\ }\href@noop {} {\bibfield  {journal}
  {\bibinfo  {journal} {Nature}\ }\textbf {\bibinfo {volume} {464}},\ \bibinfo
  {pages} {45} (\bibinfo {year} {2010})}\BibitemShut {NoStop}%
\bibitem [{\citenamefont {Gisin}\ and\ \citenamefont
  {Thew}(2007)}]{gisin2007quantum}%
  \BibitemOpen
  \bibfield  {author} {\bibinfo {author} {\bibfnamefont {N.}~\bibnamefont
  {Gisin}}\ and\ \bibinfo {author} {\bibfnamefont {R.}~\bibnamefont {Thew}},\
  }\bibfield  {title} {\bibinfo {title} {Quantum communication},\ }\href@noop
  {} {\bibfield  {journal} {\bibinfo  {journal} {Nature photonics}\ }\textbf
  {\bibinfo {volume} {1}},\ \bibinfo {pages} {165} (\bibinfo {year}
  {2007})}\BibitemShut {NoStop}%
\bibitem [{\citenamefont {Pirandola}\ \emph {et~al.}(2015)\citenamefont
  {Pirandola}, \citenamefont {Eisert}, \citenamefont {Weedbrook}, \citenamefont
  {Furusawa},\ and\ \citenamefont {Braunstein}}]{pirandola2015advances}%
  \BibitemOpen
  \bibfield  {author} {\bibinfo {author} {\bibfnamefont {S.}~\bibnamefont
  {Pirandola}}, \bibinfo {author} {\bibfnamefont {J.}~\bibnamefont {Eisert}},
  \bibinfo {author} {\bibfnamefont {C.}~\bibnamefont {Weedbrook}}, \bibinfo
  {author} {\bibfnamefont {A.}~\bibnamefont {Furusawa}},\ and\ \bibinfo
  {author} {\bibfnamefont {S.~L.}\ \bibnamefont {Braunstein}},\ }\bibfield
  {title} {\bibinfo {title} {Advances in quantum teleportation},\ }\href@noop
  {} {\bibfield  {journal} {\bibinfo  {journal} {Nature photonics}\ }\textbf
  {\bibinfo {volume} {9}},\ \bibinfo {pages} {641} (\bibinfo {year}
  {2015})}\BibitemShut {NoStop}%
\bibitem [{\citenamefont {Briegel}\ \emph {et~al.}(1998)\citenamefont
  {Briegel}, \citenamefont {D{\"u}r}, \citenamefont {Cirac},\ and\
  \citenamefont {Zoller}}]{briegel1998quantum}%
  \BibitemOpen
  \bibfield  {author} {\bibinfo {author} {\bibfnamefont {H.-J.}\ \bibnamefont
  {Briegel}}, \bibinfo {author} {\bibfnamefont {W.}~\bibnamefont {D{\"u}r}},
  \bibinfo {author} {\bibfnamefont {J.~I.}\ \bibnamefont {Cirac}},\ and\
  \bibinfo {author} {\bibfnamefont {P.}~\bibnamefont {Zoller}},\ }\bibfield
  {title} {\bibinfo {title} {Quantum repeaters: the role of imperfect local
  operations in quantum communication},\ }\href@noop {} {\bibfield  {journal}
  {\bibinfo  {journal} {Physical Review Letters}\ }\textbf {\bibinfo {volume}
  {81}},\ \bibinfo {pages} {5932} (\bibinfo {year} {1998})}\BibitemShut
  {NoStop}%
\bibitem [{\citenamefont {Kimble}(2008)}]{kimble2008quantum}%
  \BibitemOpen
  \bibfield  {author} {\bibinfo {author} {\bibfnamefont {H.~J.}\ \bibnamefont
  {Kimble}},\ }\bibfield  {title} {\bibinfo {title} {The quantum internet},\
  }\href@noop {} {\bibfield  {journal} {\bibinfo  {journal} {Nature}\ }\textbf
  {\bibinfo {volume} {453}},\ \bibinfo {pages} {1023} (\bibinfo {year}
  {2008})}\BibitemShut {NoStop}%
\bibitem [{\citenamefont {Simon}(2017)}]{simon2017towards}%
  \BibitemOpen
  \bibfield  {author} {\bibinfo {author} {\bibfnamefont {C.}~\bibnamefont
  {Simon}},\ }\bibfield  {title} {\bibinfo {title} {Towards a global quantum
  network},\ }\href@noop {} {\bibfield  {journal} {\bibinfo  {journal} {Nature
  Photonics}\ }\textbf {\bibinfo {volume} {11}},\ \bibinfo {pages} {678}
  (\bibinfo {year} {2017})}\BibitemShut {NoStop}%
\bibitem [{\citenamefont {Wehner}\ \emph {et~al.}(2018)\citenamefont {Wehner},
  \citenamefont {Elkouss},\ and\ \citenamefont {Hanson}}]{wehner2018quantum}%
  \BibitemOpen
  \bibfield  {author} {\bibinfo {author} {\bibfnamefont {S.}~\bibnamefont
  {Wehner}}, \bibinfo {author} {\bibfnamefont {D.}~\bibnamefont {Elkouss}},\
  and\ \bibinfo {author} {\bibfnamefont {R.}~\bibnamefont {Hanson}},\
  }\bibfield  {title} {\bibinfo {title} {Quantum internet: A vision for the
  road ahead},\ }\href@noop {} {\bibfield  {journal} {\bibinfo  {journal}
  {Science}\ }\textbf {\bibinfo {volume} {362}},\ \bibinfo {pages} {eaam9288}
  (\bibinfo {year} {2018})}\BibitemShut {NoStop}%
\bibitem [{\citenamefont {Sangouard}\ \emph {et~al.}(2011)\citenamefont
  {Sangouard}, \citenamefont {Simon}, \citenamefont {De~Riedmatten},\ and\
  \citenamefont {Gisin}}]{sangouard2011quantum}%
  \BibitemOpen
  \bibfield  {author} {\bibinfo {author} {\bibfnamefont {N.}~\bibnamefont
  {Sangouard}}, \bibinfo {author} {\bibfnamefont {C.}~\bibnamefont {Simon}},
  \bibinfo {author} {\bibfnamefont {H.}~\bibnamefont {De~Riedmatten}},\ and\
  \bibinfo {author} {\bibfnamefont {N.}~\bibnamefont {Gisin}},\ }\bibfield
  {title} {\bibinfo {title} {Quantum repeaters based on atomic ensembles and
  linear optics},\ }\href@noop {} {\bibfield  {journal} {\bibinfo  {journal}
  {Reviews of Modern Physics}\ }\textbf {\bibinfo {volume} {83}},\ \bibinfo
  {pages} {33} (\bibinfo {year} {2011})}\BibitemShut {NoStop}%
\bibitem [{\citenamefont {Dias}\ and\ \citenamefont
  {Ralph}(2017)}]{dias2017quantum}%
  \BibitemOpen
  \bibfield  {author} {\bibinfo {author} {\bibfnamefont {J.}~\bibnamefont
  {Dias}}\ and\ \bibinfo {author} {\bibfnamefont {T.~C.}\ \bibnamefont
  {Ralph}},\ }\bibfield  {title} {\bibinfo {title} {Quantum repeaters using
  continuous-variable teleportation},\ }\href@noop {} {\bibfield  {journal}
  {\bibinfo  {journal} {Physical Review A}\ }\textbf {\bibinfo {volume} {95}},\
  \bibinfo {pages} {022312} (\bibinfo {year} {2017})}\BibitemShut {NoStop}%
\bibitem [{\citenamefont {Lucamarini}\ \emph {et~al.}(2018)\citenamefont
  {Lucamarini}, \citenamefont {Yuan}, \citenamefont {Dynes},\ and\
  \citenamefont {Shields}}]{lucamarini2018overcoming}%
  \BibitemOpen
  \bibfield  {author} {\bibinfo {author} {\bibfnamefont {M.}~\bibnamefont
  {Lucamarini}}, \bibinfo {author} {\bibfnamefont {Z.~L.}\ \bibnamefont
  {Yuan}}, \bibinfo {author} {\bibfnamefont {J.~F.}\ \bibnamefont {Dynes}},\
  and\ \bibinfo {author} {\bibfnamefont {A.~J.}\ \bibnamefont {Shields}},\
  }\bibfield  {title} {\bibinfo {title} {Overcoming the rate--distance limit of
  quantum key distribution without quantum repeaters},\ }\href@noop {}
  {\bibfield  {journal} {\bibinfo  {journal} {Nature}\ }\textbf {\bibinfo
  {volume} {557}},\ \bibinfo {pages} {400} (\bibinfo {year}
  {2018})}\BibitemShut {NoStop}%
\bibitem [{\citenamefont {Bhaskar}\ \emph {et~al.}(2020)\citenamefont
  {Bhaskar}, \citenamefont {Riedinger}, \citenamefont {Machielse},
  \citenamefont {Levonian}, \citenamefont {Nguyen}, \citenamefont {Knall},
  \citenamefont {Park}, \citenamefont {Englund}, \citenamefont {Lon{\v{c}}ar},
  \citenamefont {Sukachev},\ and\ \citenamefont
  {Lukin}}]{bhaskar2020experimental}%
  \BibitemOpen
  \bibfield  {author} {\bibinfo {author} {\bibfnamefont {M.~K.}\ \bibnamefont
  {Bhaskar}}, \bibinfo {author} {\bibfnamefont {R.}~\bibnamefont {Riedinger}},
  \bibinfo {author} {\bibfnamefont {B.}~\bibnamefont {Machielse}}, \bibinfo
  {author} {\bibfnamefont {D.~S.}\ \bibnamefont {Levonian}}, \bibinfo {author}
  {\bibfnamefont {C.~T.}\ \bibnamefont {Nguyen}}, \bibinfo {author}
  {\bibfnamefont {E.~N.}\ \bibnamefont {Knall}}, \bibinfo {author}
  {\bibfnamefont {H.}~\bibnamefont {Park}}, \bibinfo {author} {\bibfnamefont
  {D.}~\bibnamefont {Englund}}, \bibinfo {author} {\bibfnamefont
  {M.}~\bibnamefont {Lon{\v{c}}ar}}, \bibinfo {author} {\bibfnamefont {D.~D.}\
  \bibnamefont {Sukachev}},\ and\ \bibinfo {author} {\bibfnamefont {M.~D.}\
  \bibnamefont {Lukin}},\ }\bibfield  {title} {\bibinfo {title} {Experimental
  demonstration of memory-enhanced quantum communication},\ }\href
  {https://doi.org/10.1038/s41586-020-2103-5} {\bibfield  {journal} {\bibinfo
  {journal} {Nature}\ }\textbf {\bibinfo {volume} {580}},\ \bibinfo {pages}
  {60} (\bibinfo {year} {2020})}\BibitemShut {NoStop}%
\bibitem [{\citenamefont {Xia}\ \emph {et~al.}(2017)\citenamefont {Xia},
  \citenamefont {Sun}, \citenamefont {Zhang},\ and\ \citenamefont
  {Pan}}]{xia2017long}%
  \BibitemOpen
  \bibfield  {author} {\bibinfo {author} {\bibfnamefont {X.-X.}\ \bibnamefont
  {Xia}}, \bibinfo {author} {\bibfnamefont {Q.-C.}\ \bibnamefont {Sun}},
  \bibinfo {author} {\bibfnamefont {Q.}~\bibnamefont {Zhang}},\ and\ \bibinfo
  {author} {\bibfnamefont {J.-W.}\ \bibnamefont {Pan}},\ }\bibfield  {title}
  {\bibinfo {title} {Long distance quantum teleportation},\ }\href@noop {}
  {\bibfield  {journal} {\bibinfo  {journal} {Quantum Science and Technology}\
  }\textbf {\bibinfo {volume} {3}},\ \bibinfo {pages} {014012} (\bibinfo {year}
  {2017})}\BibitemShut {NoStop}%
\bibitem [{\citenamefont {Brendel}\ \emph {et~al.}(1999)\citenamefont
  {Brendel}, \citenamefont {Gisin}, \citenamefont {Tittel},\ and\ \citenamefont
  {Zbinden}}]{brendel1999pulsed}%
  \BibitemOpen
  \bibfield  {author} {\bibinfo {author} {\bibfnamefont {J.}~\bibnamefont
  {Brendel}}, \bibinfo {author} {\bibfnamefont {N.}~\bibnamefont {Gisin}},
  \bibinfo {author} {\bibfnamefont {W.}~\bibnamefont {Tittel}},\ and\ \bibinfo
  {author} {\bibfnamefont {H.}~\bibnamefont {Zbinden}},\ }\bibfield  {title}
  {\bibinfo {title} {Pulsed energy-time entangled twin-photon source for
  quantum communication},\ }\href@noop {} {\bibfield  {journal} {\bibinfo
  {journal} {Physical Review Letters}\ }\textbf {\bibinfo {volume} {82}},\
  \bibinfo {pages} {2594} (\bibinfo {year} {1999})}\BibitemShut {NoStop}%
\bibitem [{\citenamefont {Valivarthi}\ \emph {et~al.}(2016)\citenamefont
  {Valivarthi}, \citenamefont {Puigibert}, \citenamefont {Zhou}, \citenamefont
  {Aguilar}, \citenamefont {Verma}, \citenamefont {Marsili}, \citenamefont
  {Shaw}, \citenamefont {Nam}, \citenamefont {Oblak},\ and\ \citenamefont
  {Tittel}}]{Valivarthi2016}%
  \BibitemOpen
  \bibfield  {author} {\bibinfo {author} {\bibfnamefont {R.}~\bibnamefont
  {Valivarthi}}, \bibinfo {author} {\bibfnamefont {M.~G.}\ \bibnamefont
  {Puigibert}}, \bibinfo {author} {\bibfnamefont {Q.}~\bibnamefont {Zhou}},
  \bibinfo {author} {\bibfnamefont {G.~H.}\ \bibnamefont {Aguilar}}, \bibinfo
  {author} {\bibfnamefont {V.~B.}\ \bibnamefont {Verma}}, \bibinfo {author}
  {\bibfnamefont {F.}~\bibnamefont {Marsili}}, \bibinfo {author} {\bibfnamefont
  {M.~D.}\ \bibnamefont {Shaw}}, \bibinfo {author} {\bibfnamefont {S.~W.}\
  \bibnamefont {Nam}}, \bibinfo {author} {\bibfnamefont {D.}~\bibnamefont
  {Oblak}},\ and\ \bibinfo {author} {\bibfnamefont {W.}~\bibnamefont
  {Tittel}},\ }\bibfield  {title} {\bibinfo {title} {Quantum teleportation
  across a metropolitan fibre network},\ }\href
  {https://doi.org/10.1038/nphoton.2016.180} {\bibfield  {journal} {\bibinfo
  {journal} {Nature Photonics}\ }\textbf {\bibinfo {volume} {10}},\ \bibinfo
  {pages} {676} (\bibinfo {year} {2016})}\BibitemShut {NoStop}%
\bibitem [{\citenamefont {Sun}\ \emph {et~al.}(2016)\citenamefont {Sun},
  \citenamefont {Mao}, \citenamefont {Chen}, \citenamefont {Zhang},
  \citenamefont {Jiang}, \citenamefont {Zhang}, \citenamefont {Zhang},
  \citenamefont {Miki}, \citenamefont {Yamashita}, \citenamefont {Terai},
  \citenamefont {Jiang}, \citenamefont {Chen}, \citenamefont {You},
  \citenamefont {Chen}, \citenamefont {Wang}, \citenamefont {Fan},
  \citenamefont {Zhang},\ and\ \citenamefont {Pan}}]{Sun2016}%
  \BibitemOpen
  \bibfield  {author} {\bibinfo {author} {\bibfnamefont {Q.-C.}\ \bibnamefont
  {Sun}}, \bibinfo {author} {\bibfnamefont {Y.-L.}\ \bibnamefont {Mao}},
  \bibinfo {author} {\bibfnamefont {S.-J.}\ \bibnamefont {Chen}}, \bibinfo
  {author} {\bibfnamefont {W.}~\bibnamefont {Zhang}}, \bibinfo {author}
  {\bibfnamefont {Y.-F.}\ \bibnamefont {Jiang}}, \bibinfo {author}
  {\bibfnamefont {Y.-B.}\ \bibnamefont {Zhang}}, \bibinfo {author}
  {\bibfnamefont {W.-J.}\ \bibnamefont {Zhang}}, \bibinfo {author}
  {\bibfnamefont {S.}~\bibnamefont {Miki}}, \bibinfo {author} {\bibfnamefont
  {T.}~\bibnamefont {Yamashita}}, \bibinfo {author} {\bibfnamefont
  {H.}~\bibnamefont {Terai}}, \bibinfo {author} {\bibfnamefont
  {X.}~\bibnamefont {Jiang}}, \bibinfo {author} {\bibfnamefont {T.-Y.}\
  \bibnamefont {Chen}}, \bibinfo {author} {\bibfnamefont {L.-X.}\ \bibnamefont
  {You}}, \bibinfo {author} {\bibfnamefont {X.-F.}\ \bibnamefont {Chen}},
  \bibinfo {author} {\bibfnamefont {Z.}~\bibnamefont {Wang}}, \bibinfo {author}
  {\bibfnamefont {J.-Y.}\ \bibnamefont {Fan}}, \bibinfo {author} {\bibfnamefont
  {Q.}~\bibnamefont {Zhang}},\ and\ \bibinfo {author} {\bibfnamefont {J.-W.}\
  \bibnamefont {Pan}},\ }\bibfield  {title} {\bibinfo {title} {Quantum
  teleportation with independent sources and prior entanglement distribution
  over a network},\ }\href {https://doi.org/10.1038/nphoton.2016.179}
  {\bibfield  {journal} {\bibinfo  {journal} {Nature Photonics}\ }\textbf
  {\bibinfo {volume} {10}},\ \bibinfo {pages} {671} (\bibinfo {year}
  {2016})}\BibitemShut {NoStop}%
\bibitem [{\citenamefont {Takesue}\ and\ \citenamefont
  {Miquel}(2009)}]{Takesue2009}%
  \BibitemOpen
  \bibfield  {author} {\bibinfo {author} {\bibfnamefont {H.}~\bibnamefont
  {Takesue}}\ and\ \bibinfo {author} {\bibfnamefont {B.}~\bibnamefont
  {Miquel}},\ }\bibfield  {title} {\bibinfo {title} {Entanglement swapping
  using telecom-band photons generated in fibers},\ }\href@noop {} {\bibfield
  {journal} {\bibinfo  {journal} {Opt. Express}\ }\textbf {\bibinfo {volume}
  {17}},\ \bibinfo {pages} {10748} (\bibinfo {year} {2009})}\BibitemShut
  {NoStop}%
\bibitem [{\citenamefont {Liao}\ \emph {et~al.}(2017)\citenamefont {Liao},
  \citenamefont {Yong}, \citenamefont {Liu}, \citenamefont {Shentu},
  \citenamefont {Li}, \citenamefont {Lin}, \citenamefont {Dai}, \citenamefont
  {Zhao}, \citenamefont {Li}, \citenamefont {Guan}, \citenamefont {Chen},
  \citenamefont {Gong}, \citenamefont {Li}, \citenamefont {Lin}, \citenamefont
  {Pan}, \citenamefont {Pelc}, \citenamefont {Fejer}, \citenamefont {Zhang},
  \citenamefont {Liu}, \citenamefont {Yin}, \citenamefont {Ren}, \citenamefont
  {Wang}, \citenamefont {Zhang}, \citenamefont {Peng},\ and\ \citenamefont
  {Pan}}]{liao2017long}%
  \BibitemOpen
  \bibfield  {author} {\bibinfo {author} {\bibfnamefont {S.-K.}\ \bibnamefont
  {Liao}}, \bibinfo {author} {\bibfnamefont {H.-L.}\ \bibnamefont {Yong}},
  \bibinfo {author} {\bibfnamefont {C.}~\bibnamefont {Liu}}, \bibinfo {author}
  {\bibfnamefont {G.-L.}\ \bibnamefont {Shentu}}, \bibinfo {author}
  {\bibfnamefont {D.-D.}\ \bibnamefont {Li}}, \bibinfo {author} {\bibfnamefont
  {J.}~\bibnamefont {Lin}}, \bibinfo {author} {\bibfnamefont {H.}~\bibnamefont
  {Dai}}, \bibinfo {author} {\bibfnamefont {S.-Q.}\ \bibnamefont {Zhao}},
  \bibinfo {author} {\bibfnamefont {B.}~\bibnamefont {Li}}, \bibinfo {author}
  {\bibfnamefont {J.-Y.}\ \bibnamefont {Guan}}, \bibinfo {author}
  {\bibfnamefont {W.}~\bibnamefont {Chen}}, \bibinfo {author} {\bibfnamefont
  {Y.-H.}\ \bibnamefont {Gong}}, \bibinfo {author} {\bibfnamefont
  {Y.}~\bibnamefont {Li}}, \bibinfo {author} {\bibfnamefont {Z.-H.}\
  \bibnamefont {Lin}}, \bibinfo {author} {\bibfnamefont {G.-S.}\ \bibnamefont
  {Pan}}, \bibinfo {author} {\bibfnamefont {J.~S.}\ \bibnamefont {Pelc}},
  \bibinfo {author} {\bibfnamefont {M.~M.}\ \bibnamefont {Fejer}}, \bibinfo
  {author} {\bibfnamefont {W.-Z.}\ \bibnamefont {Zhang}}, \bibinfo {author}
  {\bibfnamefont {W.-Y.}\ \bibnamefont {Liu}}, \bibinfo {author} {\bibfnamefont
  {J.}~\bibnamefont {Yin}}, \bibinfo {author} {\bibfnamefont {J.-G.}\
  \bibnamefont {Ren}}, \bibinfo {author} {\bibfnamefont {X.-B.}\ \bibnamefont
  {Wang}}, \bibinfo {author} {\bibfnamefont {Q.}~\bibnamefont {Zhang}},
  \bibinfo {author} {\bibfnamefont {C.-Z.}\ \bibnamefont {Peng}},\ and\
  \bibinfo {author} {\bibfnamefont {J.-W.}\ \bibnamefont {Pan}},\ }\bibfield
  {title} {\bibinfo {title} {Long-distance free-space quantum key distribution
  in daylight towards inter-satellite communication},\ }\href
  {https://doi.org/10.1038/nphoton.2017.116} {\bibfield  {journal} {\bibinfo
  {journal} {Nature Photonics}\ }\textbf {\bibinfo {volume} {11}},\ \bibinfo
  {pages} {509} (\bibinfo {year} {2017})}\BibitemShut {NoStop}%
\bibitem [{\citenamefont {Lvovsky}\ \emph {et~al.}(2009)\citenamefont
  {Lvovsky}, \citenamefont {Sanders},\ and\ \citenamefont
  {Tittel}}]{lvovsky2009optical}%
  \BibitemOpen
  \bibfield  {author} {\bibinfo {author} {\bibfnamefont {A.~I.}\ \bibnamefont
  {Lvovsky}}, \bibinfo {author} {\bibfnamefont {B.~C.}\ \bibnamefont
  {Sanders}},\ and\ \bibinfo {author} {\bibfnamefont {W.}~\bibnamefont
  {Tittel}},\ }\bibfield  {title} {\bibinfo {title} {Optical quantum memory},\
  }\href@noop {} {\bibfield  {journal} {\bibinfo  {journal} {Nature photonics}\
  }\textbf {\bibinfo {volume} {3}},\ \bibinfo {pages} {706} (\bibinfo {year}
  {2009})}\BibitemShut {NoStop}%
\bibitem [{\citenamefont {Lauk}\ \emph {et~al.}(2020)\citenamefont {Lauk},
  \citenamefont {Sinclair}, \citenamefont {Barzanjeh}, \citenamefont {Covey},
  \citenamefont {Saffman}, \citenamefont {Spiropulu},\ and\ \citenamefont
  {Simon}}]{lauk2020perspectives}%
  \BibitemOpen
  \bibfield  {author} {\bibinfo {author} {\bibfnamefont {N.}~\bibnamefont
  {Lauk}}, \bibinfo {author} {\bibfnamefont {N.}~\bibnamefont {Sinclair}},
  \bibinfo {author} {\bibfnamefont {S.}~\bibnamefont {Barzanjeh}}, \bibinfo
  {author} {\bibfnamefont {J.~P.}\ \bibnamefont {Covey}}, \bibinfo {author}
  {\bibfnamefont {M.}~\bibnamefont {Saffman}}, \bibinfo {author} {\bibfnamefont
  {M.}~\bibnamefont {Spiropulu}},\ and\ \bibinfo {author} {\bibfnamefont
  {C.}~\bibnamefont {Simon}},\ }\bibfield  {title} {\bibinfo {title}
  {Perspectives on quantum transduction},\ }\href@noop {} {\bibfield  {journal}
  {\bibinfo  {journal} {Quantum Science and Technology}\ }\textbf {\bibinfo
  {volume} {5}},\ \bibinfo {pages} {020501} (\bibinfo {year}
  {2020})}\BibitemShut {NoStop}%
\bibitem [{\citenamefont {Lambert}\ \emph {et~al.}(2020)\citenamefont
  {Lambert}, \citenamefont {Rueda}, \citenamefont {Sedlmeir},\ and\
  \citenamefont {Schwefel}}]{lambert2020coherent}%
  \BibitemOpen
  \bibfield  {author} {\bibinfo {author} {\bibfnamefont {N.~J.}\ \bibnamefont
  {Lambert}}, \bibinfo {author} {\bibfnamefont {A.}~\bibnamefont {Rueda}},
  \bibinfo {author} {\bibfnamefont {F.}~\bibnamefont {Sedlmeir}},\ and\
  \bibinfo {author} {\bibfnamefont {H.~G.}\ \bibnamefont {Schwefel}},\
  }\bibfield  {title} {\bibinfo {title} {Coherent conversion between microwave
  and optical photons—an overview of physical implementations},\ }\href@noop
  {} {\bibfield  {journal} {\bibinfo  {journal} {Advanced Quantum
  Technologies}\ }\textbf {\bibinfo {volume} {3}},\ \bibinfo {pages} {1900077}
  (\bibinfo {year} {2020})}\BibitemShut {NoStop}%
\bibitem [{\citenamefont {Braginsky}\ and\ \citenamefont
  {Khalili}(1996)}]{braginsky1996quantum}%
  \BibitemOpen
  \bibfield  {author} {\bibinfo {author} {\bibfnamefont {V.~B.}\ \bibnamefont
  {Braginsky}}\ and\ \bibinfo {author} {\bibfnamefont {F.~Y.}\ \bibnamefont
  {Khalili}},\ }\bibfield  {title} {\bibinfo {title} {Quantum nondemolition
  measurements: the route from toys to tools},\ }\href@noop {} {\bibfield
  {journal} {\bibinfo  {journal} {Reviews of Modern Physics}\ }\textbf
  {\bibinfo {volume} {68}},\ \bibinfo {pages} {1} (\bibinfo {year}
  {1996})}\BibitemShut {NoStop}%
\bibitem [{\citenamefont {Marcikic}\ \emph {et~al.}(2003)\citenamefont
  {Marcikic}, \citenamefont {De~Riedmatten}, \citenamefont {Tittel},
  \citenamefont {Zbinden},\ and\ \citenamefont {Gisin}}]{marcikic2003long}%
  \BibitemOpen
  \bibfield  {author} {\bibinfo {author} {\bibfnamefont {I.}~\bibnamefont
  {Marcikic}}, \bibinfo {author} {\bibfnamefont {H.}~\bibnamefont
  {De~Riedmatten}}, \bibinfo {author} {\bibfnamefont {W.}~\bibnamefont
  {Tittel}}, \bibinfo {author} {\bibfnamefont {H.}~\bibnamefont {Zbinden}},\
  and\ \bibinfo {author} {\bibfnamefont {N.}~\bibnamefont {Gisin}},\ }\bibfield
   {title} {\bibinfo {title} {Long-distance teleportation of qubits at
  telecommunication wavelengths},\ }\href@noop {} {\bibfield  {journal}
  {\bibinfo  {journal} {Nature}\ }\textbf {\bibinfo {volume} {421}},\ \bibinfo
  {pages} {509} (\bibinfo {year} {2003})}\BibitemShut {NoStop}%
\bibitem [{\citenamefont {De~Riedmatten}\ \emph {et~al.}(2004)\citenamefont
  {De~Riedmatten}, \citenamefont {Marcikic}, \citenamefont {Tittel},
  \citenamefont {Zbinden}, \citenamefont {Collins},\ and\ \citenamefont
  {Gisin}}]{de2004long}%
  \BibitemOpen
  \bibfield  {author} {\bibinfo {author} {\bibfnamefont {H.}~\bibnamefont
  {De~Riedmatten}}, \bibinfo {author} {\bibfnamefont {I.}~\bibnamefont
  {Marcikic}}, \bibinfo {author} {\bibfnamefont {W.}~\bibnamefont {Tittel}},
  \bibinfo {author} {\bibfnamefont {H.}~\bibnamefont {Zbinden}}, \bibinfo
  {author} {\bibfnamefont {D.}~\bibnamefont {Collins}},\ and\ \bibinfo {author}
  {\bibfnamefont {N.}~\bibnamefont {Gisin}},\ }\bibfield  {title} {\bibinfo
  {title} {Long distance quantum teleportation in a quantum relay
  configuration},\ }\href@noop {} {\bibfield  {journal} {\bibinfo  {journal}
  {Physical Review Letters}\ }\textbf {\bibinfo {volume} {92}},\ \bibinfo
  {pages} {047904} (\bibinfo {year} {2004})}\BibitemShut {NoStop}%
\bibitem [{\citenamefont {Takesue}\ \emph {et~al.}(2015)\citenamefont
  {Takesue}, \citenamefont {Dyer}, \citenamefont {Stevens}, \citenamefont
  {Verma}, \citenamefont {Mirin},\ and\ \citenamefont
  {Nam}}]{takesue2015quantum}%
  \BibitemOpen
  \bibfield  {author} {\bibinfo {author} {\bibfnamefont {H.}~\bibnamefont
  {Takesue}}, \bibinfo {author} {\bibfnamefont {S.~D.}\ \bibnamefont {Dyer}},
  \bibinfo {author} {\bibfnamefont {M.~J.}\ \bibnamefont {Stevens}}, \bibinfo
  {author} {\bibfnamefont {V.}~\bibnamefont {Verma}}, \bibinfo {author}
  {\bibfnamefont {R.~P.}\ \bibnamefont {Mirin}},\ and\ \bibinfo {author}
  {\bibfnamefont {S.~W.}\ \bibnamefont {Nam}},\ }\bibfield  {title} {\bibinfo
  {title} {Quantum teleportation over 100 km of fiber using highly efficient
  superconducting nanowire single-photon detectors},\ }\href@noop {} {\bibfield
   {journal} {\bibinfo  {journal} {Optica}\ }\textbf {\bibinfo {volume} {2}},\
  \bibinfo {pages} {832} (\bibinfo {year} {2015})}\BibitemShut {NoStop}%
\bibitem [{\citenamefont {Landry}\ \emph {et~al.}(2007)\citenamefont {Landry},
  \citenamefont {van Houwelingen}, \citenamefont {Beveratos}, \citenamefont
  {Zbinden},\ and\ \citenamefont {Gisin}}]{landry2007quantum}%
  \BibitemOpen
  \bibfield  {author} {\bibinfo {author} {\bibfnamefont {O.}~\bibnamefont
  {Landry}}, \bibinfo {author} {\bibfnamefont {J.~A.~W.}\ \bibnamefont {van
  Houwelingen}}, \bibinfo {author} {\bibfnamefont {A.}~\bibnamefont
  {Beveratos}}, \bibinfo {author} {\bibfnamefont {H.}~\bibnamefont {Zbinden}},\
  and\ \bibinfo {author} {\bibfnamefont {N.}~\bibnamefont {Gisin}},\ }\bibfield
   {title} {\bibinfo {title} {Quantum teleportation over the swisscom
  telecommunication network},\ }\href@noop {} {\bibfield  {journal} {\bibinfo
  {journal} {JOSA B}\ }\textbf {\bibinfo {volume} {24}},\ \bibinfo {pages}
  {398} (\bibinfo {year} {2007})}\BibitemShut {NoStop}%
\bibitem [{\citenamefont {Halder}\ \emph {et~al.}(2007)\citenamefont {Halder},
  \citenamefont {Beveratos}, \citenamefont {Gisin}, \citenamefont {Scarani},
  \citenamefont {Simon},\ and\ \citenamefont {Zbinden}}]{halder2007entangling}%
  \BibitemOpen
  \bibfield  {author} {\bibinfo {author} {\bibfnamefont {M.}~\bibnamefont
  {Halder}}, \bibinfo {author} {\bibfnamefont {A.}~\bibnamefont {Beveratos}},
  \bibinfo {author} {\bibfnamefont {N.}~\bibnamefont {Gisin}}, \bibinfo
  {author} {\bibfnamefont {V.}~\bibnamefont {Scarani}}, \bibinfo {author}
  {\bibfnamefont {C.}~\bibnamefont {Simon}},\ and\ \bibinfo {author}
  {\bibfnamefont {H.}~\bibnamefont {Zbinden}},\ }\bibfield  {title} {\bibinfo
  {title} {Entangling independent photons by time measurement},\ }\href@noop {}
  {\bibfield  {journal} {\bibinfo  {journal} {Nature physics}\ }\textbf
  {\bibinfo {volume} {3}},\ \bibinfo {pages} {692} (\bibinfo {year}
  {2007})}\BibitemShut {NoStop}%
\bibitem [{\citenamefont {Bussi{\`e}res}\ \emph {et~al.}(2014)\citenamefont
  {Bussi{\`e}res}, \citenamefont {Clausen}, \citenamefont {Tiranov},
  \citenamefont {Korzh}, \citenamefont {Verma}, \citenamefont {Nam},
  \citenamefont {Marsili}, \citenamefont {Ferrier}, \citenamefont {Goldner},
  \citenamefont {Herrmann}, \citenamefont {Silberhorn}, \citenamefont {Sohler},
  \citenamefont {Afzelius},\ and\ \citenamefont
  {Gisin}}]{bussieres2014quantum}%
  \BibitemOpen
  \bibfield  {author} {\bibinfo {author} {\bibfnamefont {F.}~\bibnamefont
  {Bussi{\`e}res}}, \bibinfo {author} {\bibfnamefont {C.}~\bibnamefont
  {Clausen}}, \bibinfo {author} {\bibfnamefont {A.}~\bibnamefont {Tiranov}},
  \bibinfo {author} {\bibfnamefont {B.}~\bibnamefont {Korzh}}, \bibinfo
  {author} {\bibfnamefont {V.~B.}\ \bibnamefont {Verma}}, \bibinfo {author}
  {\bibfnamefont {S.~W.}\ \bibnamefont {Nam}}, \bibinfo {author} {\bibfnamefont
  {F.}~\bibnamefont {Marsili}}, \bibinfo {author} {\bibfnamefont
  {A.}~\bibnamefont {Ferrier}}, \bibinfo {author} {\bibfnamefont
  {P.}~\bibnamefont {Goldner}}, \bibinfo {author} {\bibfnamefont
  {H.}~\bibnamefont {Herrmann}}, \bibinfo {author} {\bibfnamefont
  {C.}~\bibnamefont {Silberhorn}}, \bibinfo {author} {\bibfnamefont
  {W.}~\bibnamefont {Sohler}}, \bibinfo {author} {\bibfnamefont
  {M.}~\bibnamefont {Afzelius}},\ and\ \bibinfo {author} {\bibfnamefont
  {N.}~\bibnamefont {Gisin}},\ }\bibfield  {title} {\bibinfo {title} {Quantum
  teleportation from a telecom-wavelength photon to a solid-state quantum
  memory},\ }\href {https://doi.org/10.1038/nphoton.2014.215} {\bibfield
  {journal} {\bibinfo  {journal} {Nature Photonics}\ }\textbf {\bibinfo
  {volume} {8}},\ \bibinfo {pages} {775} (\bibinfo {year} {2014})}\BibitemShut
  {NoStop}%
\bibitem [{\citenamefont {Gao}\ and\ \citenamefont
  {Jafferis}(2019)}]{gao2019traversable}%
  \BibitemOpen
  \bibfield  {author} {\bibinfo {author} {\bibfnamefont {P.}~\bibnamefont
  {Gao}}\ and\ \bibinfo {author} {\bibfnamefont {D.~L.}\ \bibnamefont
  {Jafferis}},\ }\href@noop {} {\bibinfo {title} {A traversable wormhole
  teleportation protocol in the syk model}} (\bibinfo {year} {2019}),\ \Eprint
  {https://arxiv.org/abs/1911.07416} {arXiv:1911.07416 [hep-th]} \BibitemShut
  {NoStop}%
\bibitem [{\citenamefont {Miyazono}\ \emph {et~al.}(2016)\citenamefont
  {Miyazono}, \citenamefont {Zhong}, \citenamefont {Craiciu}, \citenamefont
  {Kindem},\ and\ \citenamefont {Faraon}}]{miyazono2016coupling}%
  \BibitemOpen
  \bibfield  {author} {\bibinfo {author} {\bibfnamefont {E.}~\bibnamefont
  {Miyazono}}, \bibinfo {author} {\bibfnamefont {T.}~\bibnamefont {Zhong}},
  \bibinfo {author} {\bibfnamefont {I.}~\bibnamefont {Craiciu}}, \bibinfo
  {author} {\bibfnamefont {J.~M.}\ \bibnamefont {Kindem}},\ and\ \bibinfo
  {author} {\bibfnamefont {A.}~\bibnamefont {Faraon}},\ }\bibfield  {title}
  {\bibinfo {title} {Coupling of erbium dopants to yttrium orthosilicate
  photonic crystal cavities for on-chip optical quantum memories},\ }\href@noop
  {} {\bibfield  {journal} {\bibinfo  {journal} {Applied Physics Letters}\
  }\textbf {\bibinfo {volume} {108}},\ \bibinfo {pages} {011111} (\bibinfo
  {year} {2016})}\BibitemShut {NoStop}%
\bibitem [{\citenamefont {Lauritzen}\ \emph {et~al.}(2010)\citenamefont
  {Lauritzen}, \citenamefont {Min{\'a}{\v{r}}}, \citenamefont {De~Riedmatten},
  \citenamefont {Afzelius}, \citenamefont {Sangouard}, \citenamefont {Simon},\
  and\ \citenamefont {Gisin}}]{lauritzen2010}%
  \BibitemOpen
  \bibfield  {author} {\bibinfo {author} {\bibfnamefont {B.}~\bibnamefont
  {Lauritzen}}, \bibinfo {author} {\bibfnamefont {J.}~\bibnamefont
  {Min{\'a}{\v{r}}}}, \bibinfo {author} {\bibfnamefont {H.}~\bibnamefont
  {De~Riedmatten}}, \bibinfo {author} {\bibfnamefont {M.}~\bibnamefont
  {Afzelius}}, \bibinfo {author} {\bibfnamefont {N.}~\bibnamefont {Sangouard}},
  \bibinfo {author} {\bibfnamefont {C.}~\bibnamefont {Simon}},\ and\ \bibinfo
  {author} {\bibfnamefont {N.}~\bibnamefont {Gisin}},\ }\bibfield  {title}
  {\bibinfo {title} {Telecommunication-wavelength solid-state memory at the
  single photon level},\ }\href@noop {} {\bibfield  {journal} {\bibinfo
  {journal} {Physical review letters}\ }\textbf {\bibinfo {volume} {104}},\
  \bibinfo {pages} {080502} (\bibinfo {year} {2010})}\BibitemShut {NoStop}%
\bibitem [{\citenamefont {Welinski}\ \emph {et~al.}(2019)\citenamefont
  {Welinski}, \citenamefont {Woodburn}, \citenamefont {Lauk}, \citenamefont
  {Cone}, \citenamefont {Simon}, \citenamefont {Goldner},\ and\ \citenamefont
  {Thiel}}]{welinski2019}%
  \BibitemOpen
  \bibfield  {author} {\bibinfo {author} {\bibfnamefont {S.}~\bibnamefont
  {Welinski}}, \bibinfo {author} {\bibfnamefont {P.~J.~T.}\ \bibnamefont
  {Woodburn}}, \bibinfo {author} {\bibfnamefont {N.}~\bibnamefont {Lauk}},
  \bibinfo {author} {\bibfnamefont {R.~L.}\ \bibnamefont {Cone}}, \bibinfo
  {author} {\bibfnamefont {C.}~\bibnamefont {Simon}}, \bibinfo {author}
  {\bibfnamefont {P.}~\bibnamefont {Goldner}},\ and\ \bibinfo {author}
  {\bibfnamefont {C.~W.}\ \bibnamefont {Thiel}},\ }\bibfield  {title} {\bibinfo
  {title} {Electron spin coherence in optically excited states of rare-earth
  ions for microwave to optical quantum transducers},\ }\href
  {https://doi.org/10.1103/PhysRevLett.122.247401} {\bibfield  {journal}
  {\bibinfo  {journal} {Phys. Rev. Lett.}\ }\textbf {\bibinfo {volume} {122}},\
  \bibinfo {pages} {247401} (\bibinfo {year} {2019})}\BibitemShut {NoStop}%
\bibitem [{\citenamefont {Hong}\ \emph {et~al.}(1987)\citenamefont {Hong},
  \citenamefont {Ou},\ and\ \citenamefont {Mandel}}]{hong1987oumandel}%
  \BibitemOpen
  \bibfield  {author} {\bibinfo {author} {\bibfnamefont {C.~K.}\ \bibnamefont
  {Hong}}, \bibinfo {author} {\bibfnamefont {Z.~Y.}\ \bibnamefont {Ou}},\ and\
  \bibinfo {author} {\bibfnamefont {L.}~\bibnamefont {Mandel}},\ }\bibfield
  {title} {\bibinfo {title} {Measurement of subpicosecond time intervals
  between two photons by interference},\ }\href
  {https://doi.org/10.1103/PhysRevLett.59.2044} {\bibfield  {journal} {\bibinfo
   {journal} {Phys. Rev. Lett.}\ }\textbf {\bibinfo {volume} {59}},\ \bibinfo
  {pages} {2044} (\bibinfo {year} {1987})}\BibitemShut {NoStop}%
\bibitem [{\citenamefont {Altepeter}\ \emph {et~al.}(2005)\citenamefont
  {Altepeter}, \citenamefont {Jeffrey},\ and\ \citenamefont
  {Kwiat}}]{altepeter2005photonic}%
  \BibitemOpen
  \bibfield  {author} {\bibinfo {author} {\bibfnamefont {J.~B.}\ \bibnamefont
  {Altepeter}}, \bibinfo {author} {\bibfnamefont {E.~R.}\ \bibnamefont
  {Jeffrey}},\ and\ \bibinfo {author} {\bibfnamefont {P.~G.}\ \bibnamefont
  {Kwiat}},\ }\bibfield  {title} {\bibinfo {title} {Photonic state
  tomography},\ }\href@noop {} {\bibfield  {journal} {\bibinfo  {journal}
  {Advances in Atomic, Molecular, and Optical Physics}\ }\textbf {\bibinfo
  {volume} {52}},\ \bibinfo {pages} {105} (\bibinfo {year} {2005})}\BibitemShut
  {NoStop}%
\bibitem [{\citenamefont {Ma}\ \emph {et~al.}(2005)\citenamefont {Ma},
  \citenamefont {Qi}, \citenamefont {Zhao},\ and\ \citenamefont
  {Lo}}]{ma2005practical}%
  \BibitemOpen
  \bibfield  {author} {\bibinfo {author} {\bibfnamefont {X.}~\bibnamefont
  {Ma}}, \bibinfo {author} {\bibfnamefont {B.}~\bibnamefont {Qi}}, \bibinfo
  {author} {\bibfnamefont {Y.}~\bibnamefont {Zhao}},\ and\ \bibinfo {author}
  {\bibfnamefont {H.-K.}\ \bibnamefont {Lo}},\ }\bibfield  {title} {\bibinfo
  {title} {Practical decoy state for quantum key distribution},\ }\href@noop {}
  {\bibfield  {journal} {\bibinfo  {journal} {Physical Review A}\ }\textbf
  {\bibinfo {volume} {72}},\ \bibinfo {pages} {012326} (\bibinfo {year}
  {2005})}\BibitemShut {NoStop}%
\bibitem [{\citenamefont {Iskander}\ \emph {et~al.}(2019)\citenamefont
  {Iskander}, \citenamefont {Sinclair}, \citenamefont {Pe\~{n}a}, \citenamefont
  {Xie},\ and\ \citenamefont {Spiropulu}}]{iskander2019}%
  \BibitemOpen
  \bibfield  {author} {\bibinfo {author} {\bibfnamefont {G.}~\bibnamefont
  {Iskander}}, \bibinfo {author} {\bibfnamefont {N.}~\bibnamefont {Sinclair}},
  \bibinfo {author} {\bibfnamefont {C.}~\bibnamefont {Pe\~{n}a}}, \bibinfo
  {author} {\bibfnamefont {S.}~\bibnamefont {Xie}},\ and\ \bibinfo {author}
  {\bibfnamefont {M.}~\bibnamefont {Spiropulu}},\ }\bibfield  {title} {\bibinfo
  {title} {Stabilization of an electro-optic modulator for quantum
  communication using a low-cost microcontroller},\ }\href@noop {} {\bibfield
  {journal} {\bibinfo  {journal} {Caltech Undergraduate Research Journal}\
  }\textbf {\bibinfo {volume} {20}} (\bibinfo {year} {2019})}\BibitemShut
  {NoStop}%
\bibitem [{\citenamefont {Rarity}(1995)}]{rarity1995interference}%
  \BibitemOpen
  \bibfield  {author} {\bibinfo {author} {\bibfnamefont {J.}~\bibnamefont
  {Rarity}},\ }\bibfield  {title} {\bibinfo {title} {Interference of single
  photons from separate sources a},\ }\href@noop {} {\bibfield  {journal}
  {\bibinfo  {journal} {Annals of the New York academy of Sciences}\ }\textbf
  {\bibinfo {volume} {755}},\ \bibinfo {pages} {624} (\bibinfo {year}
  {1995})}\BibitemShut {NoStop}%
\bibitem [{\citenamefont {Marsili}\ \emph {et~al.}(2013)\citenamefont
  {Marsili}, \citenamefont {Verma}, \citenamefont {Stern}, \citenamefont
  {Harrington}, \citenamefont {Lita}, \citenamefont {Gerrits}, \citenamefont
  {Vayshenker}, \citenamefont {Baek}, \citenamefont {Shaw}, \citenamefont
  {Mirin},\ and\ \citenamefont {Nam}}]{marsili2013detecting}%
  \BibitemOpen
  \bibfield  {author} {\bibinfo {author} {\bibfnamefont {F.}~\bibnamefont
  {Marsili}}, \bibinfo {author} {\bibfnamefont {V.~B.}\ \bibnamefont {Verma}},
  \bibinfo {author} {\bibfnamefont {J.~A.}\ \bibnamefont {Stern}}, \bibinfo
  {author} {\bibfnamefont {S.}~\bibnamefont {Harrington}}, \bibinfo {author}
  {\bibfnamefont {A.~E.}\ \bibnamefont {Lita}}, \bibinfo {author}
  {\bibfnamefont {T.}~\bibnamefont {Gerrits}}, \bibinfo {author} {\bibfnamefont
  {I.}~\bibnamefont {Vayshenker}}, \bibinfo {author} {\bibfnamefont
  {B.}~\bibnamefont {Baek}}, \bibinfo {author} {\bibfnamefont {M.~D.}\
  \bibnamefont {Shaw}}, \bibinfo {author} {\bibfnamefont {R.~P.}\ \bibnamefont
  {Mirin}},\ and\ \bibinfo {author} {\bibfnamefont {S.~W.}\ \bibnamefont
  {Nam}},\ }\bibfield  {title} {\bibinfo {title} {Detecting single infrared
  photons with 93{\%} system efficiency},\ }\href
  {https://doi.org/10.1038/nphoton.2013.13} {\bibfield  {journal} {\bibinfo
  {journal} {Nature Photonics}\ }\textbf {\bibinfo {volume} {7}},\ \bibinfo
  {pages} {210} (\bibinfo {year} {2013})}\BibitemShut {NoStop}%
\bibitem [{pho()}]{photonspot}%
  \BibitemOpen
  \href {https://www.photonspot.com/} {\bibinfo {title} {Photon spot}},\
  \bibinfo {howpublished} {\url{https://www.photonspot.com/}}\BibitemShut
  {NoStop}%
\bibitem [{\citenamefont {L{\"u}tkenhaus}\ \emph {et~al.}(1999)\citenamefont
  {L{\"u}tkenhaus}, \citenamefont {Calsamiglia},\ and\ \citenamefont
  {Suominen}}]{lutkenhaus1999bell}%
  \BibitemOpen
  \bibfield  {author} {\bibinfo {author} {\bibfnamefont {N.}~\bibnamefont
  {L{\"u}tkenhaus}}, \bibinfo {author} {\bibfnamefont {J.}~\bibnamefont
  {Calsamiglia}},\ and\ \bibinfo {author} {\bibfnamefont {K.-A.}\ \bibnamefont
  {Suominen}},\ }\bibfield  {title} {\bibinfo {title} {Bell measurements for
  teleportation},\ }\href@noop {} {\bibfield  {journal} {\bibinfo  {journal}
  {Physical Review A}\ }\textbf {\bibinfo {volume} {59}},\ \bibinfo {pages}
  {3295} (\bibinfo {year} {1999})}\BibitemShut {NoStop}%
\bibitem [{\citenamefont {Marcikic}\ \emph {et~al.}(2002)\citenamefont
  {Marcikic}, \citenamefont {de~Riedmatten}, \citenamefont {Tittel},
  \citenamefont {Scarani}, \citenamefont {Zbinden},\ and\ \citenamefont
  {Gisin}}]{marcikic2002femtosecond}%
  \BibitemOpen
  \bibfield  {author} {\bibinfo {author} {\bibfnamefont {I.}~\bibnamefont
  {Marcikic}}, \bibinfo {author} {\bibfnamefont {H.}~\bibnamefont
  {de~Riedmatten}}, \bibinfo {author} {\bibfnamefont {W.}~\bibnamefont
  {Tittel}}, \bibinfo {author} {\bibfnamefont {V.}~\bibnamefont {Scarani}},
  \bibinfo {author} {\bibfnamefont {H.}~\bibnamefont {Zbinden}},\ and\ \bibinfo
  {author} {\bibfnamefont {N.}~\bibnamefont {Gisin}},\ }\bibfield  {title}
  {\bibinfo {title} {Time-bin entangled qubits for quantum communication
  created by femtosecond pulses},\ }\href@noop {} {\bibfield  {journal}
  {\bibinfo  {journal} {Physical Review A}\ }\textbf {\bibinfo {volume} {66}},\
  \bibinfo {pages} {062308} (\bibinfo {year} {2002})}\BibitemShut {NoStop}%
\bibitem [{\citenamefont {Mandel}\ and\ \citenamefont
  {Wolf}(1995)}]{mandel1995optical}%
  \BibitemOpen
  \bibfield  {author} {\bibinfo {author} {\bibfnamefont {L.}~\bibnamefont
  {Mandel}}\ and\ \bibinfo {author} {\bibfnamefont {E.}~\bibnamefont {Wolf}},\
  }\href@noop {} {\emph {\bibinfo {title} {Optical coherence and quantum
  optics}}}\ (\bibinfo  {publisher} {Cambridge university press},\ \bibinfo
  {year} {1995})\BibitemShut {NoStop}%
\bibitem [{\citenamefont {Zhong}\ and\ \citenamefont
  {Wong}(2013)}]{zhong2013nonlocal}%
  \BibitemOpen
  \bibfield  {author} {\bibinfo {author} {\bibfnamefont {T.}~\bibnamefont
  {Zhong}}\ and\ \bibinfo {author} {\bibfnamefont {F.~N.}\ \bibnamefont
  {Wong}},\ }\bibfield  {title} {\bibinfo {title} {Nonlocal cancellation of
  dispersion in franson interferometry},\ }\href@noop {} {\bibfield  {journal}
  {\bibinfo  {journal} {Physical Review A}\ }\textbf {\bibinfo {volume} {88}},\
  \bibinfo {pages} {020103} (\bibinfo {year} {2013})}\BibitemShut {NoStop}%
\bibitem [{the()}]{theory_nikolai}%
  \BibitemOpen
  \bibinfo {title} {Manuscript in preparation}\BibitemShut {NoStop}%
\bibitem [{\citenamefont {Werner}(1989)}]{werner1989quantum}%
  \BibitemOpen
\bibfield  {title} {  }\bibfield  {author} {\bibinfo {author} {\bibfnamefont
  {R.~F.}\ \bibnamefont {Werner}},\ }\bibfield  {title} {\bibinfo {title}
  {Quantum states with einstein-podolsky-rosen correlations admitting a
  hidden-variable model},\ }\href@noop {} {\bibfield  {journal} {\bibinfo
  {journal} {Physical Review A}\ }\textbf {\bibinfo {volume} {40}},\ \bibinfo
  {pages} {4277} (\bibinfo {year} {1989})}\BibitemShut {NoStop}%
\bibitem [{\citenamefont {Clauser}\ \emph {et~al.}(1969)\citenamefont
  {Clauser}, \citenamefont {Horne}, \citenamefont {Shimony},\ and\
  \citenamefont {Holt}}]{clauser1969proposed}%
  \BibitemOpen
  \bibfield  {author} {\bibinfo {author} {\bibfnamefont {J.~F.}\ \bibnamefont
  {Clauser}}, \bibinfo {author} {\bibfnamefont {M.~A.}\ \bibnamefont {Horne}},
  \bibinfo {author} {\bibfnamefont {A.}~\bibnamefont {Shimony}},\ and\ \bibinfo
  {author} {\bibfnamefont {R.~A.}\ \bibnamefont {Holt}},\ }\bibfield  {title}
  {\bibinfo {title} {Proposed experiment to test local hidden-variable
  theories},\ }\href@noop {} {\bibfield  {journal} {\bibinfo  {journal}
  {Physical review letters}\ }\textbf {\bibinfo {volume} {23}},\ \bibinfo
  {pages} {880} (\bibinfo {year} {1969})}\BibitemShut {NoStop}%
\bibitem [{\citenamefont {Takesue}(2007)}]{takesue2007hom}%
  \BibitemOpen
  \bibfield  {author} {\bibinfo {author} {\bibfnamefont {H.}~\bibnamefont
  {Takesue}},\ }\bibfield  {title} {\bibinfo {title} {1.5 $\mu$ m band
  hong-ou-mandel experiment using photon pairs generated in two independent
  dispersion shifted fibers},\ }\href@noop {} {\bibfield  {journal} {\bibinfo
  {journal} {Applied physics letters}\ }\textbf {\bibinfo {volume} {90}},\
  \bibinfo {pages} {204101} (\bibinfo {year} {2007})}\BibitemShut {NoStop}%
\bibitem [{\citenamefont {Massar}\ and\ \citenamefont
  {Popescu}(1995)}]{massar2005optimal}%
  \BibitemOpen
  \bibfield  {author} {\bibinfo {author} {\bibfnamefont {S.}~\bibnamefont
  {Massar}}\ and\ \bibinfo {author} {\bibfnamefont {S.}~\bibnamefont
  {Popescu}},\ }\bibfield  {title} {\bibinfo {title} {Optimal extraction of
  information from finite quantum ensembles},\ }\href
  {https://doi.org/10.1103/PhysRevLett.74.1259} {\bibfield  {journal} {\bibinfo
   {journal} {Phys. Rev. Lett.}\ }\textbf {\bibinfo {volume} {74}},\ \bibinfo
  {pages} {1259} (\bibinfo {year} {1995})}\BibitemShut {NoStop}%
\bibitem [{\citenamefont {Specht}\ \emph {et~al.}(2011)\citenamefont {Specht},
  \citenamefont {N{\"o}lleke}, \citenamefont {Reiserer}, \citenamefont
  {Uphoff}, \citenamefont {Figueroa}, \citenamefont {Ritter},\ and\
  \citenamefont {Rempe}}]{specht2011single}%
  \BibitemOpen
  \bibfield  {author} {\bibinfo {author} {\bibfnamefont {H.~P.}\ \bibnamefont
  {Specht}}, \bibinfo {author} {\bibfnamefont {C.}~\bibnamefont {N{\"o}lleke}},
  \bibinfo {author} {\bibfnamefont {A.}~\bibnamefont {Reiserer}}, \bibinfo
  {author} {\bibfnamefont {M.}~\bibnamefont {Uphoff}}, \bibinfo {author}
  {\bibfnamefont {E.}~\bibnamefont {Figueroa}}, \bibinfo {author}
  {\bibfnamefont {S.}~\bibnamefont {Ritter}},\ and\ \bibinfo {author}
  {\bibfnamefont {G.}~\bibnamefont {Rempe}},\ }\bibfield  {title} {\bibinfo
  {title} {A single-atom quantum memory},\ }\href@noop {} {\bibfield  {journal}
  {\bibinfo  {journal} {Nature}\ }\textbf {\bibinfo {volume} {473}},\ \bibinfo
  {pages} {190} (\bibinfo {year} {2011})}\BibitemShut {NoStop}%
\bibitem [{\citenamefont {Sinclair}\ \emph {et~al.}(2014)\citenamefont
  {Sinclair}, \citenamefont {Saglamyurek}, \citenamefont {Mallahzadeh},
  \citenamefont {Slater}, \citenamefont {George}, \citenamefont {Ricken},
  \citenamefont {Hedges}, \citenamefont {Oblak}, \citenamefont {Simon},
  \citenamefont {Sohler},\ and\ \citenamefont {Tittel}}]{sinclair2014spectral}%
  \BibitemOpen
  \bibfield  {author} {\bibinfo {author} {\bibfnamefont {N.}~\bibnamefont
  {Sinclair}}, \bibinfo {author} {\bibfnamefont {E.}~\bibnamefont
  {Saglamyurek}}, \bibinfo {author} {\bibfnamefont {H.}~\bibnamefont
  {Mallahzadeh}}, \bibinfo {author} {\bibfnamefont {J.~A.}\ \bibnamefont
  {Slater}}, \bibinfo {author} {\bibfnamefont {M.}~\bibnamefont {George}},
  \bibinfo {author} {\bibfnamefont {R.}~\bibnamefont {Ricken}}, \bibinfo
  {author} {\bibfnamefont {M.~P.}\ \bibnamefont {Hedges}}, \bibinfo {author}
  {\bibfnamefont {D.}~\bibnamefont {Oblak}}, \bibinfo {author} {\bibfnamefont
  {C.}~\bibnamefont {Simon}}, \bibinfo {author} {\bibfnamefont
  {W.}~\bibnamefont {Sohler}},\ and\ \bibinfo {author} {\bibfnamefont
  {W.}~\bibnamefont {Tittel}},\ }\bibfield  {title} {\bibinfo {title} {Spectral
  multiplexing for scalable quantum photonics using an atomic frequency comb
  quantum memory and feed-forward control},\ }\href
  {https://doi.org/10.1103/PhysRevLett.113.053603} {\bibfield  {journal}
  {\bibinfo  {journal} {Phys. Rev. Lett.}\ }\textbf {\bibinfo {volume} {113}},\
  \bibinfo {pages} {053603} (\bibinfo {year} {2014})}\BibitemShut {NoStop}%
\bibitem [{\citenamefont {Bruss}\ \emph {et~al.}(1998)\citenamefont {Bruss},
  \citenamefont {Ekert},\ and\ \citenamefont
  {Macchiavello}}]{bruss1998optimal}%
  \BibitemOpen
  \bibfield  {author} {\bibinfo {author} {\bibfnamefont {D.}~\bibnamefont
  {Bruss}}, \bibinfo {author} {\bibfnamefont {A.}~\bibnamefont {Ekert}},\ and\
  \bibinfo {author} {\bibfnamefont {C.}~\bibnamefont {Macchiavello}},\
  }\bibfield  {title} {\bibinfo {title} {Optimal universal quantum cloning and
  state estimation},\ }\href@noop {} {\bibfield  {journal} {\bibinfo  {journal}
  {Physical review letters}\ }\textbf {\bibinfo {volume} {81}},\ \bibinfo
  {pages} {2598} (\bibinfo {year} {1998})}\BibitemShut {NoStop}%
\bibitem [{\citenamefont {Bu{\v{z}}ek}\ and\ \citenamefont
  {Hillery}(1996)}]{buvzek1996quantum}%
  \BibitemOpen
  \bibfield  {author} {\bibinfo {author} {\bibfnamefont {V.}~\bibnamefont
  {Bu{\v{z}}ek}}\ and\ \bibinfo {author} {\bibfnamefont {M.}~\bibnamefont
  {Hillery}},\ }\bibfield  {title} {\bibinfo {title} {Quantum copying: Beyond
  the no-cloning theorem},\ }\href@noop {} {\bibfield  {journal} {\bibinfo
  {journal} {Physical Review A}\ }\textbf {\bibinfo {volume} {54}},\ \bibinfo
  {pages} {1844} (\bibinfo {year} {1996})}\BibitemShut {NoStop}%
\bibitem [{\citenamefont {Takeoka}\ \emph {et~al.}(2015)\citenamefont
  {Takeoka}, \citenamefont {Jin},\ and\ \citenamefont {Sasaki}}]{Takeoka2015}%
  \BibitemOpen
  \bibfield  {author} {\bibinfo {author} {\bibfnamefont {M.}~\bibnamefont
  {Takeoka}}, \bibinfo {author} {\bibfnamefont {R.-B.}\ \bibnamefont {Jin}},\
  and\ \bibinfo {author} {\bibfnamefont {M.}~\bibnamefont {Sasaki}},\
  }\bibfield  {title} {\bibinfo {title} {Full analysis of multi-photon pair
  effects in spontaneous parametric down conversion based photonic quantum
  information processing},\ }\href
  {https://doi.org/10.1088/1367-2630/17/4/043030} {\bibfield  {journal}
  {\bibinfo  {journal} {New Journal of Physics}\ }\textbf {\bibinfo {volume}
  {17}},\ \bibinfo {pages} {043030} (\bibinfo {year} {2015})}\BibitemShut
  {NoStop}%
\bibitem [{\citenamefont {Weedbrook}\ \emph {et~al.}(2012)\citenamefont
  {Weedbrook}, \citenamefont {Pirandola}, \citenamefont {Garc\'{\i}a-Patr\'on},
  \citenamefont {Cerf}, \citenamefont {Ralph}, \citenamefont {Shapiro},\ and\
  \citenamefont {Lloyd}}]{Weedbrook2012}%
  \BibitemOpen
  \bibfield  {author} {\bibinfo {author} {\bibfnamefont {C.}~\bibnamefont
  {Weedbrook}}, \bibinfo {author} {\bibfnamefont {S.}~\bibnamefont
  {Pirandola}}, \bibinfo {author} {\bibfnamefont {R.}~\bibnamefont
  {Garc\'{\i}a-Patr\'on}}, \bibinfo {author} {\bibfnamefont {N.~J.}\
  \bibnamefont {Cerf}}, \bibinfo {author} {\bibfnamefont {T.~C.}\ \bibnamefont
  {Ralph}}, \bibinfo {author} {\bibfnamefont {J.~H.}\ \bibnamefont {Shapiro}},\
  and\ \bibinfo {author} {\bibfnamefont {S.}~\bibnamefont {Lloyd}},\ }\bibfield
   {title} {\bibinfo {title} {Gaussian quantum information},\ }\href
  {https://doi.org/10.1103/RevModPhys.84.621} {\bibfield  {journal} {\bibinfo
  {journal} {Rev. Mod. Phys.}\ }\textbf {\bibinfo {volume} {84}},\ \bibinfo
  {pages} {621} (\bibinfo {year} {2012})}\BibitemShut {NoStop}%
\bibitem [{\citenamefont {Zhu}\ \emph {et~al.}(2020)\citenamefont {Zhu},
  \citenamefont {Colangelo}, \citenamefont {Chen}, \citenamefont {Korzh},
  \citenamefont {Wong}, \citenamefont {Shaw},\ and\ \citenamefont
  {Berggren}}]{PMID:32271591}%
  \BibitemOpen
  \bibfield  {author} {\bibinfo {author} {\bibfnamefont {D.}~\bibnamefont
  {Zhu}}, \bibinfo {author} {\bibfnamefont {M.}~\bibnamefont {Colangelo}},
  \bibinfo {author} {\bibfnamefont {C.}~\bibnamefont {Chen}}, \bibinfo {author}
  {\bibfnamefont {B.~A.}\ \bibnamefont {Korzh}}, \bibinfo {author}
  {\bibfnamefont {F.~N.}\ \bibnamefont {Wong}}, \bibinfo {author}
  {\bibfnamefont {M.~D.}\ \bibnamefont {Shaw}},\ and\ \bibinfo {author}
  {\bibfnamefont {K.~K.}\ \bibnamefont {Berggren}},\ }\bibfield  {title}
  {\bibinfo {title} {Resolving photon numbers using a superconducting nanowire
  with impedance-matching taper},\ }\href@noop {} {\bibfield  {journal}
  {\bibinfo  {journal} {Nano Letters}\ }\textbf {\bibinfo {volume} {20}},\
  \bibinfo {pages} {3858} (\bibinfo {year} {2020})}\BibitemShut {NoStop}%
\bibitem [{\citenamefont {Krovi}\ \emph {et~al.}(2016)\citenamefont {Krovi},
  \citenamefont {Guha}, \citenamefont {Dutton}, \citenamefont {Slater},
  \citenamefont {Simon},\ and\ \citenamefont {Tittel}}]{krovi2016practical}%
  \BibitemOpen
  \bibfield  {author} {\bibinfo {author} {\bibfnamefont {H.}~\bibnamefont
  {Krovi}}, \bibinfo {author} {\bibfnamefont {S.}~\bibnamefont {Guha}},
  \bibinfo {author} {\bibfnamefont {Z.}~\bibnamefont {Dutton}}, \bibinfo
  {author} {\bibfnamefont {J.~A.}\ \bibnamefont {Slater}}, \bibinfo {author}
  {\bibfnamefont {C.}~\bibnamefont {Simon}},\ and\ \bibinfo {author}
  {\bibfnamefont {W.}~\bibnamefont {Tittel}},\ }\bibfield  {title} {\bibinfo
  {title} {Practical quantum repeaters with parametric down-conversion
  sources},\ }\href@noop {} {\bibfield  {journal} {\bibinfo  {journal} {Applied
  Physics B}\ }\textbf {\bibinfo {volume} {122}},\ \bibinfo {pages} {52}
  (\bibinfo {year} {2016})}\BibitemShut {NoStop}%
\bibitem [{\citenamefont {Mosley}\ \emph {et~al.}(2008)\citenamefont {Mosley},
  \citenamefont {Lundeen}, \citenamefont {Smith}, \citenamefont {Wasylczyk},
  \citenamefont {U’Ren}, \citenamefont {Silberhorn},\ and\ \citenamefont
  {Walmsley}}]{mosley2008heralded}%
  \BibitemOpen
  \bibfield  {author} {\bibinfo {author} {\bibfnamefont {P.~J.}\ \bibnamefont
  {Mosley}}, \bibinfo {author} {\bibfnamefont {J.~S.}\ \bibnamefont {Lundeen}},
  \bibinfo {author} {\bibfnamefont {B.~J.}\ \bibnamefont {Smith}}, \bibinfo
  {author} {\bibfnamefont {P.}~\bibnamefont {Wasylczyk}}, \bibinfo {author}
  {\bibfnamefont {A.~B.}\ \bibnamefont {U’Ren}}, \bibinfo {author}
  {\bibfnamefont {C.}~\bibnamefont {Silberhorn}},\ and\ \bibinfo {author}
  {\bibfnamefont {I.~A.}\ \bibnamefont {Walmsley}},\ }\bibfield  {title}
  {\bibinfo {title} {Heralded generation of ultrafast single photons in pure
  quantum states},\ }\href@noop {} {\bibfield  {journal} {\bibinfo  {journal}
  {Physical Review Letters}\ }\textbf {\bibinfo {volume} {100}},\ \bibinfo
  {pages} {133601} (\bibinfo {year} {2008})}\BibitemShut {NoStop}%
\bibitem [{\citenamefont {Rassoul}\ \emph {et~al.}(1997)\citenamefont
  {Rassoul}, \citenamefont {Ivanov}, \citenamefont {Freysz}, \citenamefont
  {Ducasse},\ and\ \citenamefont {Hache}}]{rassoul1997second}%
  \BibitemOpen
  \bibfield  {author} {\bibinfo {author} {\bibfnamefont {R.~M.}\ \bibnamefont
  {Rassoul}}, \bibinfo {author} {\bibfnamefont {A.}~\bibnamefont {Ivanov}},
  \bibinfo {author} {\bibfnamefont {E.}~\bibnamefont {Freysz}}, \bibinfo
  {author} {\bibfnamefont {A.}~\bibnamefont {Ducasse}},\ and\ \bibinfo {author}
  {\bibfnamefont {F.}~\bibnamefont {Hache}},\ }\bibfield  {title} {\bibinfo
  {title} {Second-harmonic generation under phase-velocity and group-velocity
  mismatch: influence of cascading self-phase and cross-phase modulation},\
  }\href@noop {} {\bibfield  {journal} {\bibinfo  {journal} {Optics letters}\
  }\textbf {\bibinfo {volume} {22}},\ \bibinfo {pages} {268} (\bibinfo {year}
  {1997})}\BibitemShut {NoStop}%
\bibitem [{\citenamefont {Korzh}\ \emph {et~al.}(2020)\citenamefont {Korzh},
  \citenamefont {Zhao}, \citenamefont {Allmaras}, \citenamefont {Frasca},
  \citenamefont {Autry}, \citenamefont {Bersin}, \citenamefont {Beyer},
  \citenamefont {Briggs}, \citenamefont {Bumble}, \citenamefont {Colangelo}
  \emph {et~al.}}]{korzh2020demonstration}%
  \BibitemOpen
  \bibfield  {author} {\bibinfo {author} {\bibfnamefont {B.}~\bibnamefont
  {Korzh}}, \bibinfo {author} {\bibfnamefont {Q.-Y.}\ \bibnamefont {Zhao}},
  \bibinfo {author} {\bibfnamefont {J.~P.}\ \bibnamefont {Allmaras}}, \bibinfo
  {author} {\bibfnamefont {S.}~\bibnamefont {Frasca}}, \bibinfo {author}
  {\bibfnamefont {T.~M.}\ \bibnamefont {Autry}}, \bibinfo {author}
  {\bibfnamefont {E.~A.}\ \bibnamefont {Bersin}}, \bibinfo {author}
  {\bibfnamefont {A.~D.}\ \bibnamefont {Beyer}}, \bibinfo {author}
  {\bibfnamefont {R.~M.}\ \bibnamefont {Briggs}}, \bibinfo {author}
  {\bibfnamefont {B.}~\bibnamefont {Bumble}}, \bibinfo {author} {\bibfnamefont
  {M.}~\bibnamefont {Colangelo}}, \emph {et~al.},\ }\bibfield  {title}
  {\bibinfo {title} {Demonstration of sub-3 ps temporal resolution with a
  superconducting nanowire single-photon detector},\ }\href@noop {} {\bibfield
  {journal} {\bibinfo  {journal} {Nature Photonics}\ }\textbf {\bibinfo
  {volume} {14}},\ \bibinfo {pages} {250} (\bibinfo {year} {2020})}\BibitemShut
  {NoStop}%
\bibitem [{\citenamefont {Valivarthi}\ \emph {et~al.}(2014)\citenamefont
  {Valivarthi}, \citenamefont {Lucio-Martinez}, \citenamefont {Rubenok},
  \citenamefont {Chan}, \citenamefont {Marsili}, \citenamefont {Verma},
  \citenamefont {Shaw}, \citenamefont {Stern}, \citenamefont {Slater},
  \citenamefont {Oblak}, \citenamefont {Nam},\ and\ \citenamefont
  {Tittel}}]{valivarthi2014efficient}%
  \BibitemOpen
  \bibfield  {author} {\bibinfo {author} {\bibfnamefont {R.}~\bibnamefont
  {Valivarthi}}, \bibinfo {author} {\bibfnamefont {I.}~\bibnamefont
  {Lucio-Martinez}}, \bibinfo {author} {\bibfnamefont {A.}~\bibnamefont
  {Rubenok}}, \bibinfo {author} {\bibfnamefont {P.}~\bibnamefont {Chan}},
  \bibinfo {author} {\bibfnamefont {F.}~\bibnamefont {Marsili}}, \bibinfo
  {author} {\bibfnamefont {V.~B.}\ \bibnamefont {Verma}}, \bibinfo {author}
  {\bibfnamefont {M.~D.}\ \bibnamefont {Shaw}}, \bibinfo {author}
  {\bibfnamefont {J.~A.}\ \bibnamefont {Stern}}, \bibinfo {author}
  {\bibfnamefont {J.~A.}\ \bibnamefont {Slater}}, \bibinfo {author}
  {\bibfnamefont {D.}~\bibnamefont {Oblak}}, \bibinfo {author} {\bibfnamefont
  {S.~W.}\ \bibnamefont {Nam}},\ and\ \bibinfo {author} {\bibfnamefont
  {W.}~\bibnamefont {Tittel}},\ }\href@noop {} {\bibfield  {journal} {\bibinfo
  {journal} {Optics express}\ }\textbf {\bibinfo {volume} {22}},\ \bibinfo
  {pages} {24497} (\bibinfo {year} {2014})}\BibitemShut {NoStop}%
\bibitem [{\citenamefont {Valivarthi}\ \emph {et~al.}(2017)\citenamefont
  {Valivarthi}, \citenamefont {Zhou}, \citenamefont {John}, \citenamefont
  {Marsili}, \citenamefont {Verma}, \citenamefont {Shaw}, \citenamefont {Nam},
  \citenamefont {Oblak},\ and\ \citenamefont {Tittel}}]{valivarthi2017cost}%
  \BibitemOpen
  \bibfield  {author} {\bibinfo {author} {\bibfnamefont {R.}~\bibnamefont
  {Valivarthi}}, \bibinfo {author} {\bibfnamefont {Q.}~\bibnamefont {Zhou}},
  \bibinfo {author} {\bibfnamefont {C.}~\bibnamefont {John}}, \bibinfo {author}
  {\bibfnamefont {F.}~\bibnamefont {Marsili}}, \bibinfo {author} {\bibfnamefont
  {V.~B.}\ \bibnamefont {Verma}}, \bibinfo {author} {\bibfnamefont {M.~D.}\
  \bibnamefont {Shaw}}, \bibinfo {author} {\bibfnamefont {S.~W.}\ \bibnamefont
  {Nam}}, \bibinfo {author} {\bibfnamefont {D.}~\bibnamefont {Oblak}},\ and\
  \bibinfo {author} {\bibfnamefont {W.}~\bibnamefont {Tittel}},\ }\bibfield
  {title} {\bibinfo {title} {A cost-effective measurement-device-independent
  quantum key distribution system for quantum networks},\ }\href@noop {}
  {\bibfield  {journal} {\bibinfo  {journal} {Quantum Science and Technology}\
  }\textbf {\bibinfo {volume} {2}},\ \bibinfo {pages} {04LT01} (\bibinfo {year}
  {2017})}\BibitemShut {NoStop}%
\bibitem [{\citenamefont {Sun}\ \emph {et~al.}(2017)\citenamefont {Sun},
  \citenamefont {Mao}, \citenamefont {Jiang}, \citenamefont {Zhao},
  \citenamefont {Chen}, \citenamefont {Zhang}, \citenamefont {Zhang},
  \citenamefont {Jiang}, \citenamefont {Chen}, \citenamefont {You},
  \citenamefont {Li}, \citenamefont {Huang}, \citenamefont {Chen},
  \citenamefont {Wang}, \citenamefont {Ma}, \citenamefont {Zhang},\ and\
  \citenamefont {Pan}}]{sun2017entanglement}%
  \BibitemOpen
  \bibfield  {author} {\bibinfo {author} {\bibfnamefont {Q.-C.}\ \bibnamefont
  {Sun}}, \bibinfo {author} {\bibfnamefont {Y.-L.}\ \bibnamefont {Mao}},
  \bibinfo {author} {\bibfnamefont {Y.-F.}\ \bibnamefont {Jiang}}, \bibinfo
  {author} {\bibfnamefont {Q.}~\bibnamefont {Zhao}}, \bibinfo {author}
  {\bibfnamefont {S.-J.}\ \bibnamefont {Chen}}, \bibinfo {author}
  {\bibfnamefont {W.}~\bibnamefont {Zhang}}, \bibinfo {author} {\bibfnamefont
  {W.-J.}\ \bibnamefont {Zhang}}, \bibinfo {author} {\bibfnamefont
  {X.}~\bibnamefont {Jiang}}, \bibinfo {author} {\bibfnamefont {T.-Y.}\
  \bibnamefont {Chen}}, \bibinfo {author} {\bibfnamefont {L.-X.}\ \bibnamefont
  {You}}, \bibinfo {author} {\bibfnamefont {L.}~\bibnamefont {Li}}, \bibinfo
  {author} {\bibfnamefont {Y.-D.}\ \bibnamefont {Huang}}, \bibinfo {author}
  {\bibfnamefont {X.-F.}\ \bibnamefont {Chen}}, \bibinfo {author}
  {\bibfnamefont {Z.}~\bibnamefont {Wang}}, \bibinfo {author} {\bibfnamefont
  {X.}~\bibnamefont {Ma}}, \bibinfo {author} {\bibfnamefont {Q.}~\bibnamefont
  {Zhang}},\ and\ \bibinfo {author} {\bibfnamefont {J.-W.}\ \bibnamefont
  {Pan}},\ }\bibfield  {title} {\bibinfo {title} {Entanglement swapping with
  independent sources over an optical-fiber network},\ }\href
  {https://doi.org/10.1103/PhysRevA.95.032306} {\bibfield  {journal} {\bibinfo
  {journal} {Phys. Rev. A}\ }\textbf {\bibinfo {volume} {95}},\ \bibinfo
  {pages} {032306} (\bibinfo {year} {2017})}\BibitemShut {NoStop}%
\bibitem [{\citenamefont {Braunstein}\ and\ \citenamefont
  {Pirandola}(2012)}]{braunstein2012side}%
  \BibitemOpen
  \bibfield  {author} {\bibinfo {author} {\bibfnamefont {S.~L.}\ \bibnamefont
  {Braunstein}}\ and\ \bibinfo {author} {\bibfnamefont {S.}~\bibnamefont
  {Pirandola}},\ }\bibfield  {title} {\bibinfo {title} {Side-channel-free
  quantum key distribution},\ }\href@noop {} {\bibfield  {journal} {\bibinfo
  {journal} {Physical review letters}\ }\textbf {\bibinfo {volume} {108}},\
  \bibinfo {pages} {130502} (\bibinfo {year} {2012})}\BibitemShut {NoStop}%
\bibitem [{\citenamefont {Gottesman}\ and\ \citenamefont
  {Chuang}(1999)}]{gottesman1999demonstrating}%
  \BibitemOpen
  \bibfield  {author} {\bibinfo {author} {\bibfnamefont {D.}~\bibnamefont
  {Gottesman}}\ and\ \bibinfo {author} {\bibfnamefont {I.~L.}\ \bibnamefont
  {Chuang}},\ }\bibfield  {title} {\bibinfo {title} {Demonstrating the
  viability of universal quantum computation using teleportation and
  single-qubit operations},\ }\href@noop {} {\bibfield  {journal} {\bibinfo
  {journal} {Nature}\ }\textbf {\bibinfo {volume} {402}},\ \bibinfo {pages}
  {390} (\bibinfo {year} {1999})}\BibitemShut {NoStop}%
\end{thebibliography}%

%%%%%%%%%% If preparing manually:
% \newcommand{\enquote}[1]{``#1''}

% \bibitem{Zhang:14}
% Y.~Zhang, S.~Qiao, L.~Sun, Q.~W. Shi, W.~Huang, L.~Li, and Z.~Yang,
%   \enquote{Photoinduced active terahertz metamaterials with nanostructured
%   vanadium dioxide film deposited by sol-gel method,}
%   {\protect\JournalTitle{Optics Express}} \textbf{22}, 11070--11078 (2014).

% {Optical Society}, \enquote{{OSA Publishing},}
%   \url{http://www.osapublishing.org}.

% \bibitem{FORSTER2007}
% P.~Forster, V.~Ramaswamy, P.~Artaxo, T.~Bernsten, R.~Betts, D.~Fahey,
%   J.~Haywood, J.~Lean, D.~Lowe, G.~Myhre, J.~Nganga, R.~Prinn, G.~Raga,
%   M.~Schulz, and R.~V. Dorland, \enquote{Changes in atmospheric consituents and
%   in radiative forcing,} in \enquote{Climate Change 2007: The Physical Science
%   Basis. Contribution of Working Group 1 to the Fourth assesment report of
%   Intergovernmental Panel on Climate Change,}  S.~Solomon, D.~Qin, M.~Manning,
%   Z.~Chen, M.~Marquis, K.~B. Averyt, M.~Tignor, and H.~L. Miler, eds.
%   (Cambridge University Press, 2007).

% \end{thebibliography}


\end{document}